\documentclass[11pt,draftcls,onecolumn]{IEEEtran}

\usepackage{amsmath,amsfonts,amssymb,amsthm}
\usepackage{mathrsfs}
\usepackage{comment,blkarray}
\usepackage{multirow,bigdelim}
\usepackage{mathrsfs}
\usepackage{cite}
\usepackage[ruled,commentsnumbered, vlined]{algorithm2e}
\usepackage{tikz}
\usepackage[latin1]{inputenc}
\usepackage{verbatim}
\usepackage{graphicx}
\usepackage{booktabs}
\usepackage{caption}													
\usepackage{subcaption}									

\includecomment{itwfull}
\excludecomment{removeEX4}
\excludecomment{itw2016}
\excludecomment{journalonly}

\usepackage{url}
\usepackage{multicol}

\usepackage{ifpdf}															%
\ifpdf																		%
	\usepackage{hyperref}
\else																		%
\fi	
\usetikzlibrary{arrows,decorations.pathmorphing,decorations.footprints,fadings,calc,
trees,mindmap,shadows,decorations.text,patterns,positioning,shapes,matrix,fit}

\makeatletter

\newcommand{\Rmnum}[1]{\expandafter\@slowromancap\romannumeral #1@}

\newif\if@borderstar
\def\bordermatrix{\@ifnextchar*{%
  \@borderstartrue\@bordermatrix@i}{\@borderstarfalse\@bordermatrix@i*}%
}
\def\@bordermatrix@i*{\@ifnextchar[{\@bordermatrix@ii}{\@bordermatrix@ii[()]}}
\def\@bordermatrix@ii[#1]#2{%
\begingroup
  \m@th\@tempdima8.75\p@\setbox\z@\vbox{%
    \def\cr{\crcr\noalign{\kern 2\p@\global\let\cr\endline }}%
    \ialign {$##$\hfil\kern 2\p@\kern\@tempdima & \thinspace %
    \hfil $##$\hfil && \quad\hfil $##$\hfil\crcr\omit\strut %
    \hfil\crcr\noalign{\kern -\baselineskip}#2\crcr\omit %
    \strut\cr}}%
  \setbox\tw@\vbox{\unvcopy\z@\global\setbox\@ne\lastbox}%
  \setbox\tw@\hbox{\unhbox\@ne\unskip\global\setbox\@ne\lastbox}%
  \setbox\tw@\hbox{%
    $\kern\wd\@ne\kern -\@tempdima\left\@firstoftwo#1%
    \if@borderstar\kern2pt\else\kern -\wd\@ne\fi%
    \global\setbox\@ne\vbox{\box\@ne\if@borderstar\else\kern 2\p@\fi}%
    \vcenter{\if@borderstar\else\kern -\ht\@ne\fi%
    \unvbox\z@\kern -\if@borderstar2\fi\baselineskip}%
    \if@borderstar\kern -2\@tempdima\kern2\p@\else\,\fi\right\@secondoftwo#1 $%
  }\null \;\vbox{\kern\ht\@ne\box\tw@}%
\endgroup
}
\makeatother

\newtheorem{thm}{Theorem}
\newtheorem{lemma}[thm]{Lemma}
\newtheorem{cor}[thm]{Corollary}

\newtheorem{prop}[thm]{Proposition}
\newtheorem{example}{Example}
\newtheorem{defn}{Definition}
\newtheorem{remark}{Remark}
\allowdisplaybreaks

\newcommand{\ein}{\mathcal{E}_{\mathrm{i}}}
\newcommand{\eout}{\mathcal{E}_{\mathrm{o}}}
\newcommand{\tail}{\mathrm{tail}}
\newcommand{\head}{\mathrm{head}}
\newcommand{\mincut}{\textrm{min-cut}}

\newcommand{\mC}{\mathcal{C}}
\newcommand{\mbC}{\mathbf{C}}
\newcommand{\mA}{\mathcal{A}}
\newcommand{\mB}{\mathcal{B}}
\newcommand{\mN}{\mathcal{N}}
\newcommand{\mE}{\mathcal{E}}
\newcommand{\mO}{\mathcal{O}}
\newcommand{\mP}{\mathcal{P}}
\newcommand{\mL}{\mathcal{L}}
\newcommand{\Fq}{\mathbb{F}_q}
\newcommand{\Fqn}{\mathbb{F}_{q^n}}
\newcommand{\cl}{{\mathrm{cl}}}
\newcommand{\Cl}{{\mathrm{Cl}}}

\newcommand{\Rank}{{\mathrm{Rank}}}

\tikzstyle{vertex}=[draw,circle,fill=gray!30,minimum size=6pt, inner sep=0pt]

\begin{document}
%
\title{Improved Upper Bound on the Network Function Computing Capacity}

\author{Xuan~Guang,~Raymond~W.~Yeung,~Shenghao Yang,~and~Congduan Li
\thanks{This paper was presented in part at 2016 IEEE Information
    Theory Workshop (ITW), Cambridge, UK.}}



\maketitle

\begin{abstract}
The problem of network function computation over a directed acyclic network is investigated in this paper. In such a network, a sink node desires to compute with zero error a {\em target function}, of which the inputs are generated at multiple source nodes. The edges in the network are assumed to be error-free and have limited capacity. The nodes in the network are assumed to have unbounded computing capability and be able to perform network coding. The {\em computing rate} of a network code that can compute the target function over the network is the average number of times that the target function is computed with zero error for one use of the network. In this paper, we obtain an improved upper bound on the computing capacity, which is applicable to arbitrary target functions and  arbitrary network topologies. This improved upper bound not only is an enhancement of the previous upper bounds but also is the first tight upper bound on the computing capacity for computing an arithmetic sum over a certain non-tree network, which has been widely studied in the literature. We also introduce a multi-dimensional array approach that facilitates evaluation of the improved upper bound. Furthermore, we apply this bound to the problem of computing a vector-linear function over a network. With this bound, we are able to not only enhance a previous result on computing a vector-linear function over a network but also simplify the proof significantly. Finally, we prove that for computing the binary maximum function over the reverse butterfly network, our improved upper bound is not achievable. This result establishes that in general our improved upper bound is non achievable, but whether it is asymptotically achievable or not remains open.
\end{abstract}

\IEEEpeerreviewmaketitle

\section{Introduction}

In this paper, we consider the problem of function computation over a directed acyclic communication network, called {\it network function computation}. A general setup of the problem can be as follows. A directed acyclic graph is used to model a communication network, where the edges model the communication links (noiseless or noisy) with capacity constraints and the nodes are assumed to have unlimited computing capability and infinite storage. In such a network, a set of nodes, referred to as the {\it source nodes}, generate possibly correlated messages, while another set of nodes, referred to as the {\it sink nodes}, are required to compute possibly different functions of the source messages with fidelity constraints. In particular, the network transmission problem, where the sink nodes are required to reconstruct certain subsets of the source messages, is a special case of network function computation with the sink nodes computing the corresponding identity functions.

The straightforward approach to network function computation is to transmit the required source messages to the sink nodes over the network and then compute the desired functions at the sink nodes. Instead of first transmitting the source messages to the sink nodes, network function computation can in general be done more efficiently in a distributed manner by means of network coding \cite{flow}. In recent years, network function computation has received considerable attention due to its important applications in sensor networks \cite{Giridhar05, Kowshik-Kumar-TIT12}, Big Data processing \cite{Infocom16}, Internet of Things (IoT) \cite{IEEEAccess16}, machine learning~\cite{IEEEAccess16},~etc.

\subsection{Related Works}

From the information theoretic point of view,\footnote{This problem has also been studied widely from the computational complexity point of view (e.g., \cite{Yao-STOC-1979,Ahlswede-Cai-TIT94,book-CommComp-1997}).} we are interested in the achievable rate region for the sink nodes to reliably compute their desired functions over the network. However, the problem with the general setup described in the foregoing is very difficult, because it encompasses various topics in information theory, including multi-terminal source coding, multi-terminal channel coding, network coding, separation of these three types of coding, etc. There are well-known open problems in each of these topics. We refer the reader to the comprehensive book by El~Gamal	and Kim \cite{EIGamal_Kim_12b}. The overwhelming complexity and difficulty of network function computation necessitate the consideration of different simplifications of the setup in order to be able to make progress.

One simplification is to consider the problem under the setting that the messages generated by the source nodes are correlated but the network topology is very simple. This line of research can be traced back to Shannon's seminal works in which the transmission of a source message over a point-to-point channel was discussed. These include the classical source coding theorem \cite{Shannon48}, channel coding theorem \cite{Shannon48}, separation of source coding and channel coding \cite{Shannon48}, and rate-distortion theorem \cite{Shannon59-rate-distorsion} (see also~\cite{Cover91b}). Witsenhausen~\cite{Witsenhausen-IT-76} considered the source coding problem with side information at the decoder. In this model, the encoder compresses a source variable $X$. The decoder, in addition to receiving the output of the encoder, also observes a source variable $Y$ which is correlated with $X$. The decoder is required to reconstruct $X$ with zero error. Orlitsky and Roche \cite{Orlitsky-Roche-IT01-Coding_Compu} generalized  Witsenhausen's model by requiring the decoder to compute an arbitrary function of $X$ and $Y$.

Multi-terminal source coding was launched by Slepian and Wolf \cite{Slepian-Wolf-IT73}, in which the following model was considered. Two correlated sources are compressed separately by two encoders. The decoder, which receives the output of both encoders, is required to reconstruct the two sources almost perfectly. Building on the Slepian-Wolf model, K\"{o}rner and Marton~\cite{Korner-Marton-IT73} investigated the computation of the modulo $2$ sum of two correlated binary sources.
To our knowledge, the K\"{o}rner-Marton problem was the first non-identity function computation problem over a network. Doshi~\textit{et al.}~\cite{doshi10} generalized the K\"{o}rner-Marton model by requiring the decoder to compute an arbitrary function of two correlated sources. More recently, Feizi and M\'{e}dard~\cite{Feizi-Medard-netw_funct_compre-IT-2014} investigated the computation of an arbitrary function of multiple correlated sources over a tree network. In such a network, the leaf nodes are the source nodes where correlated sources are generated, and the root node is the unique sink node where an arbitrary function of the sources is computed.

Another simplification is to consider network function computation under the setting that the messages generated by all the source nodes are mutually independent but the network topology can be general (an arbitrary directed network). Under this setting, when the sink nodes are required to reconstruct different subsets of the source messages, the network function computation problem degenerates to network coding \cite{flow,linear,alg} (see also \cite{Zhang-book,yeung08b}). For single-source network coding, i.e., the message generated by the single source node is required to be transmitted to every sink node, the capacity is completely characterized by a max-flow min-cut bound theorem \cite{flow}, and linear network coding is sufficient to achieve the capacity \cite{linear,alg}. For multi-source network coding, i.e., the source nodes generate mutually independent messages and each one of them is multicast to a certain subset of the sink nodes, the capacity region can only be characterized implicitly in terms of achievable entropy functions when the network is acyclic \cite{Yan-Yeung-Zhang-IT12-NC-region}. More explicit characterizations of the capacity region for some special cases can be found in \cite{Yan-Yang-Zhang-IT06-NetSharingBound, DFZ-IT07, Wang-Shroff-IT10-PairwInters-NC, Chan-IT16-Cut-setbounds-NC,Kamath-IT17-2unicast}.

To our knowledge, the first non-identity function computation problem over a directed acyclic network is the following so-called {\it sum-network} problem~\cite{Koetter-CISS2004,Ramamoorthy-ISIT08-sum-networks, Ramamoorthy-Langberg-JSAC13-sum-networks, Shenvi-Dey-ISIT10_3-3-sum-networks, 
Rai-Dey-TIT-2012, Rai-Das-IEEEComm13}. 
In a directed acyclic network, the multiple sink nodes are required to compute an algebraic sum of the messages observed by all the source nodes over a finite field (e.g., the foregoing modulo $2$ sum is an algebraic sum over the finite field $\mathbb{F}_2$). When there exists only one sink node, linear network coding achieves the computing capacity \cite{Koetter-CISS2004}. Ramamoorthy \cite{Ramamoorthy-ISIT08-sum-networks} first proved that if the number of source nodes and the number of sink nodes are at most $2$, all the sink nodes can compute the algebraic sum of the source messages with zero error by using scalar linear network coding if and only if there exists a directed path for every pair of source and sink nodes. Subsequently, Ramamoorthy and Langberg \cite{Ramamoorthy-Langberg-JSAC13-sum-networks} proved that if there are $3$ source nodes and $3$ sink nodes in the network, the existence of a single path for every pair of source and sink nodes is in general not sufficient for computing the algebraic sum of the source messages by using the network only once.\footnote{Using the network once means that each edge is used at most once.} Instead, it is sufficient if every pair of source and sink nodes can be connected by $2$ edge-disjoint paths.\footnote{The similar results were obtained independently by Shenvi and Dey \cite{Shenvi-Dey-ISIT10_3-3-sum-networks}.} However, Rai and Das~\cite{Rai-Das-IEEEComm13} showed by a counterexample that even this condition is not always sufficient if there are $7$ source nodes and $7$ sink nodes in the network.

In \cite{Appuswamy11,Appuswamy13,Appuswamy14,huang15}, the following network function computation model was considered. In a directed acyclic network, the single sink node is required to compute with zero error a function of the source messages separately observed by multiple source nodes. The network topology and the function are arbitrary. Appuswamy~\textit{et~al.}~\cite{Appuswamy11} investigated the fundamental {\it computing capacity}, i.e., the maximum average number of times that the function can be computed with zero error for one use of the network, and gave a cut-set based upper bound that is valid under certain constraints on either the network topology or the target function. Huang~\textit{et~al.}~\cite{huang15} obtained an enhancement of Appuswamy~\textit{et~al.}'s upper bound that can be applied for arbitrary functions and arbitrary network topologies. Specifically, for the case of computing an arbitrary function of the source messages over a multi-edge tree network and the case of computing the identity function or the algebraic sum function of the source messages over an arbitrary network topology, the above two upper bounds coincide and are tight (see \cite{Appuswamy11} and \cite{huang15}). However, both of these bounds are in general quite loose. Building on this model, Appuswamy~\textit{et~al.}~\cite{Appuswamy13} introduced the notions of routing, linear, and nonlinear computing capacities that respectively correspond to performing routing operations, linear network coding and nonlinear network coding at the nodes, and then compared the three different computing capacities. Recently, Appuswamy and Franceschetti~\cite{Appuswamy14} investigated the solvability (rate-$1$ achievable) of linear network codes when the single sink node is required to compute a vector-linear function of the source messages over a directed acyclic network.

\subsection{Contributions and Organization of the Paper}

In this paper, we consider the network function computation model discussed in \cite{Appuswamy11,Appuswamy13,Appuswamy14,huang15,Guang_NFC_ITW16}. To be specific, in a directed acyclic network, a single sink node is required to compute with zero error a function, called the {\it target function}, of which the arguments are the source messages generated by the multiple source nodes. The edges in the network are assumed to be error-free and have limited (unit) capacity. The nodes in the network are assumed to have unlimited computing capability and perform network coding, i.e., each node can encode the messages it receives or generates and then transmit the output of the encoding function. From the information-theoretic point of view, we are interested in the fundamental computing capacity, which is the average number of times that the target function can be computed with zero error for one use of the network.

One main contribution of this work is an improved upper bound on the computing capacity, which is applicable to arbitrary target functions and arbitrary network topologies. Our improved upper bound not only is an enhancement of the previous upper bounds (cf.~\cite{Appuswamy11,huang15}), but also is the first tight upper bound on the computing capacity for computing an arithmetic sum over a certain ``non-tree'' network (cf.~Example~\ref{eg:1} in Section~\ref{sec:preliminaries} of the current paper).\footnote{This example, first introduced by Appuswamy \textit{et al.} in \cite{Appuswamy11}, is used to show that both their upper bound and lower bounds proposed are not always tight and illustrate the combinatorial nature of the computing problem.}

An important application of our improved upper bound is in computing a vector-linear function of the source messages over an arbitrary directed acyclic network, which has been considered by Appuswamy and Franceschetti~\cite{Appuswamy14}. One of the main results in \cite{Appuswamy14} is that the {\it min-cut condition} (cf.~\cite{Appuswamy14} or Section~\ref{subsec:computing_linear_func} of the current paper), inherited from network coding, is not always sufficient for computing a vector-linear function over a network by using any rate-$1$ {\em linear} network code. The proof of this result in \cite{Appuswamy14} is rather complicated and relies on the use of some advanced algebraic tools. In contrast, by applying our improved upper bound, we can provide a simple proof of the stronger result that the min-cut condition is not always sufficient to compute a vector-linear function over a network by using any rate-$1$ network code (linear or nonlinear).

For all previously considered network function computation problems whose computing capacities are known, our improved upper bound is achievable if the computing capacity is rational, or is asymptotically achievable if the computing capacity is irrational. Another main contribution of this work is to prove that for computing the binary maximum function over the ``reverse butterfly network'', our improved upper bound is not achievable. Here, a novel network splitting approach is used to prove this result and the proof is highly nontrivial.

The paper is organized as follows. In Section~\ref{sec:preliminaries}, we formally present the network function computation model considered throughout the paper and the existing upper bounds on the computing capacity, and then give an example that suggests how the existing upper bounds can be improved. The improved upper bound is stated in Section~\ref{sec:improved_upp_bou}, followed by two discussions. The first is about evaluating the improved upper bound by using multi-dimensional arrays. The second is an application of the improved upper bound to enhance a result in \cite{Appuswamy14} on computing a vector-linear function over a network, as discussed in the foregoing. Section~\ref{sec:proof} is devoted to the proof of the improved upper bound.  We show in Section~\ref{sec:non_tight} that the improved upper bound for computing the binary maximum function over the reverse butterfly network is not achievable. In Section~\ref{sec:concl}, we conclude with a summary of our results and a remark on future research.

\begin{itw2016}
\textbf{ITW Introduction} In Big Data processing, sensor networks~\cite{Giridhar05} and many
other scenarios, a target function is calculated repeatedly crossing a
network, where the input symbols of the function are generated
at multiple source nodes, and the function value (output) is calculated
cooperatively by all network nodes and desired at a sink node. The bottleneck of the capacity of the network function computation is the network bandwidth, rather than (or in addition to) the computing capability of the network nodes. This network function
computation problem is a generalization of the network communication
problem, where the latter is just considered as the computation of the
\emph{identity function}.

Various models and special cases of the network function computation
problem have been studied in the literature. One line of related works
studies functional data compression, where the input symbols are
generated by the source nodes according to a joint distribution, and the
sink node, directly and reliably connected with all the source nodes,
recovers a function of the input symbols (see references in
\cite{doshi10}). For a plenty of settings, the achievable rate region of
functional data compression is related to graph entropy, which is in
general difficult to be characterized in a single-letter form.

We are interested in the \emph{network coding model} for network function
computation, which has been studied in
\cite{Appuswamy11,Kowshik12,Rai12,Ramam13,Appuswamy14,huang15}. Specifically,
we consider a \emph{directed acyclic network} where the network links
have limited (unit) capacity and are error-free. Each source node
generates multiple input symbols, and the network codes can perform {\em vector
network coding} by using the network multiple times, where one use of a network means the use of each link at most once. Each intermediate network node is assumed to have the unbounded computing ability, and can transmit the output of a certain fixed function of
the symbols it receives. We do not assume any particular distribution
on the input symbols and require computing the target function correctly for all input combinations. The \emph{computing rate} is the average number of times that the target function can be computed correctly for one use of the network. The maximum computing rate is called the \emph{computing capacity}.

When computing the identity function, the problem degenerates to the
extensively studied network coding problem \cite{flow}, and it is known
that in general linear network codes are sufficient to achieve the
multicast capacity when each sink node requires all sources \cite{linear, alg}. For linear target functions over a finite field, a complete characterization of the computing
capacity is not available for networks with one sink node. Certain
necessary and sufficient conditions have been obtained for sufficiency of linear network codes in calculating a linear target function \cite{Ramam13, Appuswamy14}. But linear network codes are in general not sufficient to achieve the capacity of linear
target functions \cite{Rai12}.

Networks with a single sink node are discussed in this paper, while
both the target function and the network code can be nonlinear. In
this case, the computing capacity is known when the network is a
multi-edge tree \cite{Appuswamy11} or when the target function is the
identity function. For the general case, lower and upper bounds on
the computing capacity based on cut sets have been studied
\cite{Appuswamy11,Kowshik12,huang15}. In this paper, we characterize a
general upper bound on the computing capacity, which is
strictly better than the existing ones.

For the best existing upper bound, Huang \textit{et al.} \cite{huang15} define
an equivalence relation associated with the subsets of the inputs of
the target function 
and propose a cut-set bound on the computing capacity by counting the
number of the equivalence classes related to each cut set. This upper
bound is not tight in general, which has been demonstrated by a counter example. In this paper, by examining partitions of a cut set, we find that
certain different inputs that are in the same equivalence class of
Huang \textit{et al.}, should be represented by different transmission messages on the cut set. The cut set partitions motivate us to define another equivalence relation
and then derive a better upper bound. We also show via an example that our bound is strictly better than the one in \cite{huang15}.
\end{itw2016}

\section{Model and Preliminaries}\label{sec:preliminaries}

\subsection{Network Function Computation Model}
\label{sec:net-funct-comp-model}

Let $G=(\mathcal{V},\mathcal{E})$ be a directed acyclic graph with a finite vertex set $\mathcal{V}$ and an edge set $\mathcal{E}$,
where multiple edges are allowed between two nodes. A {\em
  network} over $G$ is denoted by $\mathcal{N}=(G,S,\rho)$, where
$S\subset \mathcal{V}$ is the set of {\em source nodes}, say $S=\{\sigma_1, \sigma_2, \cdots, \sigma_s \}$ with $|S|=s$, and $\rho\in
\mathcal{V}\backslash S$ is the single {\em sink node}. The tail and the head of an edge $e$ are denoted by $\tail(e)$ and $\head(e)$, respectively. Moreover, for each node $u\in\mathcal{V}$, let $\ein(u) = \{e\in \mathcal{E}: \head(e)=u\}$ and $\eout(u)=\{e\in\mathcal{E}:\tail(e)=u\}$, both of which are the set of incoming edges and the set of outgoing edges of $u$, respectively. Without loss of generality, we assume that every source node has no incoming edges, because otherwise we can introduce a new source node and install a directed edge from the new source node to the original source node which is now regarded as a non-source node. We further assume that there exists a directed path from every node $u\in \mathcal{V}\setminus \{\rho\}$ to $\rho$ in $G$. Then it follows from the acyclicity of $G$ that the sink node $\rho$ has no outgoing edges. Let $\mathcal{B}$ be a finite alphabet, and we assume that a symbol in $\mathcal{B}$ can be transmitted on each edge reliably for each use.

Let $\mathcal{A}$ and $\mathcal{O}$ be finite alphabets, and $f:\mathcal{A}^s\to \mathcal{O}$ be the {\em target function}. For the target function $f$, the $i$th argument is generated at the $i$th source node $\sigma_i$ and all outputs of the function are demanded by the sink node $\rho$. We will compute $f$ over the network $\mN$ by using the network multiple times. Computation units with unbounded computing capability are available at all nodes in the network. However, the computing capability of the whole network is constrained by the network transmission capability.

Assume that the $i$th source node $\sigma_i$ generates $k$ symbols in $\mA$, denoted by
$\vec{x}_i=(x_{i,1},x_{i,2},\cdots,x_{i,k})^\top$, which is called the
\textit{source vector} generated by $\sigma_i$. The symbols generated
by all the source nodes constitute the \textit{source matrix}
$\vec{x}_S=(\vec{x}_{1}, \vec{x}_{2}, \cdots, \vec{x}_{s})$ of size
$k\times s$. Let
\begin{equation*}
  f(\vec{x}_S) = \big(f(x_{1,j},x_{2,j},\cdots,x_{s,j}):\ j=1,2,\ldots,k\big)^{\top}
\end{equation*}
be the $k$ outputs of the target function $f$ corresponding to the $k$ inputs of the source nodes. For any subset $J\subseteq S$, we let
$\vec{x}_J=(\vec{x}_{i}: \sigma_i\in J)$ and use $\mA^{k\times J}$ (instead of $\mA^{k\times |J|}$ for simplicity) to denote the set of all possible $k\times |J|$ matrices taken by $\vec{x}_J$.
In particular, for $k=1$, we omit the symbol ``\;$\vec{\cdot}$\;'' for notational simplicity, e.g., $x_J\in \mA^J$. Moreover, whenever we write $\vec{x}_J$ as $x_J$, we implicitly assume that $k=1$. Throughout this paper, we adopt the convention that $\mA^0$ is the singleton that contains an empty vector of dimension $0$ taking value in $\mA$. As such, for $J=\emptyset$, we have $\mA^J=\mA^{|J|}=\mA^0$. It also follows that for $J=\emptyset$, $\mA^{k\times J}=\mA^{k\times |J|}=(\mA^k)^0$.

For two positive integers $k$ and $n$, a $(k,n)$ \textit{(function-computing) network code} over the network $\mathcal{N}$ with the target function $f$ is defined as follows. Let $\vec{x}_S\in \mA^{k\times S}$ be the source matrix generated by all the source nodes. The purpose of such a network code is to compute $f(\vec{x}_S)$ by transmitting at most $n$ symbols in $\mB$ on each edge in $\mathcal{E}$, i.e., using the network at most $n$ times. A $(k,n)$ (function-computing) network code consists of a {\em local encoding function} $\theta_{e}$ for each edge $e$, where
\begin{equation}\label{defn_local_function}
  \theta_{e}:
  \begin{cases}
    \qquad \mA^k \rightarrow \mathcal{B}^n, & \text{if }\ e\in \eout(\sigma) \text{ for some $\sigma\in S$}; \\
    \prod\limits_{d\in \ein(\mathrm{tail}(e))} \mathcal{B}^n \rightarrow
    \mathcal{B}^n, & \text{otherwise.}
  \end{cases}
\end{equation}
With the encoding mechanism as described, the local encoding functions $\theta_{e}$, $e\in \mathcal{E}$ derive recursively the symbols transmitted over all edges $e$, denoted by $g_{e}(\vec{x}_S)$, which can be considered as vectors in $\mB^n$. Specifically, if $e$ is an outgoing edge of the $i$th source node $\sigma_{i}$, then
$g_{e}(\vec{x}_S) = \theta_{e}(\vec{x}_i)$;
if $e$ is an outgoing edge of some non-source node $u$ in $\mathcal{V}$, then
$g_{e}(\vec{x}_S) = \theta_{e}\big(g_{\ein(u)}(\vec{x}_S)\big)$. Similar to the classical network codes (see \cite{Zhang-book, yeung08b}), for each edge $e$, we call $g_{e}$ the {\em global encoding function} for $e$. For an edge set $E\subset\mathcal{E}$, we let
$$g_{E}(\vec{x}_S)=\big( g_{e}(\vec{x}_S):\ e\in E \big).$$

Furthermore, the $(k,n)$ network code consists of a decoding function
\begin{equation*}
  \varphi: \prod_{e\in \ein(\rho)} \mathcal{B}^n \rightarrow
    \mathcal{O}^k
\end{equation*}
at the sink node $\rho$. Define
$\psi(\vec{x}_S) = \varphi\big(g_{\ein(\rho)}(\vec{x}_S)\big)$.
If the network code can {\em compute} $f$, i.e., $\psi(\vec{x}_S)=f(\vec{x}_S)$ for all source matrices $\vec{x}_S\in\mA^{k\times S}$, then $\frac{k}{n}\log_{|\mathcal{B}|}|\mathcal{A}|$ is called an {\em achievable
computing rate}. Further, a nonnegative real number $r$ is called {\it asymptotically achievable} if $\forall~\epsilon>0$, there exists a $(k,n)$ network code that can compute $f$ such that
\begin{align*}
\frac{k}{n}\log_{|\mathcal{B}|}|\mathcal{A}|>r-\epsilon.
\end{align*}
Clearly, any achievable computing rate must be asymptotically achievable. The {\em rate region} for computing $f$ over $\mN$ is defined as
\begin{align}\label{def:rate_region}
\mathfrak{R}(\mN, f)=\Big\{ r:\ r \text{ is asymptotically achievable for computing $f$ over $\mN$} \Big\},
\end{align}
which is evidently closed and bounded. The {\em computing capacity} of the network $\mN$ with respect
to the target function $f$ is defined as
\begin{align}\label{def:comp_cap}
 \mC(\mN,f)=\max~\mathfrak{R}(\mN, f).
\end{align}
Without loss of generality, we assume throughout the paper that $\mA=\mB$, so that $\frac{k}{n}\log_{|\mathcal{B}|}|\mathcal{A}|$ in the above is simplified to $\frac{k}{n}$. Although the definition of the computing capacity $\mC(\mN,f)$ here is a little different from the one used in \cite{Appuswamy11} and \cite{huang15}, i.e.,
$\sup\big\{ k/n:\ k/n\textrm{ is achievable} \big\}$, it is easy to see that they are equivalent. Our definition has the advantage that it is more consistent with the usual concept of rate region in information theory problems. In this paper we are interested in general upper bounds on $\mC(\mN,f)$, where ``general'' means that the upper bounds are applicable to arbitrary network $\mN$ and arbitrary function~$f$.

\subsection{Existing Upper Bounds}\label{subsec:pre_B}

Let us first discuss a simple upper bound. For two nodes $u$ and $v$ in $\mathcal{V}$, if there exists a directed path from $u$ to $v$ in $G$, denote this relation by $u\rightarrow v$. If there exists no such directed path from $u$ to $v$, we say that $u$ is
\emph{separated} from $v$. Given a set of edges $C\subseteq
\mathcal{E}$, define $I_C$ as the set of the source nodes that are separated from the sink node $\rho$ if $C$ is deleted from $\mathcal{E}$, i.e.,
\begin{align*}
I_C=\left\{ \sigma\in S:\ \sigma \text{ is separated from } \rho \text{ upon deleting the edges in $C$ from $\mathcal{E}$} \right\}.
\end{align*}
Equivalently, $I_C$ is the set of source nodes from which all directed paths to the sink node $\rho$ pass through~$C$. For two cut sets $C_1$ and $C_2$ in $\Lambda(\mathcal{N})$, it is clear that $I_{C_i}\subseteq I_{C_1\cup\, C_2}$, $i=1,2$. Thus,
\begin{align}\label{eq:I_C_subseteq}
I_{C_1}\cup I_{C_2}\subseteq I_{C_1\cup\, C_2}.
\end{align}
However, $I_{C_1}\cup I_{C_2}\neq I_{C_1\cup\, C_2}$ in general.

An edge set $C$ is said to be a {\em cut set} if $I_C\neq \emptyset$, and let $\Lambda(\mathcal{N})$ be the family of all cut sets in the network~$\mathcal{N}$, i.e.,
$$\Lambda(\mathcal{N})=\{ C\subseteq \mE:\ I_C \neq \emptyset \}.$$
In particular, we say a cut set $C$ with $I_C=S$ as a {\em global cut set}.

Denote by $f(\mA^s)$ the set of all possible images of $f$ on $\mathcal{O}$, i.e.,
\begin{align*}
f(\mA^s)=\big\{o\in \mO:\ o=f(x_S) \text{ for some } x_S\in A^S \big\}.
\end{align*}
A $(k,n)$ network code that can compute $f$ has to distinguish all images in $f(\mA^s)$ on every global cut set $C$. We elaborate this as follows. Let $\{g_e: e\in \mE \}$ be the set of all global encoding functions of a given $(k,n)$ network code. By the acyclicity of $G$, since $C$ is a global cut set, $g_{\ein(\rho)}(\vec{x}_S)$ is a function of $g_C(\vec{x}_S)$. For any two source matrices $\vec{a}_S$ and $\vec{b}_S$ in $\mA^{k\times S}$, if $f(\vec{a}_S)\neq f(\vec{b}_S)$, then $g_C(\vec{a}_S)\neq g_C(\vec{b}_S)$, because otherwise we have $g_{\ein(\rho)}(\vec{a}_S)=g_{\ein(\rho)}(\vec{b}_S)$, a contradiction to the assumption that this $(k,n)$ network code can compute $f$ over $\mN$. Hence, the following inequality is satisfied:
\begin{align*}
|\mA|^{n\cdot|C|}\geq |f(\mA^s)|^k.
\end{align*}
This implies the following upper bound (also see \cite[Proposition~2]{huang15}):
\begin{equation}\label{eq:1}
  \mathcal{C}(\mathcal{N},f) \leq \min_{C\in
    \Lambda(\mathcal{N}): I_C=S}
  \frac{|C|}{\log_{|\mathcal{A}|}|f(\mathcal{A}^s)|}.
\end{equation}


In \cite{Appuswamy11}, Appuswamy~\textit{et~al.} gave a proof of an enhanced upper bound by considering an equivalence relation defined on the input vectors of the target function $f$ with respect to the cut sets. However, it was subsequently pointed out by Huang~\textit{et~al.}~\cite{huang15} that the proof in \cite{Appuswamy11} is incorrect and in fact the claimed upper bound is valid only for either arbitrary target functions but special network topologies or arbitrary network topologies but special target functions. Instead, they fixed the upper bound in \cite{Appuswamy11} by modifying the equivalence relation considered in \cite{Appuswamy11}.  However, it was pointed out in \cite{huang15} that this enhanced upper bound is not tight for an example first studied in \cite{Appuswamy11}. A main contribution of this work is a further enhanced upper bound that is tight for this example. These will be discussed in detail in the rest of the paper.

Next, we review the upper bound obtained in \cite{huang15}. Define a set $K_C$ for a cut set $C\in \Lambda(\mN)$ as
\begin{align}\label{def_K_C}
K_C=\left\{ \sigma\in S:\ \exists\ e\in C \text{ s.t. } \sigma\rightarrow\tail(e) \right\}.
\end{align}
Recall that $u\rightarrow \rho$ for all $u\in\mathcal{V}\setminus\{\rho\}$. In particular, $\tail(e)\rightarrow \rho$ for all $e\in C$. Then we can easily see that $K_C$ is the set of source nodes from which there exists a directed path to the sink node $\rho$ that passes through $C$. Evidently, $I_C\subseteq K_C$. Further, let $J_C=K_C\backslash I_C$, and hence $K_C=I_C\cup J_C$ and $I_C \cap J_C=\emptyset$. Note that once $C$ is given, $K_C$, $I_C$ and $J_C$ are determined.

For notational convenience in the rest of the paper, we suppose that the argument of the target function $f$ with subscript $i$ always stands for the symbol generated by the $i$th source node $\sigma_i$, so that we can ignore the order of the arguments of $f$. For example, let $S=\{\sigma_1, \sigma_2, \sigma_3, \sigma_4\}$, $I=\{\sigma_2, \sigma_4\}$, and $J=\{\sigma_1, \sigma_3\}$ (clearly, $S=I\cup J$ and $I\cap J=\emptyset$). Then we regard $f(x_I, x_J)$ and $f(x_J, x_I)$ as being the same as $f(x_S)$, i.e.,
\begin{align*}
f(x_2,x_4,x_1,x_3)=f(x_1,x_3,x_2,x_4)=f(x_1,x_2,x_3,x_4).
\end{align*}
This abuse of notation should cause no ambiguity and would greatly simplify the notation.


\begin{defn}\label{def:ec}
Consider two disjoint sets $I,J\subseteq S$ and a fixed $\vec{a}_{J}\in \mA^{k\times J}$ for a positive integer $k$. For any $\vec{b}_I, \vec{b}'_I \in \mA^{k\times I}$, we say $\vec{b}_I$ and $\vec{b}'_I$ are $(I,\vec{a}_J)$-equivalent if
\begin{align*}
f(\vec{b}_I, \vec{a}_{J}, \vec{d})=f(\vec{b}'_I, \vec{a}_{J},\vec{d}),\quad \forall\ \vec{d} \in \mA^{k\times S\setminus (I\cup J)}.
\end{align*}
\end{defn}


We remark that Definition~\ref{def:ec} depends only on the target function $f$ but not on the network $\mN$. It is easily seen that the above relation is an equivalence relation. Now, we consider a fixed cut set $C\in \Lambda(\mN)$ and let $I$ and $J$ in Definition~\ref{def:ec} be $I_C$ and $J_C$, respectively (evidently, $I\cap J=\emptyset$ by definition). Fix $\vec{a}_J\in \mA^{k\times J}$. Then the $(I,\vec{a}_J)$-equivalence relation induces a partition of $\mA^{k\times I}$ and the blocks in the partition are called {\em $(I,\vec{a}_J)$-equivalence classes}.

Let $\{g_e:\ e\in \mE\}$ be the set of all global encoding functions of a given $(k,n)$ network code that can compute $f$ over $\mN$. Then this network code has to distinguish all the $(I,\vec{a}_J)$-equivalence classes on the cut set $C$. Intuitively, for any two source matrices $\vec{b}_I$ and $\vec{b}'_I$ in
$\mA^{k\times I}$ that are not $(I, \vec{a}_J)$-equivalent, it is
necessary that
\begin{align}\label{neq_g_C}
g_C(\vec{b}_I, \vec{a}_J)\neq g_C(\vec{b}'_I, \vec{a}_J).
\end{align}
This can be formally proved as follows. First, since no directed path exist from any source node in $S\setminus (I\cup J)$ to any node in $\{\tail(e): e\in C\}$, the input symbols $\vec{x}_{S\setminus(I\cup J)}$ do not contribute to the values of $g_C=(g_e: e\in C)$. Hence, we write $g_C(\vec{x}_I, \vec{x}_J, \vec{x}_{S\setminus(I\cup J)})$ as $g_C(\vec{x}_I, \vec{x}_J)$. Consider any $\vec{b}_I$ and $\vec{b}'_I$ in $\mA^{k\times I}$ that are not $(I, \vec{a}_J)$-equivalent, i.e., $\exists~\vec{d}\in \mA^{k \times S\setminus (I\cup J)}$ such that
\begin{align}\label{equ:pf_lem1_1}
f(\vec{b}_I, \vec{a}_J, \vec{d}) \neq f(\vec{b}'_I, \vec{a}_J, \vec{d}).
\end{align}
Let $D=\bigcup_{\sigma\in (S\setminus I)}\eout(\sigma)$, an edge subset of $\mE$. Then $\widehat{C}=C\cup D$ is a global cut set, i.e., $I_{\widehat{C}}=S$. Since $g_{\ein(\rho)}(\vec{x}_S)$ is a function of $g_{\widehat{C}}(\vec{x}_S)$ and the network code can compute $f$, \eqref{equ:pf_lem1_1} implies that
\begin{align}\label{neq_g_C_1}
g_{\widehat{C}}(\vec{b}_I, \vec{a}_J, \vec{d})\neq g_{\widehat{C}}(\vec{b}'_I, \vec{a}_J, \vec{d}).
\end{align}
Together with $K_C=I\cup J$ and $K_{D}=S\setminus I$, we have
\begin{align*}
\big( g_{C}(\vec{b}_I, \vec{a}_{J}),\ g_{D}(\vec{a}_J, \vec{d})\big)
=g_{\widehat{C}}(\vec{b}_I, \vec{a}_J, \vec{d})
\neq
g_{\widehat{C}}(\vec{b}'_I, \vec{a}_J, \vec{d})
=\big( g_{C}(\vec{b}'_I, \vec{a}_J),\ g_{D}(\vec{a}_J, \vec{d})\big).
\end{align*}
By comparing the ordered pairs on the left and right above, we obtain \eqref{neq_g_C}.

Let $W_{C,f}^{(\vec{a}_J)}$ denote the number of all $(I, \vec{a}_J)$-equivalence classes. Then it follows from the above discussion that $|\mA|^{n\cdot|C|}\geq W_{C,f}^{(\vec{a}_J)}$, and furthermore that
\begin{align}\label{ineq}
|\mA|^{n\cdot|C|}\geq \max_{ \vec{a}_J\in \mA^{k\times J}} W_{C,f}^{(\vec{a}_J)}.
\end{align}
In \eqref{ineq}, for $k=1$, $\max\limits_{ \vec{a}_J\in \mA^{k\times J}} W_{C,f}^{(\vec{a}_J)}$ becomes $\max\limits_{ {a}_J\in \mA^{J}} W_{C,f}^{({a}_J)}$, and we denote it by $w_{C,f}$. Together with the claim that $\max\limits_{ \vec{a}_J\in \mA^{k\times J}} W_{C,f}^{(\vec{a}_J)}=(w_{C,f})^k$ in \cite{huang15}, we obtain the upper bound therein:
\begin{equation}
  \label{eq:2}
  \mathcal{C}(\mathcal{N},f)\leq
  \min_{C\in\Lambda(\mathcal{N})}\dfrac{|C|}{\log_{|\mathcal{A}|}w_{C,f}}.
\end{equation}



When the cut set $C$ is a global cut set, i.e., $I=I_C=S$, we have $J=J_C=\emptyset$. Then we can see that two source inputs $a_S$ and $b_S$ in $\mA^S$ are $(I,{a}_J)$-equivalent provided that $f(a_S)=f(b_S)$ (note that $\forall~{a}_J\in \mA^J$, $a_J$ is an empty vector). This implies that $w_{C,f}=|f(\mA^s)|$. Considering the right hand side of \eqref{eq:2}, we have
\begin{align*}
\min_{C\in\Lambda(\mathcal{N})}\dfrac{|C|}{\log_{|\mathcal{A}|}w_{C,f}}\leq \min_{C\in\Lambda(\mathcal{N}): I_C=S}\dfrac{|C|}{\log_{|\mathcal{A}|}w_{C,f}}=
\min_{C\in\Lambda(\mathcal{N}): I_C=S} \frac{|C|}{\log_{|\mathcal{A}|}|f(\mathcal{A}^s)|}.
\end{align*}
Hence, the upper bound in \eqref{eq:2} is an enhancement of the one in \eqref{eq:1}. It was shown in \cite{huang15} that this bound in fact is tighter than the one in \eqref{eq:1} and is tight for {\em multi-edge tree networks}, where a multi-edge tree network is a tree with multiple edges allowed between two adjacent nodes (see \cite{Appuswamy11} and \cite{huang15}). However, it was also demonstrated in \cite{huang15} that this bound is not tight for an example that was first studied in \cite{Appuswamy11}.

\begin{example}[\!\!\cite{Appuswamy11,huang15}]\label{eg:1}

\begin{figure}[t]
  \centering
{
 \begin{tikzpicture}[x=0.6cm]
    \draw (0,0) node[vertex] (2) [label=above:$\sigma_2$] {};
    \draw (-2,-1.5) node[vertex] (1') [label=left:] {};
    \draw (2,-1.5) node[vertex] (2') [label=right:] {};
    \draw (0,-3) node[vertex] (0) [label=below:$\rho$] {};
    \draw[->,>=latex] (2) -- (1') node[midway, auto,swap, left=0mm] {$e_2$};
    \draw[->,>=latex] (2) -- (2') node[midway, auto, right=0mm] {$e_3$};
    \draw[->,>=latex] (1') -- (0) node[midway, auto,swap,  left=0mm] {$e_5$};
    \draw[->,>=latex] (2') -- (0) node[midway, auto, right=0mm] {$e_6$};

    \draw node[vertex,label=above:$\sigma_1$] at (-4,0) (1) {};
    \draw[->,>=latex] (1) -- (1') node[midway, auto,swap,  left=0mm] {$e_1$};
    \draw node[vertex,label=above:$\sigma_3$] at (4,0) (3) {};
    \draw[->,>=latex] (3) -- (2') node[midway, auto, right=0mm] {$e_4$};
    \end{tikzpicture}
}
\caption{The network $\mathcal{N}$ has three binary sources
  $\sigma_1$, $\sigma_2$, $\sigma_3$ and one sink $\rho$ that
  computes the {\em arithmetic sum} of the source messages as the target function $f$, i.e., $f(x_1,x_2,x_3)=x_1+x_2+x_3$, with $\mA=\{0,1\}$ and $\mO=\{0,1,2,3\}$.}
  \label{fig:1}
\end{figure}
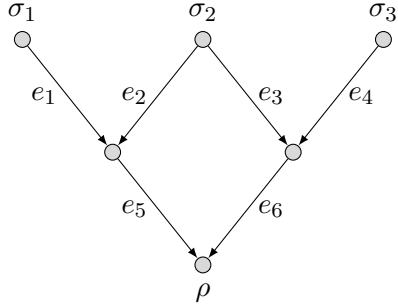

For the network function computation problem in Fig.~\ref{fig:1}, denoted by $(\mN, f)$, both the upper bounds in \eqref{eq:1} and \eqref{eq:2} are equal to $1$, giving $\mathcal{C}(\mathcal{N},f)\leq 1$. To be specific, the right hand side of \eqref{eq:1} is minimized by the global cut set $C=\{e_5, e_6\}$ with cardinality $2$ and $|f(\mathcal{A}^s)|=4$, giving the upper bound $1$. It was shown in \cite{huang15} that the right hand side of \eqref{eq:2} is minimized by the cut set $C=\{e_5, e_6\}$. Denote $I_C$ and $J_C$ by $I$ and $J$, respectively. Evidently, $I=S$ and $J=\emptyset$. Then, $\forall~{a}_J\in \mA^J$, ${a}_J$ is an empty vector. Thus, the $(I, {a}_J)$-equivalence classes are
\begin{align*}
&\Cl_1= \{(0,0,0)\}, \quad \Cl_2=\{(0,0,1),(0,1,0),(1,0,0) \}, \\
&\Cl_3= \{(0,1,1),(1,0,1),(1,1,0)\}, \quad \Cl_4=\{(1,1,1)\},
\end{align*}
and for the input vectors in the same equivalence class $\Cl_i$, $1\leq i \leq 4$, the function $f$ takes the same value.\footnote{In fact, for any network computation problem $(\mN,f)$, we can easily see that for every global cut set $C$, i.e., $I=S$ and $J=\emptyset$, two source inputs $b_S$ and $b_S'$ in $\mA^S$ are $(I, a_J)$-equivalent if and only if $f(b_S)=f(b_S')$.} Hence, we have $w_{C,f}=4$ and $|C|/\log_{|\mA|} w_{C,f}=1$, giving the upper bound $1$.

However, it was shown in \cite{Appuswamy11} that the exact computing capacity of $(\mN,f)$ is $2/(1+\log_2 3)\approx 0.77$, which is considerably smaller than $1$. The proof of this computing capacity is non-trivial.

Note that the network in Fig.~\ref{fig:1} has a very simple ``non-tree'' structure. Thus, this example indicates that the existing upper bounds are far from being tight for general non-tree network topologies.
\end{example}

We now use Example~\ref{eg:1} to give an intuitive (but not complete) explanation why the upper bound $1$ on $\mC(N,f)$ in \eqref{eq:2} is not tight. Suppose this upper bound is tight so that the rate $1$ is achievable, i.e., there exists a $(k,k)$ network code for some positive integer $k$. Let us for the time being assume that $k=1$, and let the set of global encoding functions be $\{ g_{e_i}: 1\leq i \leq 6 \}$. Since this code can compute $f$, according to the upper bound \eqref{eq:2}, it is necessary for it to distinguish the four $(I,a_J)$-equivalence classes $\Cl_1$, $\Cl_2$, $\Cl_3$ and $\Cl_4$ at the cut set $C=\{e_5, e_6\}$.

However, we now show that this condition is not sufficient for the code to compute $f$. Suppose $g_C$ takes the same value for all the inputs in $\Cl_2$. Since the two inputs $(0,0,1)$ and $(1,0,0)$ are in $\Cl_2$, we have
\begin{align*}
 g_C(0,0,1)=&\big( g_{e_5}(x_1=0,x_2=0), g_{e_6}(x_2=0,x_3=1) \big)\\
=&\big( g_{e_5}(x_1=1,x_2=0), g_{e_6}(x_2=0,x_3=0) \big)=g_C(1,0,0),
\end{align*}
implying that
$$g_{e_5}(x_1=0,x_2=0)=g_{e_5}(x_1=1,x_2=0).$$
On the other hand, by considering the input $(1,0,1)$ in $\Cl_3$ and the input $(0,0,1)$ in $\Cl_2$, we obtain
\begin{align*}
g_C(1,0,1)=&\big(g_{e_5}(x_1=1,x_2=0), g_{e_6}(x_2=0,x_3=1) \big)\\
=&\big(g_{e_5}(x_1=0, x_2=0), g_{e_6}(x_2=0,x_3=1) \big)=g_C(0,0,1),
\end{align*}
implying that the code cannot distinguish these $2$ inputs and hence cannot compute $f$ because $2=f(1,0,1) \neq f(0,0,1)=1$.

In other words, the necessary condition that has been used to obtain \eqref{eq:2} is not strong enough to be also sufficient, and hence the upper bound \eqref{eq:2} is not tight. Nevertheless, based on the intuition obtained in the above discussion, we will propose a new upper bound that is applicable to arbitrary network topology and target function. This upper bound not only is an enhancement of the upper bound in \eqref{eq:2}, but also is tight for the network function computation problem in Example~\ref{eg:1}.

\section{Improved Upper Bound}\label{sec:improved_upp_bou}

In this section, we state our improved upper bound with some discussions. The proof of this bound is deferred to Section~\ref{sec:proof}.

\subsection{The Improved Upper Bound}

\begin{defn}\label{def:strong_parti}
Let $C\in \Lambda(\mN)$ be a cut set and $\mP_C=\{C_1,C_2,\cdots, C_m \}$ be a partition of the cut set $C$. The partition $\mP_C$ is said to be a strong partition of $C$ if the following two conditions are satisfied:
\begin{enumerate}
  \item $I_{C_l} \neq \emptyset$, $\forall~1\leq l \leq m$;
  \item $I_{C_i}\cap I_{C_j}=\emptyset$, $\forall~1\leq i, j \leq m$ and $i\neq j$.
\end{enumerate}
\end{defn}

For any cut set $C$ in $\Lambda(\mN)$, the partition $\{C \}$ is called the \textit{trivial strong partition} of $C$.

\begin{defn}
\label{defn:Par_Equ_Relation_ScalarCase}
Let $I$ and $J$ be two disjoint subsets of $S$. Let $I_l$, $l=1,2, \cdots, m$, be $m$ disjoint subsets of $I$ and let $L=I\setminus(\bigcup_{l=1}^m I_l)$. For given ${a}_J\in \mA^{J}$ and ${a}_L\in \mA^{L}$, we say that ${b}_{I_l}$ and $b'_{I_l}$ in $\mA^{I_l}$ are $(I_l, {a}_L, {a}_J)$-equivalent for $1 \leq l \leq m$, if for each ${c}_{I_j} \in \mA^{I_j}$ with $1 \leq j \leq m$ and $j\neq l$, $({b}_{I_l},{a}_{L}, {c}_{I_j},\ 1 \leq j \leq m, j\neq l)$ and $({b}'_{I_l}, {a}_{L}, {c}_{I_j},\ 1 \leq j \leq m, j\neq l)$ in $\mA^{I}$ are $(I, {a}_J)$-equivalent.
\end{defn}

It is easily seen that the above relation for every $l$ is an equivalence relation and thus partitions $\mA^{I_l}$ into $(I_l, {a}_L, {a}_J)$-equivalence classes. Similar to the $(I, {a}_J)$-equivalence relation, the $(I_l, {a}_L, {a}_J)$-equivalence relation does not depend on any cut set or the network topology.

Note that Definition~\ref{defn:Par_Equ_Relation_ScalarCase} subsumes Definition~\ref{def:ec} because the former reduces to the latter when $m=1$ and $I_1=I$ ($L=\emptyset$ and $a_L$ is an empty vector), i.e., the $(I_1, {a}_L, {a}_J)$-equivalence relation becomes the $(I, {a}_J)$-equivalence relation and the $(I_1, {a}_L, {a}_J)$-equivalence classes become the $(I, {a}_J)$-equivalence classes.

Fix a cut set $C\in \Lambda(\mN)$ and let $\mP_C=\{ C_1,C_2,\cdots, C_m \}$ be a strong partition of $C$. For notational simplicity, let $I=I_C$, $J=J_C$ and $I_l=I_{C_l}$ for $l=1,2, \cdots, m$. By \eqref{eq:I_C_subseteq}, $\bigcup_{l=1}^m I_l\subseteq I_{\bigcup_{l=1}^m C_l}= I$, and accordingly we let $L=I\setminus(\bigcup_{l=1}^m I_l)$. Then we see that $\{ I_1, I_2, \cdots, I_m, L \}$ forms a partition of $I$.

We use $\Cl[{a}_J]$ to denote an $(I,{a}_J)$-equivalence class. For $l=1,2,\cdots,m$, we use $\cl_{I_l}[{a}_{L}, {a}_J]$ to denote an $(I_l,{a}_L,{a}_J)$-equivalence class, and use $V_{I_l}^{[{a}_{L},{a}_J]}$ to denote the number of the $(I_l,{a}_L,{a}_J)$-equivalence classes. In particular, when ${a}_{L}$ and ${a}_J$ are clear from the context, we write $\cl_{I_l}$ and $V_{I_l}$ to simplify the notation. Now, we define the set
\begin{align}\label{def:langle_set_rangle}
\big\langle \cl_{I_1}, \cl_{I_2}, \cdots, \cl_{I_m}, {a}_{L}\big \rangle
\triangleq \Big\{ ({b}_{I_1}, {b}_{I_2}, \cdots, {b}_{I_m}, {a}_{L}):\ {b}_{I_l}\in \cl_{I_l}, l=1,2,\cdots,m \Big\}\subseteq \mA^{I}
\end{align}
and state the following lemma. The proof is deferred to Section~\ref{sec:proof}.

\begin{lemma}\label{prop_cl}
For any set of $(I_l, {a}_{L}, {a}_J)$-equivalence classes $\cl_{I_l}$, $l=1, 2, \cdots, m$, all source inputs $({b}_{I_1}, {b}_{I_2}, \cdots, {b}_{I_m}, {a}_{L})$ in $\big\langle \cl_{I_1}, \cl_{I_2}, \cdots, \cl_{I_m}, {a}_{L}\big \rangle$ are $(I,{a}_J)$-equivalent. In other words, there exists an $(I,{a}_J)$-equivalence class $\Cl[{a}_J]$ such that
\begin{align}\label{equ:lem:prop_cl}
\big\langle \cl_{I_1}, \cl_{I_2}, \cdots, \cl_{I_m}, {a}_{L} \big\rangle \subseteq \Cl[{a}_J].
\end{align}
\end{lemma}

With Lemma~\ref{prop_cl}, we can define a function $h$ that maps $(\cl_{I_1}, \cl_{I_2}, \cdots, \cl_{I_m})$ to the corresponding $(I, {a}_J)$-equivalence class for given $a_L\in \mA^L$ and $a_J\in \mA^J$.

For each $(I,{a}_J)$-equivalence class $\Cl[{a}_J]$, we define
\begin{align}\label{no_finer_eq_cl}
 N\big({a}_{L}, \Cl[{a}_J]\big)=&
\# \Big\{ \big( \cl_{I_1}, \cl_{I_2}, \cdots, \cl_{I_m} \big):\
       \cl_{I_l} \text{ is an $(I_l, {a}_{L}, {a}_J)$-equivalence class, }
       l=1,2,\cdots,m,\nonumber\\
&\qquad \qquad \qquad \qquad \qquad \quad \textrm{and } h(\cl_{I_1}, \cl_{I_2}, \cdots, \cl_{I_m})=\Cl[{a}_J]\Big\}\nonumber\\
=&\# \Big\{ \big( \cl_{I_1}, \cl_{I_2}, \cdots, \cl_{I_m} \big):\
       \cl_{I_l} \text{ is an $(I_l, {a}_{L}, {a}_J)$-equivalence class, }
       l=1,2,\cdots,m,\nonumber\\
&\qquad \qquad \qquad \qquad \qquad \quad \textrm{and } \big\langle \cl_{I_1}, \cl_{I_2}, \cdots, \cl_{I_m}, {a}_{L} \big\rangle \subseteq \Cl[{a}_J]\Big\},
\end{align}
where we use ``$\#\{\cdot\}$'' to stand for the cardinality of the set. Further, let
\begin{align}\label{equ:N_Cl}
N\big(\Cl[{a}_J]\big)=\max_{a_{L}\in \mA^{L}} N\big(a_{L}, \Cl[{a}_J]\big).
\end{align}

Note that $N\big(a_{L}, \Cl[{a}_J]\big)$ can be equal to $0$. On the other hand, $N\big(\Cl[{a}_J]\big)$ is always positive, which is explained as follows. Note that $\Cl[{a}_J]$ is an $(I, a_J)$-equivalence class and hence non-empty. Therefore, there exists $b_I$ in $\mA^{I}$ such that $b_I\in \Cl[{a}_J]$, and we write
$$b_I=({b}_{I_1}, {b}_{I_2}, \cdots, {b}_{I_m}, {b}_{L}),$$
where ${b}_{L}$ is equal to some $a_L\in \mA^L$. For $l=1,2,\cdots,m$, since $\mA^{I_l}$ is partitioned into $(I_l,a_L,a_J)$-equivalence classes, $b_{I_l}$ is in some $\cl_{I_l}[a_L,a_J]$, abbreviated as $\cl_{I_l}$. Also, since $b_I\in \Cl[{a}_J]$, we have $h(\cl_{I_1}, \cl_{I_2}, \cdots, \cl_{I_m})=\Cl[{a}_J]$ by Lemma~\ref{prop_cl}. Therefore, we see that $N\big(a_{L}, \Cl[{a}_J]\big)\geq 1$.

Next, we consider the summation of $N\big(\Cl[{a}_J]\big)$ over all the $(I,{a}_J)$-equivalence classes, i.e.,
\begin{align}\label{eq:sum_N_Cl}
\sum_{\text{all }\Cl[{a}_J]} N\big(\Cl[{a}_J]\big).
\end{align}
Let
\begin{align}\label{def:{a}_J_star0}
{a}_J^* \in \arg\max_{{a}_J\in \mA^{J}} \sum_{\text{all }\Cl[{a}_J]} N\big(\Cl[{a}_J]\big),
\end{align}
i.e.,
\begin{align}\label{def:{a}_J_star}
\sum_{\text{all }\Cl[{a}^*_J]} N\big(\Cl[{a}^*_J]\big)=\max_{{a}_J\in \mA^{J}} \sum_{\text{all }\Cl[{a}_J]} N\big(\Cl[{a}_J]\big),
\end{align}
and further
\begin{align}\label{n_C_Parti_1st}
n_C(\mP_C) = \sum_{\text{all }\Cl[{a}_J^*]}
N\big(\Cl[{a}_J^*]\big).
\end{align}
Denote by $n_{C,f}$ the maximum of $n_C(\mP_C)$ over all strong partitions $\mP_C$ of $C$,
i.e.,
\begin{align}\label{n_C_f_1st}
n_{C,f}=\max_{\text{all strong partitions }\mP_C \text{ of } C}  n_C(\mP_C).
\end{align}

Based on the above, 
we give in the following theorem our improved upper bound which is applicable to arbitrary networks and target functions.

\begin{thm}
  \label{thm:upper_bound}
  Let $\mathcal{N}$ be a network and $f$ be a target function. Then
\begin{align}\label{equ:improved_upper_bound}
\mathcal{C}(\mathcal{N},f)\leq \min_{C\in\Lambda(\mN)}\dfrac{|C|}{\log_{|\mA|}n_{C,f}}.
\end{align}
\end{thm}

For a cut set $C\in \Lambda(\mN)$, if we consider its trivial strong partition $\mP_C=\{ C \}$, then we have $m=1$ and $I_1=I$ ($L=\emptyset$ and $a_L$ is an empty vector). Following the discussion in the second paragraph below Definition~\ref{defn:Par_Equ_Relation_ScalarCase}, we see from \eqref{no_finer_eq_cl} that $N\big({a}_{L}, \Cl[{a}_J]\big)=1$, and from \eqref{equ:N_Cl} that
$$N\big(\Cl[{a}_J]\big)=N\big({a}_{L}, \Cl[{a}_J]\big)=1.$$
Then the summation in \eqref{eq:sum_N_Cl} becomes $W_{C,f}^{(a_J)}$ and the right hand side of \eqref{def:{a}_J_star} becomes $w_{C,f}$. Finally, it follows from \eqref{n_C_Parti_1st} and \eqref{n_C_f_1st} that $n_{C}(\{C\})=w_{C,f}\leq n_{C,f}$. Hence, the upper bound in \eqref{equ:improved_upper_bound} is an enhancement of the one in \eqref{eq:2}.


\subsection{Evaluation of the Improved Upper Bound}

An important step toward evaluating the upper bound \eqref{equ:improved_upper_bound} in Theorem~\ref{thm:upper_bound} is to calculate the value of $N\big(a_L, \Cl[{a}_J]\big)$ in \eqref{no_finer_eq_cl} for $a_{L} \in \mA^L$ and an $(I,{a}_J)$-equivalence class $\Cl[{a}_J]$. In this subsection, we introduce a {\em multi-dimensional array} to facilitate this calculation.

For $l=1,2,\cdots, m$, let $\cl_{I_l}$ be an $(I_l,a_L,
{a}_J)$-equivalence class. By Lemma \ref{prop_cl},
$(\cl_{I_1},\cl_{I_2},\cdots,\cl_{I_m})$ uniquely
determines an $(I, {a}_J)$-equivalence class $\Cl[{a}_J]$ through the function $h$, namely, $$h(\cl_{I_1},\cl_{I_2},\cdots,\cl_{I_m})=\Cl[{a}_J].$$
We can then define the following $m$-dimensional array (when $m=2$, this array can be regarded as a matrix):
\begin{align}\label{equ:matrix_M}
M(a_L,{a}_J)=
\Big[h\big(\cl_{I_1,i_1}[a_L,{a}_J], \cl_{I_2,i_2}[a_L,{a}_J], \cdots, \cl_{I_m,i_m}[a_L,{a}_J]\big)\Big]
_{
\substack{1\leq i_1 \leq V_{I_1}^{[a_L,{a}_J]}\\
1\leq i_2 \leq V_{I_2}^{[a_L,{a}_J]}\\
\vdots\quad\quad\ \\
1\leq i_m \leq V_{I_m}^{[a_L,{a}_J]}
}}
\end{align}
where $\cl_{I_l,i_l}[a_L,{a}_J]$, $1\leq i_l \leq V_{I_l}^{[a_L,{a}_J]}$ are all the $(I_l, a_L, {a}_J)$-equivalence classes partitioning $\mA^{I_l}$ for $1\leq l \leq m$. We observe that $N\big(a_{L}, \Cl[{a}_J]\big)$ is simply the number of the entries equal to $\Cl[{a}_J]$ in the array $M(a_L, {a}_J)$.

We continue to use the setup in Example~\ref{eg:1} to illustrate the computation of the upper bound in Theorem~\ref{thm:upper_bound} by using the array $M(a_L,{a}_J)$. We will also see that our improved upper bound is tight for the network function computation problem in Example~\ref{eg:1}.

\begin{journalonly}
Let $k=1$ and then we obtain the corresponding results for the scalar case. Concretely, for the given source vectors ${a}_J\in \mA^{J}$ and $a_L\in \mA^{L}$, we have the following matrices:
\begin{align*}
M(a_L)=\Big[M_{i,j}(a_L)\Big]_{1\leq i \leq V_{I_1}(a_{L}, {a}_J),\ 1\leq j \leq V_{I_2}(a_{L}, {a}_J)}.
\end{align*}
Subsequently, we write
\begin{align}
\vec{a}_J&=({a}_J^{(1)}, {a}_J^{(2)}, \cdots, {a}_J^{(k)})^\top,\label{nota_{a}_J}\\
\vec{a}_{L}&=(a_{L}^{(1)}, a_{L}^{(2)}, \cdots, a_{L}^{(k)})^\top,\label{nota_a_L}
\end{align}
and the corresponding equivalence class $\Cl^{(k)}[\vec{a}_J]$ is written as
\begin{align}
\Cl^{(k)}[\vec{a}_J]=\big(\Cl^{(1)}({a}_J^{(1)}),\ \Cl^{(1)}({a}_J^{(2)}),\ \cdots,\ \Cl^{(1)}({a}_J^{(k)})\big)^\top,\label{nota_Cl_J}
\end{align}
which is considered as a column vector constituted by $k$ scalar equivalence classes. For all $p$, $1\leq p \leq k$, one has the $k$ corresponding matrices as follows
\begin{align*}
M(a_L^{(p)})=\Big[M_{i,j}(a_L^{(p)})\Big]_{1\leq i \leq V_{I_1}(a_{L}^{(p)}, c_{J,p}),\ 1\leq j \leq V_{I_2}(a_{L}^{(p)}, c_{J,p})},
\end{align*}
and the value $N(a_{L}^{(p)}, \Cl^{(1)}(c_{J,p}))$ is the number of the entries $M_{i,j}(a_L^{(p)})$ equal to $\Cl^{(1)}(c_{J,p})$ in the matrix $M(a_L^{(p)})$. Furthermore, it is not difficult to see that
\begin{align}\label{prod}
N(\vec{a}_{L}, \Cl^{(k)}[\vec{a}_J])=\prod_{p=1}^{k} N(a_{L}^{(p)}, \Cl^{(1)}(c_{J,p})).
\end{align}

Therefore, we derive that for $1\leq i \leq W_{C,f}^{(\vec{a}_J)}$,
\begin{align}\label{ineq7}
(\ref{ineq6})=N(\vec{a}_{L,(i)}, \Cl^{(k)}_i(\vec{a}_J))=\prod_{p=1}^{k} N(a_{L,(i)}^{(p)}, \Cl^{(1)}_i(c_{J,p})).
\end{align}

In order to clear the proof, before discussion further we review all the inequalities from (\ref{ineq1}) to (\ref{ineq7}) as follows
\begin{align*}
&|\mA|^{n|C|}\\
\geq &\#\{ g_C(\vec{x}_I, \vec{x}_J):\ \text{\ all } \vec{x}_I\in \mA^{k\times I}, \vec{x}_J\in \mA^{k\times J}\} \tag*{// by (\ref{ineq1})}\\
\geq & \#\{ g_C(\vec{x}_I, \vec{a}_J): \text{ all } \vec{x}_I\in \mA^{k\times I}\} \tag*{ // for every  $\vec{a}_J\in \mA^{k\times J}$ by (\ref{ineq2})}\\
=&\sum_{i=1}^{W_{C,f}^{(\vec{a}_J)}}\#\{ g_C(\vec{b}_I, \vec{a}_J):\text{ all } \vec{b}_I\in \Cl^{(k)}_i(\vec{a}_J)\}\tag*{// by (\ref{ineq3})}\\
=&\sum_{i=1}^{W_{C,f}^{(\vec{a}_J)}}\#\{ \big(g_{C_1}(\vec{a}_{I_1}, \vec{a}_{L}, \vec{a}_J), g_{C_2}(\vec{a}_{I_2}, \vec{a}_{L}, \vec{a}_J) \big): \\
& \qquad\qquad\qquad\quad \text{ all } \vec{b}_I=(\vec{a}_{I_1}, \vec{a}_{L}, \vec{a}_{I_2})\in \Cl^{(k)}_i(\vec{a}_J)\} \tag*{// by (\ref{ineq4})}\\
=&\sum_{i=1}^{W_{C,f}^{(\vec{a}_J)}}\#\{ \big(g_{C_1}(\vec{a}_{I_1}, \vec{a}_{L}), g_{C_2}(\vec{a}_{I_2}, \vec{a}_{L}) \big): \\
&\qquad\qquad\qquad\quad \text{ all } \vec{b}_I=(\vec{a}_{I_1}, \vec{a}_{L}, \vec{a}_{I_2})\in \Cl^{(k)}_i(\vec{a}_J)\} \tag*{// just for notation simplicity by (\ref{ineq5}) }\\
\geq &\sum_{i=1}^{W_{C,f}^{(\vec{a}_J)}}\#\mathop{\cup}_{\text{all }\vec{a}_{L}\in \mA^{k\times L}} \{ \big(g_{C_1}(\vec{a}_{I_1}, \vec{a}_{L}), g_{C_2}(\vec{a}_{I_2}, \vec{a}_{L}) \big):\ \text{all } \\
& \vec{a}_{I_1}\in \mA^{k\times I_1}, \vec{a}_{I_2}\in \mA^{k\times I_2} \textrm{ s.t. } (\vec{a}_{I_1}, \vec{a}_{L}, \vec{a}_{I_2})\in \Cl^{(k)}_i(\vec{a}_J)\}\nonumber\\
\geq & \sum_{i=1}^{W_{C,f}^{(\vec{a}_J)}}\#\{ \big(g_{C_1}(\vec{a}_{I_1}, \vec{a}_{L,(i)}), g_{C_2}(\vec{a}_{I_2}, \vec{a}_{L,(i)}) \big):\ \\
&\quad \vec{a}_{I_l}\in \mA^{k\times I_l}, l=1,2, \textrm{ s.t. } (\vec{a}_{I_1}, \vec{a}_{L,(i)}, \vec{a}_{I_2})\in \Cl^{(k)}_i(\vec{a}_J)\}\tag*{ // for all possible  $\vec{a}_{L,(i)}$ by (\ref{ineq_for_every})}\\
\geq & \sum_{i=1}^{W_{C,f}^{(\vec{a}_J)}} \#\{ \big(\cl^{(k)}_{I_1}(\vec{a}_{L,(i)},\vec{a}_J),\ \cl^{(k)}_{I_2}(\vec{a}_{L,(i)}, \vec{a}_J)\big): \\
&\qquad \text{ all } \cl^{(k)}_{I_1}(\vec{a}_{L,(i)},\vec{a}_J), \text{ and } \cl^{(k)}_{I_2}(\vec{a}_{L,(i)},\vec{a}_J), \textrm{ s.t. } \\
& \big(\cl^{(k)}_{I_1}(\vec{a}_{L,(i)},\vec{a}_J),\ \vec{a}_{L,(i)},\ \cl^{(k)}_{I_2}(\vec{a}_{L,(i)}, \vec{a}_J)\big)\subseteq \Cl^{(k)}_i(\vec{a}_J)\}\tag*{}\\
= & \sum_{i=1}^{W_{C,f}^{(\vec{a}_J)}} N(\vec{a}_{L,(i)}, \Cl^{(k)}_i(\vec{a}_J)) \tag*{// by (\ref{ineq6})}\\
= & \sum_{i=1}^{W_{C,f}^{(\vec{a}_J)}} \prod_{p=1}^{k} N(a_{L,(i)}^{(p)}, \Cl^{(1)}_i(c_{J,p})),  \tag*{// by (\ref{ineq7})}
\end{align*}
where recall (\ref{nota_{a}_J}), (\ref{nota_a_L}), and (\ref{nota_Cl_J}) for the meanings of notation $a_{L,(i)}^{(p)}$, $c_{J,p}$ and $\Cl^{(1)}_i(c_{J,p})$. Consequently, we summarize that
\begin{align}\label{ineq_final}
|\mA|^{n|C|}\geq \sum_{i=1}^{W_{C,f}^{(\vec{a}_J)}} \prod_{p=1}^{k} N(a_{L,(i)}^{(p)}, \Cl^{(1)}_i(c_{J,p})),
\end{align}
for arbitrary ${a}_J^{(p)}\in \mA^{J}$, and arbitrary $a_{L,(i)}^{(p)}\in \mA^{L}$ corresponding to the $(I, J, {a}_J^{(p)})$-equivalence class $\Cl^{(1)}_i(c_{J,p})$, $1\leq p \leq k$, $1 \leq i \leq W_{C,f}^{(\vec{a}_J)}$.

On the other hand, notice that for any ${a}_J\in \mA^{J}$, the $(I,{a}_J)$-equivalence relation induces different equivalence classes $\Cl^{(1)}_j({a}_J)$, $1 \leq j \leq W_{C,f}^{({a}_J)}$, constituting a partition of $\mA^{I}$. Then for every equivalence class $\Cl^{(1)}_j({a}_J)$, let
\begin{align*}
a_L^*(\Cl^{(1)}_j({a}_J))=\arg\max_{a_L\in \mA^{L}}N(a_{L}, \Cl^{(1)}_j({a}_J)).
\end{align*}
Let further
\begin{align*}
{a}_J^*=\arg\max_{{a}_J\in \mA^{J}} \sum_{j=1}^{W_{C,f}^{({a}_J)}} N(a_L^*(\Cl^{(1)}_j({a}_J)), \Cl^{(1)}_j({a}_J)),
\end{align*}
and thus we can obtain
\begin{align*}
&\max_{{a}_J\in \mA^{J}} \max_{a_{L,j}\in \mA^{L}:\atop 1\leq j \leq W_{C,f}^{({a}_J)}} \sum_{j=1}^{W_{C,f}^{({a}_J)}} N(a_{L,j}, \Cl^{(1)}_j({a}_J))\\
=&\sum_{j=1}^{W_{C,f}^{({a}_J^*)}} N(a_{L}^*(\Cl^{(1)}_j({a}_J^*)), \Cl^{(1)}_j({a}_J^*))\\
\triangleq &\ n_C(C_1,C_2).
\end{align*}
Together with (\ref{ineq_final}), we deduce that
\begin{align}\label{ineq_k_n}
|\mA|^{n|C|}\geq & \prod_{p=1}^{k} \sum_{j=1}^{W_{C,f}^{({a}_J^*)}} N(a_{L}^*(\Cl^{(1)}_j({a}_J^*)), \Cl^{(1)}_j({a}_J^*))\nonumber\\
= & \Big[ \sum_{j=1}^{W_{C,f}^{({a}_J^*)}} N(a_{L}^*(\Cl^{(1)}_j({a}_J^*)), \Cl^{(1)}_j({a}_J^*))\Big]^k \nonumber\\
= & \big[ n_C(C_1,C_2) \big]^k.
\end{align}

Furthermore, notice that the above inequalities follow for all partitions $C=C_1\cup C_2$ with $I_{C_i}\neq \emptyset$, $i=1,2$. Thus, we choose that partition such that $n_C(C_1,C_2)$ achieves the maximum and denote such a partition by $C=C_1^*\cup C_2^*$, i.e.,
$$n_C(C_1^*, C_2^*)=\max_{C=C_1\cup C_2 \text{ with} \atop I_{C_l}\neq \emptyset,\ l=1,2} n_C(C_1,C_2),$$
further denoted by $n_{C,f}$ because this maximum value $n_C(C_1^*, C_2^*)$ only depends on the cut set $C$ for the given $f$.
Then by (\ref{ineq_k_n}), we can show that
\begin{align*}
|\mA|^{n|C|}\geq \big[ n_C(C_1^*,C_2^*) \big]^k=n_{C,f}^k,
\end{align*}
and equivalently,
\begin{align}\label{ineq_upp_bound_k_n}
\frac{k}{n}\leq \dfrac{|C|}{\log_{|\mA|}n_{C,f}}.
\end{align}

At last, the above inequality (\ref{ineq_upp_bound_k_n}) should be satisfied for all cuts $C$ with $I_C \neq \emptyset$, that is, for all $C\in \Lambda{\mN}$. Therefore, we obtain the upper bound
\begin{align*}
\frac{k}{n}\leq \min_{C\in\Lambda(\mN)}\dfrac{|C|}{\log_{|\mA|}n_{C,f}}.
\end{align*}
Now, we can give the main result below.

\begin{thm}
  \label{thm:upper_bound}
  Let $\mathcal{N}=(G, S, \rho)$ be a network and $f$ be a target function. Then
\begin{align*}
\mathcal{C}(\mathcal{N},f)\leq \min_{C\in\Lambda(\mN)}\dfrac{|C|}{\log_{|\mA|}n_{C,f}}.
\end{align*}
\end{thm}


The upper bound herein is better than the previous one (\ref{eq:2}), since $n_{C,f}\geq w_{C,f}$ always holds for every $C\in \Lambda(\mN)$. Subsequently, we continue using the Example \ref{eg:1} to depict this obtained bound and show that it is tight for this example, i.e., achieves the capacity.
\end{journalonly}

\begin{example}\label{eg:2}
For the network function computation problem $(\mN,f)$ depicted in Fig.~\ref{fig:1}, consider the cut set $C=\{e_5, e_6\}$, and let $I=I_C$ and $J=J_C$. Then $I=S$, $J=\emptyset$, and $a_J$ is an empty vector. Recall in Example~\ref{eg:1} that all the $(I, {a}_J)$-equivalence classes are
\begin{align*}
&\Cl_1= \{(0,0,0)\}, \quad \Cl_2=\{(0,0,1),(0,1,0),(1,0,0) \}, \\
&\Cl_3= \{(0,1,1),(1,0,1),(1,1,0)\}, \quad \Cl_4=\{(1,1,1)\}.
\end{align*}

Furthermore, the only nontrivial (strong) partition of $C$ is $\mP_C=\{C_1=\{e_5\},
C_2=\{e_6\}\}$. Let $I_1= I_{C_1}=\{\sigma_1\}$,
$I_2= I_{C_2}=\{\sigma_3\}$ and accordingly $L = I_C\setminus (I_1\cup I_2)=\{ \sigma_2 \}$.

When $\sigma_2$ generates $0$ (i.e., $a_L=0$), since $(0,0,0)\in\Cl_1$ and $(1,0,0)\in
\Cl_2$, i.e., $0$ and $1$ in $\mA^{I_1}=\{0,1\}$ are not $(I_1, a_L=0, a_J)$-equivalent,   $\mA^{I_1}$ is partitioned into two $(I_1,a_L=0, {a}_J)$-equivalence classes
$$\cl_{I_1,1}[0]=\{0\} \quad \text{ and }\quad  \cl_{I_1,2}[0]=\{1\}.$$
Here, we have simplified $\cl_{I_1,1}[0,a_J]$ to $\cl_{I_1,1}[0]$ since $a_J$ is an empty vector, so on and so forth. Symmetrically, since $(0,0,0)\in\Cl_1$ and $(0,0,1)\in \Cl_2$, $\mA^{I_2}=\{0,1\}$ is also partitioned
into two $(I_2, a_L=0, {a}_J)$-equivalence classes
$$\cl_{I_2,1}[0]=\{0\} \quad  \text{ and } \quad \cl_{I_2,2}[0]=\{1\}.$$
Similarly, when $\sigma_2$ generates $1$ (i.e., $a_L=1$), $\mA^{I_1}$ is partitioned into two
$(I_1, a_L=1, {a}_J)$-equivalence classes
$$\cl_{I_1,1}[1]=\{0\} \quad \text{ and } \quad \cl_{I_1,2}[1]=\{1\},$$
and $\mA^{I_2}$ is partitioned into two
$(I_2, a_L=1, {a}_J)$-equivalence classes
$$\cl_{I_2,1}[1]=\{0\} \quad \text{ and } \quad \cl_{I_2,2}[1]=\{1\}.$$

Denote the matrices $M(a_L=0, {a}_J)$ and $M(a_L=1, {a}_J)$ respectively by $M(0)$ and $M(1)$ for simplicity. By \eqref{equ:matrix_M}, we have
\begin{align}\label{matrix_EX2_1}
M(0)=\bordermatrix[{[]}]{%
              &\text{\footnotesize{$\cl_{I_2,1}[0]$}}&\text{\footnotesize{$\cl_{I_2,2}[0]$}} \cr
\text{\footnotesize{$\cl_{I_1,1}[0]$}}&     \Cl_1      & \Cl_2          \cr
\text{\footnotesize{$\cl_{I_1,2}[0]$}}&     \Cl_2      & \Cl_3          \cr
},
\end{align}
and
\begin{align}\label{matrix_EX2_2}
M(1)=\bordermatrix[{[]}]{%
              &\text{\footnotesize{$\cl_{I_2,1}[1]$}}&\text{\footnotesize{$\cl_{I_2,2}[1]$}} \cr
\text{\footnotesize{$\cl_{I_1,1}[1]$}}&     \Cl_2      & \Cl_3          \cr
\text{\footnotesize{$\cl_{I_1,2}[1]$}}&     \Cl_3      & \Cl_4          \cr
}.
\end{align}
As an explanation, for the $\big(\cl_{I_1,1}[0], \cl_{I_2,1}[0]\big)$-th entry of $M(0)$, since $\cl_{I_1,1}[0]=\{0\}$ and $\cl_{I_2,1}[0]=\{0\}$, and $a_L=0$, we have
$\big\langle \cl_{I_1,1}[0], \cl_{I_2,1}[0], a_L=0 \big\rangle \subseteq \Cl_1$, and so this entry is equal to $\Cl_1$.

From \eqref{matrix_EX2_1} and \eqref{matrix_EX2_2}, we see that
\begin{align*}
N(0,\Cl_1)=1,\ N(0,\Cl_2)=2,\ N(0,\Cl_3)=1,\ N(0,\Cl_4)=0,\\
N(1,\Cl_1)=0,\ N(1,\Cl_2)=1,\ N(1,\Cl_3)=2,\ N(1,\Cl_4)=1,
\end{align*}
and further $N(\Cl_1)=1$, $N(\Cl_2)=2$, $N(\Cl_3)=2$, and $N(\Cl_4)=1$ by \eqref{equ:N_Cl}. Since $a_J$ is an empty vector, $\forall~a_J\in \mA^J$, it follows from \eqref{n_C_Parti_1st} that
$$n_C(\mP_C)=N(\Cl_1)+N(\Cl_2)+N(\Cl_3)+N(\Cl_4)=6.$$
By Theorem~\ref{thm:upper_bound}, we have
\begin{align*}
\mC(\mN,f)\leq \frac{|C|}{\log_{|\mA|} n_{C,f}}=\frac{|C|}{\log_{|\mA|} n_C(\mP_C)}=\frac{2}{\log_2 6}=\frac{2}{1+\log_2 3}.
\end{align*}

On the other hand, the network code designed in \cite{Appuswamy11} achieves the rate $2/(1+\log_2 3)$. Hence, the upper bound in Theorem \ref{thm:upper_bound} is tight for the network function computation problem $(\mN, f)$.

\end{example}

\subsection{Computing a Linear Function over a Network}\label{subsec:computing_linear_func}

In \cite{Appuswamy14}, Appuswamy and Franceschetti considered the achievability of rate $1$ for computing a {\em linear function} over a network, where a linear function is defined as follows. A target function $f:\mathcal{A}^s\to \mathcal{O}$ is {\em linear}, if
\begin{enumerate}
  \item the alphabet $\mA$ is a finite field $\Fq$, where $q$ is a prime power;
  \item the alphabet $\mO$ is $\Fq^{l}$, where $l$ is a positive integer;
  \item there exists an $l\times s$ matrix $T$ over $\mA$ such that $f(x_S)=T\cdot x_S^{\top}$, $\forall~x_S\in \mA^S$, where `$\top$' denotes matrix transposition.
\end{enumerate}
Without loss of generality we assume that $T$ is full-rank over $\Fq$ and has no all-zero columns. Hence, we can regard the size of $T$ as $l\times s$, where $1\leq l \leq s$. Note that if $l=s$, i.e., $T$ is a full-rank matrix of size $s\times s$, this network function computation problem reduces to a network coding problem.

Let $A$ and $B$ be two matrices in $\Fq^{l\times s}$. We write $A\sim B$ if there exists an $l\times l$ invertible matrix $Q$ over $\Fq$ and an $s\times s$  permutation matrix $\Pi$ such that $Q\cdot A\cdot \Pi=B$. Now, we consider a special linear target function $f$ corresponding to a matrix $T\in \Fq^{l\times s}$ with $T\sim (I~P)$, where $I$ is an $l\times l$ identity matrix and at least one element of $P\in\Fq^{l\times (s-l)}$ is zero. For this target function $f$, denote the columns of the matrix $T$ by $T_1,T_2,\cdots,T_s$, i.e., $T=(T_1~T_2~\cdots~T_s)$. Then the following so-called {\em min-cut condition} was given in \cite{Appuswamy14}:
\begin{align}\label{mincut_T}
\mincut(\mN, T)\triangleq \min_{C\in \Lambda(\mN)}\dfrac{|C|}{\Rank\big(\big[T_i: \sigma_i\in I_C\big]\big)}=1,
\end{align}
which is considered as a necessary condition of importance throughout \cite{Appuswamy14} for determining the rate-$1$ achievability of a linear function over a network. Theorem~\Rmnum{3}.5 in \cite{Appuswamy14}, one of main results in \cite{Appuswamy14}, showed that there always exists a network $\mN$ such that, even if the min-cut condition \eqref{mincut_T} is satisfied, there does not exist a rate-$1$ linear network code for computing $f$ over $\mN$, where this rate-$1$ linear network code is allowed to be over any extension field of $\Fq$. More specifically, we consider a linear network code over an extension field $\Fqn$ of $\Fq$ ($n$ is a positive integer) and the source matrices $\vec{x}_S\in \Fq^{n\times S}$ can be regarded as an $s$-dimensional row vector in $\Fqn^{S}$.

\begin{figure}[!t]
\centering
\begin{minipage}[b]{0.5\textwidth}
\centering
 \begin{tikzpicture}[x=0.7cm]
    \draw (0,0) node[vertex] (2) [label=above:$\sigma_2$] {};
    \draw (-2,-1.5) node[vertex] (1') [label=left:] {};
    \draw (2,-1.5) node[vertex] (2') [label=right:] {};
    \draw (0,-3) node[vertex] (0) [label=below:$\rho$] {};
    \draw[->,>=latex] (2) -- (1') node[pos=0.3, left=0mm] {$e_2$};
    \draw[->,>=latex] (2) -- (2') node[pos=0.3, right=0mm] {$e_3$};
    \draw[->,>=latex] (1') -- (0) node[pos=0.3, right=0mm] {$e_5$};
    \draw[->,>=latex] (2') -- (0) node[pos=0.3, left=0mm] {$e_6$};

    \draw node[vertex,label=above:$\sigma_1$] at (-4,0) (1) {};
    \draw[->,>=latex] (1) -- (1') node[pos=0.3, right=0mm] {$e_1$};
    \draw node[vertex,label=above:$\sigma_3$] at (4,0) (3) {};
    \draw[->,>=latex] (3) -- (2') node[pos=0.3, left=0mm] {$e_4$};

    \draw[->,>=latex] (2) -- (1') node[pos=0.7, right=0mm] {$g_2$};
    \draw[->,>=latex] (2) -- (2') node[pos=0.7, left=0mm] {$g_3$};
    \draw[->,>=latex] (1') -- (0) node[pos=0.7, left=0mm] {$g_5$};
    \draw[->,>=latex] (2') -- (0) node[pos=0.7, right=0mm] {$g_6$};
    \draw[->,>=latex] (1) -- (1') node[pos=0.7, left=0mm]{$g_1$};
    \draw[->,>=latex] (3) -- (2') node[pos=0.7, right=0mm] {$g_4$};
    \end{tikzpicture}
\vspace{5mm}
\caption{The problem $(\widehat{\mN}, \widehat{T})$.\newline\newline}
\label{fig:vector_fig}
\end{minipage}%
\begin{minipage}[b]{0.5\textwidth}
\centering
\begin{tabular}{p{4mm}p{2cm}p{0mm}p{4mm}p{2cm}}
\hline
\specialrule{0em}{2pt}{2pt}
$\sigma_1$: & $(x_{1,1}, x_{1,2})$ & & $\sigma_2$: & $(x_{2,1}, x_{2,2})$ \\
\specialrule{0em}{2pt}{2pt}
$\sigma_3$: & $(x_{3,1}, x_{3,2})$ & &  \\
\specialrule{0em}{2pt}{2pt}
$g_1$: & ${\scriptsize \begin{bmatrix} x_{1,1} \\ x_{1,2} \end{bmatrix}}$ & & $g_4$: & ${\scriptsize \begin{bmatrix} x_{3,1} \\ x_{3,2} \end{bmatrix}}$\\
\specialrule{0em}{2pt}{2pt}
$g_2$: & $x_{2,1}$ & & $g_3$: & $x_{2,2}$\\
\specialrule{0em}{2pt}{2pt}
$g_5$: & ${\scriptsize \begin{bmatrix} x_{1,1} \\ x_{1,2} \\ x_{2,1} \end{bmatrix}}$ & & $g_6$: & ${\scriptsize \begin{bmatrix} x_{3,1} \\ x_{3,2} \\ x_{2,2} \end{bmatrix}}$\\
\specialrule{0em}{2pt}{2pt}
\hline
\end{tabular}
\caption{A trivial rate-$\frac{2}{3}$ network code $\{g_i:1\leq i \leq 6\}$, where the linear function $\hat{f}$ can be computed at the sink node $\rho$ from its inputs.}
\label{fig:vector_scheme}
\end{minipage}
\end{figure}

To prove this result, \cite{Appuswamy14} restricts attention to the rate-$1$ achievability of the linear function $\hat{f}$ corresponding to the matrix
\begin{align}\label{eq:T1}
\widehat{T}=\begin{pmatrix}
1 & 0 & \gamma \\
0 & 1 & 0 \\
\end{pmatrix}
, \quad \gamma\neq 0,
\end{align}
over $\Fq$ on the network $\widehat{\mN}$ as shown in Fig.~\ref{fig:vector_fig}. Further, this specific network function computation problem $(\widehat{\mN}, \widehat{T})$ is used as a building block to establish a general network function computation problem $(\mN, T)$. Then, it is proved that the existence of a rate-$1$ linear network code for computing $T$ over $\mN$ implies the existence of a rate-$1$ linear network code for computing $\widehat{T}$ over $\widehat{\mN}$. Equivalently, $(\mN, T)$ is not rate-$1$ achievable by a linear network code provided that $(\widehat{\mN}, \widehat{T})$ is not rate-$1$ achievable by a linear network code. Hence, the key here is to prove that $(\widehat{\mN}, \widehat{T})$ is not rate-$1$ achievable by a linear network code (i.e., \cite[Lemma~\Rmnum{3}.4]{Appuswamy14}).

In \cite{Appuswamy14}, the proof that $(\widehat{\mN}, \widehat{T})$ is not rate-$1$ achievable by a linear network code is complicated, and it relies on the use of some advanced algebraic tools. To be specific, by applying the Gr\"{o}bner basis of an ideal generated by a subset of a polynomial ring over this polynomial ring itself and Hilbert's Nullstellensatz (theorem of zeros), a necessary and sufficient condition for the existence of a rate-$1$ linear network code over the algebraic closure $\bar{\mathbb{F}}_q$ of $\Fq$ for computing a linear function $f$ over the field $\Fq$ on a network $\mN$ is given (see Theorem~\Rmnum{2}.4 in \cite{Appuswamy14}). Then, it was proved that the condition is not satisfied for the network function computation problem $(\widehat{\mN}, \widehat{T})$.

In contrast, by applying our upper bound in Theorem~\ref{thm:upper_bound}, we can easily prove that $\mC(\widehat{\mN}, \widehat{T})\leq 2/3$ (the upper bound on $\mC(\widehat{\mN}, \widehat{T})$ in \eqref{eq:2} is $1$), and in fact $\mC(\widehat{\mN}, \widehat{T})= 2/3$ (see Example~\ref{linear_vector_case} below). This not only implies that no rate-$1$ linear network codes exist for $(\widehat{\mN}, \widehat{T})$ but also that no rate-$1$ network codes (linear or nonlinear) exist for $(\widehat{\mN}, \widehat{T})$, which enhances Lemma~\Rmnum{3}.4 in \cite{Appuswamy14}. This further implies that no linear or nonlinear rate-$1$ network codes exist for $(\mN,T)$, as stated in the next proposition.

\begin{prop}
Consider a linear target function $f$ corresponding to a matrix $T\in \Fq^{l\times s}$ with $T\sim (I~P)$ so that at least one element of $P\in\Fq^{l\times (s-l)}$ is zero. Then there exists a network $\mN$ such that no rate-$1$ network codes (linear or nonlinear) exist for computing $f$ over $\mN$.
\end{prop}

\begin{example}\label{linear_vector_case}
In Fig.~\ref{fig:vector_fig}, consider the cut set $C=\{e_5, e_6\}$, and let $I=I_C$ and $J=J_C$ with $I=S$ and $J=\emptyset$. Then, ${a}_J$ is an empty vector. For the linear target function $\hat{f}$ corresponding to the matrix $\widehat{T}$ in \eqref{eq:T1}, the $(I, {a}_J)$-equivalence classes are:
\begin{align}\label{Cl_alpha_beta}
\Cl_{\alpha, \beta}&=\big\{x_S=(x_1,x_2,x_3)\in \Fq^3:\ \widehat{T}\cdot x_S^{\top}=\big(\begin{smallmatrix}\alpha \\ \beta \end{smallmatrix}\big) \big\}\nonumber\\
&=\big\{x_S=(x_1,x_2,x_3)\in \Fq^3:\ x_1+\gamma x_3=\alpha\text{ and } x_2=\beta  \big\}\nonumber\\
&=\big\{x_S=(x_1,\beta,x_3)\in \Fq^3:\ x_1+\gamma x_3=\alpha \big\}
\end{align}
for all pairs $(\alpha, \beta)\in \Fq\times \Fq$. Then the total number of $(I, {a}_J)$-equivalence classes is $q^2$.

Let $\mP_C=\{C_1=\{e_5\}, C_2=\{e_6\}\}$, the only nontrivial (strong) partition of $C$, and further let $I_1= I_{C_1}=\{\sigma_1\}$, $I_2= I_{C_2}=\{\sigma_3\}$ and accordingly $L=I\setminus (I_1\cup I_2)=\{ \sigma_2 \}$. First, fix $a_L=x_2=\beta$. Since for any two distinct elements $\xi$ and $\eta$ in $\Fq^{I_1}=\Fq$,
\begin{align*}
\xi+\gamma x_3\neq \eta+\gamma x_3, \quad \forall\ x_3\in \Fq,
\end{align*}
every element in $\Fq$ ($=\Fq^{I_1}$) itself constitutes an $(I_1, a_L=\beta, {a}_J)$-equivalence class so that the total number of $(I_1, a_L=\beta, {a}_J)$-equivalence classes is $q$. Further, for any two distinct elements $\xi$ and $\eta$ in $\Fq^{I_2}=\Fq$, since $\gamma\neq 0$, we have
\begin{align*}
x_1+\gamma \xi \neq x_1+\gamma \eta, \quad \forall\ x_1\in \Fq.
\end{align*}
This implies that every element in $\Fq$ ($=\Fq^{I_2}$) itself constitutes an $(I_2, a_L=\beta, {a}_J)$-equivalence class and so the total number of $(I_2, a_L=\beta, {a}_J)$-equivalence classes is also $q$.

Furthermore, note that
\begin{align}\label{equ_x1x3}
\big|\big\{ (x_1,x_3)\in \Fq^2: x_1+\gamma x_3=\alpha \big\}\big|=q,\ \forall\ \alpha\in \Fq.
\end{align}
Therefore, from the above discussion, we obtain that for any pair $(\alpha, \beta)\in \Fq\times \Fq$,
\begin{align*}
N(a_L=\tau, \Cl_{\alpha, \beta})=
\begin{cases}
q, & \text{if } \tau=\beta,\\
0, & \text{otherwise;}
\end{cases}
\end{align*}
({\rm cf}.~\eqref{no_finer_eq_cl}) and consequently,
\begin{align*}
N(\Cl_{\alpha, \beta})=\max_{\tau\in \Fq}N(a_L=\tau, \Cl_{\alpha, \beta})=q.
\end{align*}
Hence,
\begin{align*}
n_C(\mP_C)=\sum_{{\rm all}\ \Cl_{\alpha, \beta}}N(\Cl_{\alpha, \beta})=q^3.
\end{align*}
By Theorem~\ref{thm:upper_bound}, we have
\begin{align}\label{ineq:C_N1_T1_up_bound}
\mC(\widehat{\mN},\widehat{T})\leq \frac{|C|}{\log_{|\Fq|} n_{C,\widehat{T}}}=\frac{|C|}{\log_{|\Fq|} n_C(\mP_C)}=\frac{2}{\log_q q^3}=\dfrac{2}{3}.
\end{align}

On the other hand, a trivial rate-$\frac{2}{3}$ linear network code for $(\widehat{\mN}, \widehat{T})$ is given in Fig.~\ref{fig:vector_scheme}. Together with \eqref{ineq:C_N1_T1_up_bound}, we obtain $\mC(\widehat{\mN},\widehat{T})=2/3$.
\end{example}


\section{Proof of Theorem~\ref{thm:upper_bound}}\label{sec:proof}

Consider $\{g_e(\vec{x}_S)\in \mA^n:\ e\in \mE\}$, the set of global encoding functions of a given $(k,n)$ network code that can compute the target function $f$ over $\mN$.
Fix a cut set $C\in \Lambda(\mN)$ and let $I=I_C$ and $J=J_C$, respectively. Then,
\begin{align}\label{ineq1}
|\mA|^{n|C|}& \geq \#\big\{ g_C(\vec{x}_S):\ \vec{x}_S\in \mA^{k\times S} \big\}\\
& = \#\big\{ g_C(\vec{x}_I, \vec{x}_J):\ \vec{x}_I\in \mA^{k\times I} \text{ and } \vec{x}_J\in \mA^{k\times J} \big\}\\
&=  \#\ \bigcup_{\vec{x}_J\in \mA^{k\times J}}\big\{ g_C(\vec{x}_I, \vec{x}_J):\  \vec{x}_I\in \mA^{k\times I}\big\}.
\end{align}
Hence, for every $\vec{a}_J\in \mA^{k\times J}$, we have
\begin{align}\label{ineq2}
|\mA|^{n|C|}\geq \# \big\{ g_C(\vec{x}_I, \vec{a}_J):\ \vec{x}_I\in \mA^{k\times I} \big\}.
\end{align}

We use $\Cl^{(k)}[\vec{a}_J]$ to denote an $(I,\vec{a}_J)$-equivalence class. Since all $(I,\vec{a}_J)$-equivalence classes form a partition of $\mA^{k\times I}$, we can write the right hand side of \eqref{ineq2} as
\begin{align*}
\#\bigcup_{\text{all } \Cl^{(k)}[\vec{a}_J]} \big\{ g_C(\vec{b}_I, \vec{a}_J):\ \vec{b}_I\in \Cl^{(k)}[\vec{a}_J]\big\}.
\end{align*}
Applying the assertion in the second paragraph below Definition~\ref{def:ec} that $g_C(\vec{b}_I, \vec{a}_J)\neq g_C(\vec{b}'_I, \vec{a}_J)$ for any $\vec{b}_I, \vec{b}'_I \in \mA^{k\times I}$ that are not $(I, \vec{a}_J)$-equivalent, we further obtain that
\begin{align}
|\mA|^{n|C|}\geq&\#\bigcup_{\text{all } \Cl^{(k)}[\vec{a}_J]} \big\{ g_C(\vec{b}_I, \vec{a}_J):\ \vec{b}_I\in \Cl^{(k)}[\vec{a}_J]\big\} \label{ineq_add_1}\\
=&\sum_{\text{all } \Cl^{(k)}[\vec{a}_J]}\#\{ g_C(\vec{b}_I, \vec{a}_J):\ \vec{b}_I\in \Cl^{(k)}[\vec{a}_J]\}. \label{ineq3}
\end{align}



\subsection{Partition Equivalence Relation}

For the cut set $C\in \Lambda(\mN)$, let $\mP_C=\{C_1,C_2,\cdots, C_m \}$ be a strong partition of $C$ (cf.~Definition~\ref{def:strong_parti}). Let $I_l=I_{C_l}$ for $l=1,2, \cdots, m$ and accordingly $L=I\setminus(\bigcup_{l=1}^m I_l)$. Now, we rewrite the sets in the summation in \eqref{ineq3} as follows:
\begin{align}
& \big\{ g_C(\vec{b}_I, \vec{a}_J): \ \vec{b}_I\in \Cl^{(k)}[\vec{a}_J] \big\}\label{ineq:1}\\
&= \big\{ g_C(\vec{b}_{I_1}, \vec{b}_{I_2}, \cdots, \vec{b}_{I_m}, \vec{b}_{L}, \vec{a}_J):
  \ \vec{b}_I=(\vec{b}_{I_1}, \vec{b}_{I_2}, \cdots, \vec{b}_{I_m}, \vec{b}_{L})\in \Cl^{(k)}[\vec{a}_J] \big\}\\
&= \big\{ \big(g_{C_l}(\vec{b}_{I_l}, \vec{b}_{L}, \vec{a}_J),\ l=1,2,\cdots,m \big):
\ \vec{b}_I=(\vec{b}_{I_1}, \vec{b}_{I_2}, \cdots, \vec{b}_{I_m}, \vec{b}_{L})\in \Cl^{(k)}[\vec{a}_J]\big\} \label{ineq4} \\
&= \bigcup_{\vec{b}_{L}\in \mA^{k\times L}}
\big\{ \big(g_{C_l}(\vec{b}_{I_l}, \vec{b}_{L}, \vec{a}_J),\ l=1,2,\cdots,m \big): \ \vec{b}_{I_l} \in \mA^{k\times I_l}, l=1,2,\cdots,m, \textrm{ and } \nonumber\\[-4mm]
&\quad \qquad \qquad \qquad \qquad \qquad \qquad \qquad \qquad \qquad \quad\   (\vec{b}_{I_1}, \vec{b}_{I_2}, \cdots, \vec{b}_{I_m}, \vec{b}_{L})
\in \Cl^{(k)}[\vec{a}_J]\big\},\label{ineq5_0}
\end{align}
where \eqref{ineq4} follows from that for each $l$, the value of $g_{C_l}$ does not depend on $\vec{b}_{I_j}$, $1\leq j \leq m$ and $j\neq l$. Further, for any $\vec{a}_{L}\in \mA^{k\times L}$, we have
\begin{align}
{\rm RHS\ of\ }\eqref{ineq5_0}&\supseteq \big\{ \big(g_{C_l}(\vec{b}_{I_l}, \vec{a}_{L}, \vec{a}_J),\ l=1,2,\cdots,m \big): \ \vec{b}_{I_l} \in \mA^{k\times I_l}, l=1,2,\cdots,m, \textrm{ and } \nonumber\\
& \quad \qquad \qquad \qquad \qquad \qquad \qquad \qquad \qquad\   (\vec{b}_{I_1}, \vec{b}_{I_2}, \cdots, \vec{b}_{I_m}, \vec{a}_{L})
\in \Cl^{(k)}[\vec{a}_J]\big\}\label{ineq5}
\end{align}

Next, we give the definition of {\em partition equivalence relation}, and observe that Definition~\ref{defn:Par_Equ_Relation_ScalarCase} is the special case with $k=1$. The importance of this relation will become clear in Lemma~\ref{lem_second_equi_relation}.

\begin{defn}[{Partition Equivalence Relation}]\label{defn_P_E_Relation}
Let $I$ and $J$ be two disjoint subsets of $S$. Let $I_l$, $l=1,2, \cdots, m$ be $m$ disjoint subsets of $I$ and accordingly $L=I\setminus(\bigcup_{l=1}^m I_l)$. Given $\vec{a}_J\in \mA^{k\times J}$ and $\vec{a}_L\in \mA^{k\times L}$, for $1 \leq l \leq m$, we say that $\vec{b}_{I_l}$ and $\vec{b}'_{I_l}$ in $\mA^{k\times I_l}$ are $(I_l, \vec{a}_L, \vec{a}_J)$-equivalent if for each $\vec{c}_{I_j} \in \mA^{k\times I_j}$ with $1 \leq j \leq m$ and $j\neq l$, $(\vec{b}_{I_l}, \vec{a}_{L}, \vec{c}_{I_j},\ 1 \leq j \leq m, j\neq l)$ and $(\vec{b}'_{I_l}, \vec{a}_{L}, \vec{c}_{I_j},\ 1 \leq j \leq m, j\neq l)$ in $\mA^{k \times I}$ are $(I, \vec{a}_J)$-equivalent.
\end{defn}

We remark that Definition~\ref{defn_P_E_Relation} depends only on the target function $f$ but not on the network $\mN$, and evidently, every relation above is an equivalence relation.

\begin{lemma}\label{lem_second_equi_relation}
Let $\{g_e: e\in \mE\}$ be the set of global encoding functions of a $(k,n)$ network code that can compute $f$ over $\mN$. For a cut set $C$ in $\Lambda(\mN)$ with a strong partition $\mP_C=\{C_1,C_2,\cdots, C_m \}$, let $I=I_{C}$, $J=J_{C}$, and $I_l=I_{C_l}$ for $l=1,2, \cdots, m$ and accordingly $L=I\setminus(\bigcup_{l=1}^m I_l)$. Fix $\vec{a}_J\in \mA^{k\times J}$ and $\vec{a}_L\in \mA^{k\times L}$. Then for each $1\leq l \leq m$ and any two source inputs $\vec{b}_{I_l}$ and $\vec{b}'_{I_l}$ in $\mA^{k\times I_l}$ that are not $(I_l,\vec{a}_L, \vec{a}_J)$-equivalent, it is necessary that $g_{C_l}(\vec{b}_{I_l}, \vec{a}_{L}, \vec{a}_J)\neq g_{C_l}(\vec{b}'_{I_l}, \vec{a}_{L}, \vec{a}_J)$.
\end{lemma}
\begin{IEEEproof}
Without loss of generality, it suffices to prove the lemma for $l=1$ only.
Consider two source inputs $\vec{b}_{I_1}$ and $\vec{b}'_{I_1}$ in $\mA^{k\times I_1}$ that are not $(I_1, \vec{a}_L, \vec{a}_J)$-equivalent. Then there exist $\vec{c}_{I_j}\in \mA^{k\times I_j}$ for $j=2,3,\cdots,m$ such that $\vec{b}_I\triangleq(\vec{b}_{I_1}, \vec{c}_{I_2}, \cdots, \vec{c}_{I_m}, \vec{a}_{L})$ and $\vec{b}'_I\triangleq(\vec{b}'_{I_1}, \vec{c}_{I_2}, \cdots, \vec{c}_{I_m}, \vec{a}_{L})$ are not $(I,\vec{a}_J)$-equivalent. In other words, there exists $\vec{d}\in \mA^{k\times S\backslash (I\cup J)}$ such that
\begin{align}\label{ineq_pf_lem_second_equi_relation}
f(\vec{b}_I, \vec{a}_J, \vec{d})\neq f(\vec{b}'_I, \vec{a}_J, \vec{d}).
\end{align}

Next, let $D=\bigcup_{\sigma\in (S\setminus I)}\eout(\sigma)$, an edge subset of $\mE$. Then $\widehat{C}=C\cup D$ is a global cut set, i.e., $I_{\widehat{C}}=S$. Since $g_{\ein(\rho)}(\vec{x}_S)$ is a function of $g_{\widehat{C}}(\vec{x}_S)$ and the network code can compute $f$, \eqref{ineq_pf_lem_second_equi_relation} implies that
\begin{align*}
g_{\widehat{C}}(\vec{b}_I, \vec{a}_J, \vec{d})\neq g_{\widehat{C}}(\vec{b}'_I, \vec{a}_J, \vec{d}).
\end{align*}
Equivalently,
\begin{align*}
\big( g_{C}(\vec{b}_I, \vec{a}_J),\ g_{D}(\vec{a}_J, \vec{d})\big)=g_{\widehat{C}}(\vec{b}_I, \vec{a}_J, \vec{d})\neq g_{\widehat{C}}(\vec{b}'_I, \vec{a}_J, \vec{d})
=\big( g_{C}(\vec{b}'_I, \vec{a}_J),\ g_{D}(\vec{a}_J, \vec{d})\big).
\end{align*}
By comparing the left hand side and the right hand side above, we immediately obtain $g_{C}(\vec{b}_I, \vec{a}_J)\neq g_{C}(\vec{b}'_I, \vec{a}_J)$, i.e.,
\begin{align*}
&\big(g_{C_1}(\vec{b}_{I_1}, \vec{a}_{L}, \vec{a}_J),\ g_{{C}_j}(\vec{c}_{I_j}, \vec{a}_{L}, \vec{a}_J), j=2,3,\cdots,m \big)\\
&\neq \big(g_{C_1}(\vec{b}'_{I_1}, \vec{a}_{L}, \vec{a}_J),\ g_{C_j}(\vec{c}_{I_j}, \vec{a}_{L}, \vec{a}_J), j=2,3,\cdots,m \big),
\end{align*}
which implies $g_{C_1}(\vec{b}_{I_1}, \vec{a}_{L}, \vec{a}_J)\neq g_{C_1}(\vec{b}'_{I_1}, \vec{a}_{L}, \vec{a}_J)$. The lemma is proved.
\end{IEEEproof}

For $l=1,2,\cdots,m$, we use $\cl_{I_l}[\vec{a}_{L}, \vec{a}_J]$ to denote an $(I_l, \vec{a}_L, \vec{a}_J)$-equivalence class. All $(I_l, \vec{a}_L, \vec{a}_J)$-equivalence classes form a partition of $\mA^{k\times I_l}$. When $\vec{a}_{L}$ and $\vec{a}_J$ are clear from the context, we write $\cl_{I_l}[\vec{a}_{L}, \vec{a}_J]$ as $\cl^{(k)}_{I_l}$ to simplify notation. In the following, we give a lemma that reduces to Lemma~\ref{prop_cl} for the case $k=1$.

\begin{lemma}\label{prop_cl_vector_type}
For any set of $(I_l, \vec{a}_{L}, \vec{a}_J)$-equivalence classes $\cl^{(k)}_{I_l}$, $l=1, 2, \cdots, m$, define the set
\begin{align*}
\big\langle \cl^{(k)}_{I_1}, \cl^{(k)}_{I_2}, \cdots, \cl^{(k)}_{I_m}, \vec{a}_{L}\big \rangle
\triangleq \Big\{ (\vec{b}_{I_1}, \vec{b}_{I_2}, \cdots, \vec{b}_{I_m}, \vec{a}_{L}):
\ \vec{b}_{I_l}\in \cl^{(k)}_{I_l}, l=1,2,\cdots,m \Big\}\subseteq \mA^{k\times I}.
\end{align*}
Then all source inputs $(\vec{b}_{I_1}, \vec{b}_{I_2}, \cdots, \vec{b}_{I_m}, \vec{a}_{L})$ in $\big\langle \cl^{(k)}_{I_1}, \cl^{(k)}_{I_2}, \cdots, \cl^{(k)}_{I_m}, \vec{a}_{L}\big \rangle$ are $(I,\vec{a}_J)$-equivalent. In other words, there exists an $(I,\vec{a}_J)$-equivalence class $\Cl^{(k)}[\vec{a}_J]$ such that
$$\big\langle \cl^{(k)}_{I_1}, \cl^{(k)}_{I_2}, \cdots, \cl^{(k)}_{I_m}, \vec{a}_{L} \big\rangle \subseteq \Cl^{(k)}[\vec{a}_J].$$
\end{lemma}
\begin{IEEEproof}
Let $\vec{b}_{I_l}$ and $\vec{b}'_{I_l}$ be arbitrarily two source matrices in $\cl^{(k)}_{I_l}$ for $l=1,2,\cdots,m$. Throughout this proof, we write $\vec{x}_I \sim \vec{y}_I$ for $\vec{x}_I,\vec{y}_I\in \mA^{k\times I}$ if $\vec{x}_I$ and $\vec{y}_I$ are $(I, \vec{a}_J)$-equivalent.

Next, we will prove that for $1\leq l \leq m$,
\begin{align*}
(\vec{b}_{I_1}, \vec{b}_{I_2}, \cdots, \vec{b}_{I_m}, \vec{a}_{L}) \sim (\vec{b}'_{I_1}, \vec{b}'_{I_2} \cdots, \vec{b}'_{I_l}, \vec{b}_{I_{l+1}}, \cdots, \vec{b}_{I_{m}}, \vec{a}_{L})
\end{align*}
by induction on $l$. In particular, when $l=m$, we have
\begin{align*}
(\vec{b}_{I_1}, \vec{b}_{I_2}, \cdots, \vec{b}_{I_m}, \vec{a}_{L}) \sim (\vec{b}'_{I_1}, \vec{b}'_{I_2} \cdots, \vec{b}'_{I_m}, \vec{a}_{L}).
\end{align*}
This proves the lemma.

First, since $\vec{b}_{I_1}$ and $\vec{b}'_{I_1}$ are $(I_1,\vec{a}_L,\vec{a}_J)$-equivalent, by Definition~\ref{defn_P_E_Relation}, we have \begin{align*}
(\vec{b}_{I_1}, \vec{b}_{I_2}, \cdots, \vec{b}_{I_m}, \vec{a}_{L}) \sim
(\vec{b}'_{I_1}, \vec{b}_{I_2}, \cdots, \vec{b}_{I_m}, \vec{a}_{L}).
\end{align*}
Assume that
\begin{align}\label{equ:assumption}
(\vec{b}_{I_1}, \vec{b}_{I_2}, \cdots, \vec{b}_{I_m}, \vec{a}_{L}) \sim
(\vec{b}'_{I_1}, \cdots, \vec{b}'_{I_l}, \vec{b}_{I_{l+1}}, \cdots, \vec{b}_{I_{m}}, \vec{a}_{L})
\end{align}
for some $1 \leq l < m$. We now prove that
\begin{align}\label{equ:tobeproved}
(\vec{b}_{I_1}, \vec{b}_{I_2}, \cdots, \vec{b}_{I_m}, \vec{a}_{L}) \sim (\vec{b}'_{I_1}, \cdots, \vec{b}'_{I_{l+1}}, \vec{b}_{I_{l+2}}, \cdots, \vec{b}_{I_{m}}, \vec{a}_{L}).
\end{align}
Since $\vec{b}_{I_{l+1}}$ and $\vec{b}'_{I_{l+1}}$ are $(I_{l+1},\vec{a}_L,\vec{a}_J)$-equivalent, we see that
$$(\vec{b}'_{I_1}, \cdots, \vec{b}'_{I_l}, \vec{b}_{I_{l+1}}, \vec{b}_{I_{l+2}}, \cdots, \vec{b}_{I_{m}}, \vec{a}_{L}) \sim (\vec{b}'_{I_1}, \cdots, \vec{b}'_{I_l}, \vec{b}'_{I_{l+1}}, \vec{b}_{I_{l+2}}, \cdots, \vec{b}_{I_{m}}, \vec{a}_{L}).$$
Together with the assumption \eqref{equ:assumption} and the transitivity of the $(I, \vec{a}_J)$-equivalence relation ``$\sim$'', we have proved \eqref{equ:tobeproved} and hence accomplished the proof.
\end{IEEEproof}


\subsection{Derivation of the Improved Upper Bound}

From \eqref{ineq:1} to \eqref{ineq5}, we obtain that for every $\vec{a}_{L}$ in $\mA^{k\times L}$,
\begin{align}
& \# \big\{ g_C(\vec{b}_I, \vec{a}_J): \ \vec{b}_I\in \Cl^{(k)}[\vec{a}_J] \big\}\nonumber\\
& \geq \# \big\{ \big(g_{C_l}(\vec{b}_{I_l}, \vec{a}_{L}, \vec{a}_J),\ l=1,2,\cdots,m \big): \ \vec{b}_{I_l} \in \mA^{k\times I_l}, l=1,2,\cdots,m, \textrm{ and }\nonumber\\
&\quad\ \qquad \qquad \qquad \qquad \qquad \qquad \qquad \qquad \ \   (\vec{b}_{I_1}, \vec{b}_{I_2}, \cdots, \vec{b}_{I_m}, \vec{a}_{L})
\in \Cl^{(k)}[\vec{a}_J]\big\}. \label{ineq_for_every}
\end{align}
We now derive a lower bound on \eqref{ineq_for_every}; the steps are explained after the derivation.
\begin{align}
&\# \big\{ \big(g_{C_l}(\vec{b}_{I_l}, \vec{a}_{L}, \vec{a}_J),\ l=1,2,\cdots,m \big): \ \vec{b}_{I_l} \in \mA^{k\times I_l}, l=1,2,\cdots,m, \textrm{ and }\nonumber\\
&\quad\ \qquad \qquad \qquad \qquad \qquad \qquad \qquad\quad (\vec{b}_{I_1}, \vec{b}_{I_2}, \cdots, \vec{b}_{I_m}, \vec{a}_{L})
\in \Cl^{(k)}[\vec{a}_J]\big\}\nonumber\\
&=\# \big\{ \big(g_{C_l}(\vec{b}_{I_l}, \vec{a}_{L}, \vec{a}_J),\ l=1,2,\cdots,m \big): \vec{b}_{I_l} \in \cl^{(k)}_{I_l}, \text{ an $(I_l, \vec{a}_{L}, \vec{a}_J)$-equivalence class, } 1 \leq l \leq m,\nonumber\\
& \qquad\qquad \qquad \qquad \qquad\qquad\qquad\qquad\   \textrm{ and }\big\langle \cl^{(k)}_{I_1}, \cl^{(k)}_{I_2}, \cdots, \cl^{(k)}_{I_m}, \vec{a}_{L} \big\rangle \subseteq \Cl^{(k)}[\vec{a}_J]\big\}\label{ineq_for_every_2}\\
&\geq \# \big\{ \big( \cl^{(k)}_{I_1}, \cl^{(k)}_{I_2}, \cdots, \cl^{(k)}_{I_m} \big):\
       \cl^{(k)}_{I_l} \text{ is an $(I_l, \vec{a}_{L}, \vec{a}_J)$-equivalence class, } l=1,2,\cdots,m, \nonumber \\
& \qquad\qquad \qquad \qquad \qquad\qquad\qquad\qquad\   \textrm{ and } \big\langle \cl^{(k)}_{I_1}, \cl^{(k)}_{I_2}, \cdots, \cl^{(k)}_{I_m}, \vec{a}_{L} \big\rangle \subseteq \Cl^{(k)}[\vec{a}_J]\big\}.\label{ineq6}
\end{align}
\begin{itemize}
  \item The equality~\eqref{ineq_for_every_2} is justified by establishing the following:
\begin{align}
&\big\{ (\vec{b}_{I_1}, \vec{b}_{I_2}, \cdots, \vec{b}_{I_m}, \vec{a}_L): \vec{b}_{I_l} \in \mA^{k\times I_l}, l=1,2,\cdots,m, \textrm{ and } (\vec{b}_{I_1}, \vec{b}_{I_2}, \cdots, \vec{b}_{I_m}, \vec{a}_{L})\in \Cl^{(k)}[\vec{a}_J]\big\}\nonumber\\
&=\bigcup_{ \textrm{ all } \left(\cl^{(k)}_{I_1}, \cl^{(k)}_{I_2}, \cdots, \cl^{(k)}_{I_m}\right) \textrm{ s.t. } \atop \left\langle \cl^{(k)}_{I_1}, \cl^{(k)}_{I_2}, \cdots, \cl^{(k)}_{I_m}, \vec{a}_{L} \right\rangle \subseteq \Cl^{(k)}[\vec{a}_J]} \big\langle \cl^{(k)}_{I_1}, \cl^{(k)}_{I_2}, \cdots, \cl^{(k)}_{I_m}, \vec{a}_{L} \big\rangle.  \label{equ_2sets}
\end{align}
To see \eqref{equ_2sets}, we first consider an arbitrary $(\vec{b}_{I_1}, \vec{b}_{I_2}, \cdots, \vec{b}_{I_m}, \vec{a}_L)$ in LHS of \eqref{equ_2sets}, i.e.,
\begin{align}\label{equ1_pf_equ_2sets}
(\vec{b}_{I_1}, \vec{b}_{I_2}, \cdots, \vec{b}_{I_m}, \vec{a}_{L})\in \Cl^{(k)}[\vec{a}_J].
\end{align}
Let $\cl^{(k)}_{I_l}$ be the corresponding $(I_l, \vec{a}_{L}, \vec{a}_J)$-equivalence class containing $\vec{b}_{I_l}$ for $1\leq l \leq m$. Then
\begin{align}\label{equ2_pf_equ_2sets}
(\vec{b}_{I_1}, \vec{b}_{I_2}, \cdots, \vec{b}_{I_m}, \vec{a}_L)\in \big\langle \cl^{(k)}_{I_1}, \cl^{(k)}_{I_2}, \cdots, \cl^{(k)}_{I_m}, \vec{a}_{L} \big\rangle.
\end{align}
Combining \eqref{equ1_pf_equ_2sets} and \eqref{equ2_pf_equ_2sets} and by Lemma~\ref{prop_cl_vector_type}, we have
\begin{align}\label{equ3_pf_equ_2sets}
(\vec{b}_{I_1}, \vec{b}_{I_2}, \cdots, \vec{b}_{I_m}, \vec{a}_L)\in \big\langle \cl^{(k)}_{I_1}, \cl^{(k)}_{I_2}, \cdots, \cl^{(k)}_{I_m}, \vec{a}_{L} \big\rangle \subseteq \Cl^{(k)}[\vec{a}_J],
\end{align}
which shows that LHS of \eqref{equ_2sets} is a subset of RHS of \eqref{equ_2sets}. On the other hand, it is evident that RHS of \eqref{equ_2sets} is a subset of LHS of \eqref{equ_2sets}, proving \eqref{equ_2sets}. Immediately, \eqref{equ_2sets} implies \eqref{ineq_for_every_2}.

\item The inequality~\eqref{ineq6} is proved as follows. For every $\big( \cl^{(k)}_{I_1}, \cl^{(k)}_{I_2}, \cdots, \cl^{(k)}_{I_m} \big)$ in the set on the RHS of \eqref{ineq6}, we arbitrarily choose a vector $(\vec{b}_{I_1}, \vec{b}_{I_2}, \cdots, \vec{b}_{I_m})$ such that $\vec{b}_{I_l}\in \cl^{(k)}_{I_l}$, for $l=1,2,\cdots,m$. For any two distinct $\big( \cl^{(k)}_{I_1}, \cl^{(k)}_{I_2}, \cdots, \cl^{(k)}_{I_m} \big)$ and $\big( \cl'^{(k)}_{I_1}, \cl'^{(k)}_{I_2}, \cdots, \cl'^{(k)}_{I_m} \big)$, let $(\vec{b}_{I_1}, \vec{b}_{I_2}, \cdots, \vec{b}_{I_m})$ and $(\vec{b}'_{I_1}, \vec{b}'_{I_2}, \cdots, \vec{b}'_{I_m})$ be the corresponding vectors that have been chosen. Then by Lemma~\ref{lem_second_equi_relation}, we have
\begin{align}\label{equ1_pf_ineq6}
\big( g_{C_l}(\vec{b}_{I_l}, \vec{a}_{L}, \vec{a}_J),\ l=1,2,\cdots,m\big)\neq
\big( g_{C_l}(\vec{b}'_{I_l}, \vec{a}_{L}, \vec{a}_J),\ l=1,2,\cdots,m\big),
\end{align}
which implies \eqref{ineq6}.

\end{itemize}

We denote the RHS of \eqref{ineq6} by $N\big(\vec{a}_{L}, \Cl^{(k)}[\vec{a}_J]\big)$, which is consistent with the notation $N\big({a}_{L}, \Cl[{a}_J]\big)$ in \eqref{no_finer_eq_cl} for the case $k=1$.


We write $\vec{a}_J=\left(a_{J,1}, a_{J,2}, \cdots, a_{J,k}\right)^\top$, where $a_{J,p} \in \mA^{J}$, $p=1,2,\cdots,k$, are the rows of $\vec{a}_J\in \mA^{k\times J}$. Let $\vec{b}_I$ and $\vec{b}'_I$ in $\mA^{k\times I}$ be two source inputs. Similarly, we write $\vec{b}_I= \big(b_{I,1}, b_{I,2}, \cdots, b_{I,k}\big)^\top$ and $\vec{b}'_I= \big(b'_{I,1}, b'_{I,2}, \cdots, b'_{I,k}\big)^\top$ with $b_{I,p}, b'_{I,p}\in \mA^{I}$ for $1\leq p \leq k$.

By Definition~\ref{def:ec}, we see that $\vec{b}_I$ and $\vec{b}'_I$ are $(I,\vec{a}_{J})$-equivalent if and only if $b_{I,p}$ and $b'_{I,p}$ are $(I,a_{J,p})$-equivalent for all $1\leq p \leq k$. Thus, every $(I,\vec{a}_J)$-equivalence class corresponds to a set of $(I,a_{J,p})$-equivalence classes, $p=1,2,\cdots,k$. On the other hand, every set of $(I,a_{J,p})$-equivalence classes, $p=1,2,\cdots,k$, also corresponds to an $(I,\vec{a}_J)$-equivalence class.
For an $(I,\vec{a}_{J})$-equivalence class $\Cl^{(k)}[\vec{a}_J]$, denote the corresponding set of $(I,a_{J,p})$-equivalence classes by $\big\{\Cl_{p}[a_{J,p}],\ p=1,2,\cdots,k\big\}$.

Next, we consider the $(I_l, \vec{a}_{L}, \vec{a}_J)$-equivalence relation, $1\leq l\leq m$, and obtain a similar result. To be specific, we also write $\vec{a}_{L}=\left(a_{L,1}, a_{L,2}, \cdots, a_{L,k}\right)^\top$ with $a_{L,p} \in \mA^{L}$, $p=1,2,\cdots,k$ being the rows of $\vec{a}_L\in \mA^{k\times L}$, and consider two source inputs $\vec{b}_{I_l}= \big(b_{I_l,1}, b_{I_l,2}, \cdots, b_{I_l,k}\big)^\top$ and $\vec{b}'_{I_l}= \big(b'_{I_l,1}, b'_{I_l,2}, \cdots, b'_{I_l,k}\big)^\top$ in $\mA^{k\times I_l}$ with $b_{I_l,p}, b'_{I_l,p}\in \mA^{I_l}$ for $1\leq p \leq k$. Similarly, by Definition~\ref{defn_P_E_Relation}, $\vec{b}_{I_l}$ and $\vec{b}'_{I_l}$ are $(I_l,\vec{a}_{L}, \vec{a}_{J})$-equivalent if and only if $b_{I_l,p}$ and $b'_{I_l,p}$ are $(I_l,a_{L,p},a_{J,p})$-equivalent for all $1\leq p \leq k$. Thus, every $(I_l,\vec{a}_{L}, \vec{a}_{J})$-equivalence class corresponds to a set of $(I_l, a_{L,p}, a_{J,p})$-equivalence classes, $1\leq p \leq k$, and vice versa. For an $(I_l,\vec{a}_{L}, \vec{a}_{J})$-equivalence class $\cl^{(k)}_{I_l}$, denote the corresponding set of $(I_l, a_{L,p}, a_{J,p})$-equivalence classes by $\big\{\cl_{I_l,p},\ p=1,2,\cdots,k\big\}$.

We now consider $(I_l, \vec{a}_{L}, \vec{a}_J)$-equivalence classes $\cl^{(k)}_{I_l}$, $1\leq l \leq m$. Based on the above arguments, we obtain that
\begin{align}
\big\langle \cl^{(k)}_{I_1}, \cl^{(k)}_{I_2}, \cdots, \cl^{(k)}_{I_m}, \vec{a}_{L}\big \rangle \subseteq \Cl^{(k)}[\vec{a}_J]
\end{align}
if and only if
\begin{align}
\big\langle \cl_{I_1,p}, \cl_{I_2,p}, \cdots, \cl_{I_m,p}, a_{L,p}\big\rangle \subseteq \Cl_p[a_{J,p}],\quad \forall~p=1,2,\cdots,k.
\end{align}
Hence, this implies that
\begin{align}\label{prod}
N\left(\vec{a}_{L}, \Cl^{(k)}[\vec{a}_J]\right)=\prod_{p=1}^{k} N\big(a_{L,p}, \Cl_{p}[a_{J,p}]\big).
\end{align}

Considering all $\vec{a}_L$ in $\mA^{k\times L}$ and by combining \eqref{ineq_for_every}-\eqref{ineq6} with \eqref{prod}, we have
\begin{align}
& \#\Big\{ g_C(\vec{b}_I, \vec{a}_J): \ \vec{b}_I\in \Cl^{(k)}[\vec{a}_J]\Big\}\nonumber\\
& \geq \max_{\vec{a}_L \in \mA^{k\times L}} N\big(\vec{a}_{L}, \Cl^{(k)}[\vec{a}_J]\big)\label{prod-1}\\
& = \max_{\vec{a}_L \in \mA^{k\times L}} \prod_{p=1}^{k} N\big(a_{L,p}, \Cl_{p}[a_{J,p}]\big)\label{prod-2}\\
& = \max_{a_{L,1} \in \mA^{L}}~\max_{a_{L,2} \in \mA^{L}}\cdots\max_{a_{L,k} \in \mA^{L}} \prod_{p=1}^{k} N\big(a_{L,p}, \Cl_{p}[a_{J,p}]\big)\\
& = \prod_{p=1}^{k} \max_{a_{L,p} \in \mA^{L}} N\big(a_{L,p}, \Cl_{p}[a_{J,p}]\big)\\
& = \prod_{p=1}^{k} N\big(\Cl_{p}[a_{J,p}]\big),\label{a_star}
\end{align}
where \eqref{a_star} follows from the definition in \eqref{equ:N_Cl}.

We now combine \eqref{ineq_add_1}, \eqref{ineq3}, and \eqref{prod-1}-\eqref{a_star} to obtain
\begin{align}
|\mA|^{n|C|}\geq & \sum_{\text{all }\Cl^{(k)}[\vec{a}_J]}\left[\prod_{p=1}^{k} N\big(\Cl_{p}[a_{J,p}]\big)\right]\label{ineq_final-1}\\
= & \sum_{\text{all }\Cl_{1}[a_{J,1}]~}\sum_{\text{all }\Cl_{2}[a_{J,2}]}\cdots\sum_{\text{all }\Cl_{k}[a_{J,k}]}
     \left[\prod_{p=1}^{k} N\big(\Cl_{p}[a_{J,p}]\big)\right]\label{ineq_final-2}\\
= & \prod_{p=1}^{k}\left[ \sum_{\text{all }\Cl_{p}[a_{J,p}]}  N\big(\Cl_{p}[a_{J,p}]\big)\right].\label{ineq_final-3}
\end{align}
Note that the inequality \eqref{ineq_final-1} holds for an arbitrary $\vec{a}_J\in \mA^{k\times J}$, or equivalently, arbitrary $a_{J,p} \in \mA^{J}$, $p=1,2,\cdots, k$. Let
\begin{align*}
{a}_J^*\in \arg\max_{{a}_J\in \mA^{J}} \sum_{\text{all }\Cl[{a}_J]} N\big(\Cl[{a}_J]\big),
\end{align*}
i.e.,
\begin{align*}
\sum_{\text{all }\Cl[{a}_J^*]} N\big(\Cl[{a}_J^*]\big)=\max_{{a}_J\in \mA^{J}} \sum_{\text{all }\Cl[{a}_J]} N\big(\Cl[{a}_J]\big).
\end{align*}
Then it follows from \eqref{ineq_final-1}-\eqref{ineq_final-3} that
\begin{align}
&|\mA|^{n|C|}\geq \left[ \sum_{\text{all }\Cl[{a}_J^*]} N\big( \Cl[{a}_J^*]\big)\right]^k. \label{ineq_k_n}
\end{align}

For the strong partition $\mP_C=\{ C_1,C_2,\cdots,C_m \}$ of the cut set $C$, recall from \eqref{n_C_Parti_1st} and \eqref{n_C_f_1st}, the definitions of $n_C(\mP_C)$ and $n_{C,f}$, respectively. Since the inequality \eqref{ineq_k_n} is valid for all strong partitions $\mP_C$ of $C$, we have
$$|\mA|^{n|C|}\geq n_{C,f}^k,$$
or equivalently,
\begin{align}\label{ineq_upp_bound_k_n}
\frac{k}{n}\leq \dfrac{|C|}{\log_{|\mA|}n_{C,f}}.
\end{align}
Finally, considering all cut sets $C\in \Lambda(\mN)$, we obtain by \eqref{ineq_upp_bound_k_n} that
\begin{align}\label{equ:improved_upper_bound_re}
\mathcal{C}(\mathcal{N},f)\leq \min_{C\in\Lambda(\mN)}\dfrac{|C|}{\log_{|\mA|}n_{C,f}}.
\end{align}
Therefore, we have proved Theorem~\ref{thm:upper_bound}.

\begin{journalonly}
Let $k=1$ and then we obtain the corresponding results for the scalar case. Concretely, for the given source vectors ${a}_J\in \mA^{J}$ and $a_L\in \mA^{L}$, we have the following matrices:
\begin{align*}
M(a_L)=\Big[M_{i,j}(a_L)\Big]_{1\leq i \leq V_{I_1}(a_{L}, {a}_J),\ 1\leq j \leq V_{I_2}(a_{L}, {a}_J)}.
\end{align*}
Subsequently, we write
\begin{align}
\vec{a}_J&=({a}_J^{(1)}, {a}_J^{(2)}, \cdots, {a}_J^{(k)})^\top,\label{nota_{a}_J}\\
\vec{a}_{L}&=(a_{L}^{(1)}, a_{L}^{(2)}, \cdots, a_{L}^{(k)})^\top,\label{nota_a_L}
\end{align}
and the corresponding equivalence class $\Cl^{(k)}[\vec{a}_J]$ is written as
\begin{align}
\Cl^{(k)}[\vec{a}_J]=\big(\Cl^{(1)}({a}_J^{(1)}),\ \Cl^{(1)}({a}_J^{(2)}),\ \cdots,\ \Cl^{(1)}({a}_J^{(k)})\big)^\top,\label{nota_Cl_J}
\end{align}
which is considered as a column vector constituted by $k$ scalar equivalence classes. For all $p$, $1\leq p \leq k$, one has the $k$ corresponding matrices as follows
\begin{align*}
M(a_L^{(p)})=\Big[M_{i,j}(a_L^{(p)})\Big]_{1\leq i \leq V_{I_1}(a_{L}^{(p)}, c_{J,p}),\ 1\leq j \leq V_{I_2}(a_{L}^{(p)}, c_{J,p})},
\end{align*}
and the value $N(a_{L}^{(p)}, \Cl^{(1)}(c_{J,p}))$ is the number of the entries $M_{i,j}(a_L^{(p)})$ equal to $\Cl^{(1)}(c_{J,p})$ in the matrix $M(a_L^{(p)})$. Furthermore, it is not difficult to see that
\begin{align}\label{prod}
N(\vec{a}_{L}, \Cl^{(k)}[\vec{a}_J])=\prod_{p=1}^{k} N(a_{L}^{(p)}, \Cl^{(1)}(c_{J,p})).
\end{align}

Therefore, we derive that for $1\leq i \leq W_{C,f}^{(\vec{a}_J)}$,
\begin{align}\label{ineq7}
(\ref{ineq6})=N(\vec{a}_{L,(i)}, \Cl^{(k)}_i(\vec{a}_J))=\prod_{p=1}^{k} N(a_{L,(i)}^{(p)}, \Cl^{(1)}_i(c_{J,p})).
\end{align}

In order to clear the proof, before discussion further we review all the inequalities from (\ref{ineq1}) to (\ref{ineq7}) as follows
\begin{align*}
&|\mA|^{n|C|}\\
\geq &\#\{ g_C(\vec{x}_I, \vec{x}_J):\ \text{\ all } \vec{x}_I\in \mA^{k\times I}, \vec{x}_J\in \mA^{k\times J}\} \tag*{// by (\ref{ineq1})}\\
\geq & \#\{ g_C(\vec{x}_I, \vec{a}_J): \text{ all } \vec{x}_I\in \mA^{k\times I}\} \tag*{ // for every  $\vec{a}_J\in \mA^{k\times J}$ by (\ref{ineq2})}\\
=&\sum_{i=1}^{W_{C,f}^{(\vec{a}_J)}}\#\{ g_C(\vec{b}_I, \vec{a}_J):\text{ all } \vec{b}_I\in \Cl^{(k)}_i(\vec{a}_J)\}\tag*{// by (\ref{ineq3})}\\
=&\sum_{i=1}^{W_{C,f}^{(\vec{a}_J)}}\#\{ \big(g_{C_1}(\vec{a}_{I_1}, \vec{a}_{L}, \vec{a}_J), g_{C_2}(\vec{a}_{I_2}, \vec{a}_{L}, \vec{a}_J) \big): \\
& \qquad\qquad\qquad\quad \text{ all } \vec{b}_I=(\vec{a}_{I_1}, \vec{a}_{L}, \vec{a}_{I_2})\in \Cl^{(k)}_i(\vec{a}_J)\} \tag*{// by (\ref{ineq4})}\\
=&\sum_{i=1}^{W_{C,f}^{(\vec{a}_J)}}\#\{ \big(g_{C_1}(\vec{a}_{I_1}, \vec{a}_{L}), g_{C_2}(\vec{a}_{I_2}, \vec{a}_{L}) \big): \\
&\qquad\qquad\qquad\quad \text{ all } \vec{b}_I=(\vec{a}_{I_1}, \vec{a}_{L}, \vec{a}_{I_2})\in \Cl^{(k)}_i(\vec{a}_J)\} \tag*{// just for notation simplicity by (\ref{ineq5}) }\\
\geq &\sum_{i=1}^{W_{C,f}^{(\vec{a}_J)}}\#\mathop{\cup}_{\text{all }\vec{a}_{L}\in \mA^{k\times L}} \{ \big(g_{C_1}(\vec{a}_{I_1}, \vec{a}_{L}), g_{C_2}(\vec{a}_{I_2}, \vec{a}_{L}) \big):\ \text{all } \\
& \vec{a}_{I_1}\in \mA^{k\times I_1}, \vec{a}_{I_2}\in \mA^{k\times I_2} \textrm{ s.t. } (\vec{a}_{I_1}, \vec{a}_{L}, \vec{a}_{I_2})\in \Cl^{(k)}_i(\vec{a}_J)\}\nonumber\\
\geq & \sum_{i=1}^{W_{C,f}^{(\vec{a}_J)}}\#\{ \big(g_{C_1}(\vec{a}_{I_1}, \vec{a}_{L,(i)}), g_{C_2}(\vec{a}_{I_2}, \vec{a}_{L,(i)}) \big):\ \\
&\quad \vec{a}_{I_l}\in \mA^{k\times I_l}, l=1,2, \textrm{ s.t. } (\vec{a}_{I_1}, \vec{a}_{L,(i)}, \vec{a}_{I_2})\in \Cl^{(k)}_i(\vec{a}_J)\}\tag*{ // for all possible  $\vec{a}_{L,(i)}$ by (\ref{ineq_for_every})}\\
\geq & \sum_{i=1}^{W_{C,f}^{(\vec{a}_J)}} \#\{ \big(\cl^{(k)}_{I_1}(\vec{a}_{L,(i)},\vec{a}_J),\ \cl^{(k)}_{I_2}(\vec{a}_{L,(i)}, \vec{a}_J)\big): \\
&\qquad \text{ all } \cl^{(k)}_{I_1}(\vec{a}_{L,(i)},\vec{a}_J), \text{ and } \cl^{(k)}_{I_2}(\vec{a}_{L,(i)},\vec{a}_J), \textrm{ s.t. } \\
& \big(\cl^{(k)}_{I_1}(\vec{a}_{L,(i)},\vec{a}_J),\ \vec{a}_{L,(i)},\ \cl^{(k)}_{I_2}(\vec{a}_{L,(i)}, \vec{a}_J)\big)\subseteq \Cl^{(k)}_i(\vec{a}_J)\}\tag*{}\\
= & \sum_{i=1}^{W_{C,f}^{(\vec{a}_J)}} N(\vec{a}_{L,(i)}, \Cl^{(k)}_i(\vec{a}_J)) \tag*{// by (\ref{ineq6})}\\
= & \sum_{i=1}^{W_{C,f}^{(\vec{a}_J)}} \prod_{p=1}^{k} N(a_{L,(i)}^{(p)}, \Cl^{(1)}_i(c_{J,p})),  \tag*{// by (\ref{ineq7})}
\end{align*}
where recall (\ref{nota_{a}_J}), (\ref{nota_a_L}), and (\ref{nota_Cl_J}) for the meanings of notation $a_{L,(i)}^{(p)}$, $c_{J,p}$ and $\Cl^{(1)}_i(c_{J,p})$. Consequently, we summarize that
\begin{align}\label{ineq_final}
|\mA|^{n|C|}\geq \sum_{i=1}^{W_{C,f}^{(\vec{a}_J)}} \prod_{p=1}^{k} N(a_{L,(i)}^{(p)}, \Cl^{(1)}_i(c_{J,p})),
\end{align}
for arbitrary ${a}_J^{(p)}\in \mA^{J}$, and arbitrary $a_{L,(i)}^{(p)}\in \mA^{L}$ corresponding to the $(I, J, {a}_J^{(p)})$-equivalence class $\Cl^{(1)}_i(c_{J,p})$, $1\leq p \leq k$, $1 \leq i \leq W_{C,f}^{(\vec{a}_J)}$.

On the other hand, notice that for any ${a}_J\in \mA^{J}$, the $(I,{a}_J)$-equivalence relation induces different equivalence classes $\Cl^{(1)}_j({a}_J)$, $1 \leq j \leq W_{C,f}^{({a}_J)}$, constituting a partition of $\mA^{I}$. Then for every equivalence class $\Cl^{(1)}_j({a}_J)$, let
\begin{align*}
a_L^*(\Cl^{(1)}_j({a}_J))=\arg\max_{a_L\in \mA^{L}}N(a_{L}, \Cl^{(1)}_j({a}_J)).
\end{align*}
Let further
\begin{align*}
{a}_J^*=\arg\max_{{a}_J\in \mA^{J}} \sum_{j=1}^{W_{C,f}^{({a}_J)}} N(a_L^*(\Cl^{(1)}_j({a}_J)), \Cl^{(1)}_j({a}_J)),
\end{align*}
and thus we can obtain
\begin{align*}
&\max_{{a}_J\in \mA^{J}} \max_{a_{L,j}\in \mA^{L}:\atop 1\leq j \leq W_{C,f}^{({a}_J)}} \sum_{j=1}^{W_{C,f}^{({a}_J)}} N(a_{L,j}, \Cl^{(1)}_j({a}_J))\\
=&\sum_{j=1}^{W_{C,f}^{({a}_J^*)}} N(a_{L}^*(\Cl^{(1)}_j({a}_J^*)), \Cl^{(1)}_j({a}_J^*))\\
\triangleq &\ n_C(C_1,C_2).
\end{align*}
Together with (\ref{ineq_final}), we deduce that
\begin{align}\label{ineq_k_n}
|\mA|^{n|C|}\geq & \prod_{p=1}^{k} \sum_{j=1}^{W_{C,f}^{({a}_J^*)}} N(a_{L}^*(\Cl^{(1)}_j({a}_J^*)), \Cl^{(1)}_j({a}_J^*))\nonumber\\
= & \Big[ \sum_{j=1}^{W_{C,f}^{({a}_J^*)}} N(a_{L}^*(\Cl^{(1)}_j({a}_J^*)), \Cl^{(1)}_j({a}_J^*))\Big]^k \nonumber\\
= & \big[ n_C(C_1,C_2) \big]^k.
\end{align}

Furthermore, notice that the above inequalities follow for all partitions $C=C_1\cup C_2$ with $I_{C_i}\neq \emptyset$, $i=1,2$. Thus, we choose that partition such that $n_C(C_1,C_2)$ achieves the maximum and denote such a partition by $C=C_1^*\cup C_2^*$, i.e.,
$$n_C(C_1^*, C_2^*)=\max_{C=C_1\cup C_2 \text{ with} \atop I_{C_l}\neq \emptyset,\ l=1,2} n_C(C_1,C_2),$$
further denoted by $n_{C,f}$ because this maximum value $n_C(C_1^*, C_2^*)$ only depends on the cut set $C$ for the given $f$.
Then by (\ref{ineq_k_n}), we can show that
\begin{align*}
|\mA|^{n|C|}\geq \big[ n_C(C_1^*,C_2^*) \big]^k=n_{C,f}^k,
\end{align*}
and equivalently,
\begin{align}\label{ineq_upp_bound_k_n}
\frac{k}{n}\leq \dfrac{|C|}{\log_{|\mA|}n_{C,f}}.
\end{align}

At last, the above inequality (\ref{ineq_upp_bound_k_n}) should be satisfied for all cuts $C$ with $I_C \neq \emptyset$, that is, for all $C\in \Lambda{\mN}$. Therefore, we obtain the upper bound
\begin{align*}
\frac{k}{n}\leq \min_{C\in\Lambda(\mN)}\dfrac{|C|}{\log_{|\mA|}n_{C,f}}.
\end{align*}
Now, we can give the main result below.

\begin{thm}
  \label{thm:upper_bound}
  Let $\mathcal{N}=(G, S, \rho)$ be a network and $f$ be a target function. Then
\begin{align*}
\mathcal{C}(\mathcal{N},f)\leq \min_{C\in\Lambda(\mN)}\dfrac{|C|}{\log_{|\mA|}n_{C,f}}.
\end{align*}
\end{thm}


The upper bound herein is better than the previous one (\ref{eq:2}), since $n_{C,f}\geq w_{C,f}$ always holds for every $C\in \Lambda(\mN)$. Subsequently, we continue using the Example \ref{eg:1} to depict this obtained bound and show that it is tight for this example, i.e., achieves the capacity.
\end{journalonly}


\section{A Nontrivial Example}\label{sec:non_tight}

In the last section, we have proved an improved upper bound in Theorem~\ref{thm:upper_bound} on network function computing capacity. For all previously considered network function computation problems whose computing capacities are known, our improved upper bound is achievable if the computing capacity is rational, or is asymptotically achievable if the computing capacity is irrational, e.g., arbitrary target functions over a multi-edge tree network, the identity function or algebraic sum function over an arbitrary network topology, and the problem previously considered in \cite{Appuswamy11,huang15} (see Fig.~\ref{fig:1} and Example~\ref{eg:2}). Nevertheless, in this section we prove that our improved upper bound is not necessarily achievable even when its value is rational. The result is stated in the following theorem, whose proof is highly nontrivial. This is the first example showing the non-achievability of the improved upper bound when its value is rational.

\begin{figure}[t]
  \centering
{
 \begin{tikzpicture}[x=0.6cm]
    \draw (-3,0) node[vertex] (1) [label=above:$\sigma_1$] {};
    \draw ( 3,0) node[vertex] (2) [label=above:$\sigma_2$] {};
    \draw ( 0,-1.5) node[vertex] (3) [label=left:] {};
    \draw (-3,-5) node[vertex] (4) [label=left:] {};
    \draw ( 3,-5) node[vertex] (5) [label=right:] {};
    \draw ( 0,-3.5) node[vertex] (6) [label=right:] {};
    \draw ( 0,-6.5) node[vertex] (7) [label=below: $\rho$] {};

    \draw[->,>=latex] (1) -- (4) node[midway, auto,swap, left=-1mm] {$e_1$};
    \draw[->,>=latex] (1) -- (3) node[midway, auto, right=-0.5mm] {$e_2$};
    \draw[->,>=latex] (2) -- (3) node[midway, auto,swap, left=-0.5mm] {$e_3$};
    \draw[->,>=latex] (2) -- (5) node[midway, auto, right=-1mm] {$e_4$};
    \draw[->,>=latex] (3) -- (6) node[midway, auto, right=-1mm] {$e_5$};
    \draw[->,>=latex] (6) -- (4) node[midway, auto, left=-0.5mm] {$e_6$};
    \draw[->,>=latex] (6) -- (5) node[midway, auto,swap, right=-0.5mm] {$e_7$};
    \draw[->,>=latex] (4) -- (7) node[midway, auto,swap, right=-0.5mm] {$e_8$};
    \draw[->,>=latex] (5) -- (7) node[midway, auto, left=-0.5mm] {$e_9$};
    \end{tikzpicture}
}
\caption{The reverse butterfly network $\mN$ has two binary sources
  $\sigma_1$ and $\sigma_2$, and one sink $\rho$ that
  computes the binary maximum function of the source messages, i.e.,
  $f(x_1,x_2)=\max\{ x_1, x_2 \}$, where $\mA=\mO=\{0,1\}$ and the elements in $\mA$ and $\mO$ are taken as real numbers.}
  \label{fig:butterfly_network}
\end{figure}
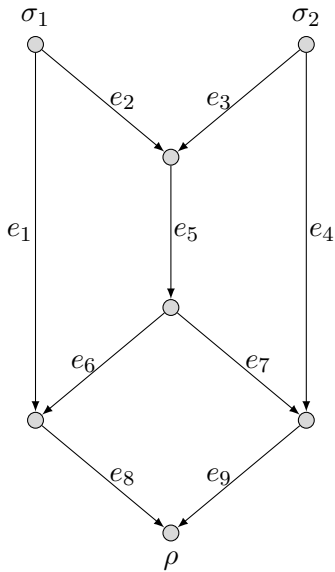

\begin{thm}\label{thm_butterfly_network_non-tight}
For the computation problem of the binary maximum function $f=\max$ over the reverse butterfly network $\mN$ (depicted in Fig.~\ref{fig:butterfly_network}), the upper bound in Theorem~\ref{thm:upper_bound} on the computing capacity $\mC(\mN, f=\max)$ is not achievable, i.e., for any $(k,n)$ network code that can compute $f=\max$ over $\mN$, the rate
\begin{align*}
\frac{k}{n}< \min_{C\in\Lambda(\mN)}\dfrac{|C|}{\log_{|\mA|}n_{C,f}}=2.
\end{align*}
\end{thm}

We first show that the upper bound in Theorem~\ref{thm:upper_bound} for this computation problem $(\mN,f)$ is equal to $2$, i.e.,
\begin{align}\label{equ:upb=2}
\min_{C\in\Lambda(\mN)}\dfrac{|C|}{\log_{|\mA|}n_{C,f}}=2.
\end{align}
We claim that for any cut set $C\in\Lambda(\mN)$,
\begin{align}\label{reverse_BF_UppB}
\left\{
  \begin{array}{ll}
    |C|\geq 4\text{ and }n_{C,f}\leq 4, & \hbox{if $C$ has a nontrivial strong partition;} \\
    |C|\geq 2\text{ and }n_{C,f}\leq 2, & \hbox{otherwise.}
  \end{array}
\right.
\end{align}
To see this, we first prove the following for an \textit{arbitrary} network function computation problem $(\mN,f)$:
\begin{align}\label{ineq:N_Cf_leq_AI}
n_{C,f}\leq \big|\mathcal{A}^{I_C}\big|=|\mathcal{A}|^{|I_C|},\quad \forall~C\in\Lambda(\mN).
\end{align}

Let $C$ be a cut set in $\Lambda(\mN)$ and $\mP_C=\{C_1,C_2,\cdots,C_m\}$, $m\geq 1$, be an arbitrary strong partition of~$C$. For notational simplicity, let $I=I_C$, $J=J_C$, $I_l=I_{C_l}$, $1\leq l \leq m$, and $L=I\setminus(\bigcup_{l=1}^m I_l)$.
Recall the definition of $N\big({a}_{L}, \Cl[{a}_J]\big)$ in \eqref{no_finer_eq_cl},
where $\Cl[{a}_J]$ stands for an arbitrary $(I,a_J)$-equivalence class. It follows from \eqref{equ_2sets} in Section~\ref{sec:proof} that for any $a_L\in \mA^{L}$ and $a_J\in \mA^{J}$,
\begin{align*}
\big|\Cl[{a}_J]\big| & \geq \# \Big\{ (b_{I_1}, b_{I_2}, \cdots, b_{I_m}, a_L): b_{I_l} \in \mA^{I_l}, l=1,2,\cdots,m, \textrm{ and } (b_{I_1}, b_{I_2}, \cdots, b_{I_m}, a_{L})\in \Cl[a_J]\Big\}\\
&= \sum_{\textrm{ all } \left(\cl_{I_1}, \cl_{I_2}, \cdots, \cl_{I_m}\right) \textrm{ s.t. } \atop \left\langle \cl_{I_1}, \cl_{I_2}, \cdots, \cl_{I_m}, a_{L} \right\rangle \subseteq \Cl[a_J]} \Big|\big\langle \cl_{I_1}, \cl_{I_2}, \cdots, \cl_{I_m}, a_{L} \big\rangle \Big|\\
&\geq \sum_{\textrm{ all } \left(\cl_{I_1}, \cl_{I_2}, \cdots, \cl_{I_m}\right) \textrm{ s.t. } \atop \left\langle \cl_{I_1}, \cl_{I_2}, \cdots, \cl_{I_m}, a_{L} \right\rangle \subseteq \Cl[a_J]} 1 = N\big({a}_{L}, \Cl[{a}_J]\big).
\end{align*}
Thus,
\begin{align}\label{ineq:N_Cf_leq_AI_pf_1}
N\big({a}_{L}, \Cl[{a}_J]\big)\leq \big|\Cl[{a}_J]\big|, \quad \forall~a_L\in \mA^{L} \text{ and }\forall~a_J\in \mA^{J}.
\end{align}
By \eqref{equ:N_Cl}, \eqref{ineq:N_Cf_leq_AI_pf_1} immediately implies that $N\big(\Cl[{a}_J]\big)\leq \big|\Cl[{a}_J]\big|$, and thus
\begin{align*}
\sum_{\text{all }\Cl[{a}_J]} N\big(\Cl[{a}_J]\big)\leq \sum_{\text{all }\Cl[{a}_J]} \big|\Cl[{a}_J]\big| = |\mA^I|,
\end{align*}
where the last equality follows from the fact that all $(I,a_J)$-equivalence classes constitute a partition of~$\mA^I$. Finally, by \eqref{def:{a}_J_star0}, \eqref{def:{a}_J_star}, and \eqref{n_C_Parti_1st}, we have
\begin{align*}
n_C(\mP_C) = \sum_{\text{all }\Cl[{a}_J^*]}
N\big(\Cl[{a}_J^*]\big) \leq |\mA^I|,
\end{align*}
and hence $n_{C,f}\leq |\mA^I|$ by \eqref{n_C_f_1st}, proving \eqref{ineq:N_Cf_leq_AI}.

Now, let us return to the proof of \eqref{reverse_BF_UppB} for the network computation problem $(\mN, f)$ in Theorem~\ref{thm_butterfly_network_non-tight} by considering the following two cases:

\noindent\textbf{Case 1:} A cut set $C\in\Lambda(\mN)$ has a nontrivial strong partition.

Let $\mP_C=\{C_1,C_2,\cdots,C_m\}$ be a nontrivial strong partition of $C$. Clearly, $m\geq 2$. Since $\mP_C$ is a strong partition (see Definition~\ref{def:strong_parti}), we have $I_{C_l} \neq \emptyset$, $\forall~1\leq l \leq m$ and $I_{C_i}\cap I_{C_j}=\emptyset$, $\forall~1\leq i, j \leq m$ and $i\neq j$. Together with $\bigcup_{l=1}^m I_{C_l}\subseteq I_C \subseteq S$, we obtain that
\begin{align*}
2\leq m\leq \sum_{l=1}^m |I_{C_l}| \leq |S| = 2,
\end{align*}
which implies $m=2$, i.e., $\mP_C$ is a two-partition given by $\{C_1, C_2\}$, and $|I_{C_1}|=|I_{C_2}|=1$.

We first prove that $|C|\geq 4$. It is readily seen from the network $\mN$ that the minimum cut capacity between $\sigma_i$ and $\rho$ is equal to $2$, $i=1,2$. Then, for any cut set $C_i'$ such that $I_{C_i'}=\{\sigma_i\}$, we have $|C_i'|\geq 2$, $i=1,2$. This implies that $|C|=|C_1|+|C_2|\geq 2+2=4$ (e.g., $C=\{e_1,e_2,e_3,e_4\}$ has a unique nontrivial strong partition $\mP_C=\big\{C_1=\{e_1,e_2\}, C_2=\{e_3,e_4\}\big\}$ with $I_{C_1}=\{\sigma_1\}$ and $I_{C_2}=\{\sigma_2\}$).

We now prove that $n_{C,f}\leq 4$. This can be obtained from \eqref{ineq:N_Cf_leq_AI} with $|\mA|=2$ and $I_C=S$, so that $|\mA|^{|I_C|}=| \mA|^{|S|}=4$.

\noindent\textbf{Case 2:} A cut set $C\in\Lambda(\mN)$ has no nontrivial strong partition.

Following the discussion in Case 1, it is easy to see that $|C|\geq 2$, $\forall~C\in\Lambda(\mN)$, because $C$ separates at least one source node from the sink node $\rho$. To obtain $n_{C,f}\leq 2$, we consider the following two subcases:
\begin{itemize}
  \item if $|I_C|=2$, i.e., $C$ is a global cut set (e.g., $C=\{e_8,e_9\}$ with $I_C=\{\sigma_1, \sigma_2\}$), then $n_{C,f}=|f(\mA^2)|=2$ (see the discussion below Theorem~\ref{thm:upper_bound} and the discussion below \eqref{eq:2} in Section~\ref{subsec:pre_B});
  \item if $|I_C|=1$ (e.g., $C=\{e_1,e_2\}$ with $I_C=\{\sigma_1\}$), then $n_{C,f}\leq |\mA|^{|I_C|}=2$ by \eqref{ineq:N_Cf_leq_AI}.
\end{itemize}

\begin{remark}
By means of an evaluation of $n_{C,f}$ specific to the network computation problem $(\mN,f)$ in Theorem~\ref{thm_butterfly_network_non-tight}, it can be shown that the upper bounds on $n_{C,f}$ in \eqref{reverse_BF_UppB} are in fact tight. Since we do not need this result in the sequel, the details are omitted here.
\end{remark}

\bigskip

Consequently, we can obtain from \eqref{reverse_BF_UppB} that for each cut set $C\in\Lambda(\mN)$,
\begin{align}\label{equ:upb=2-1}
\dfrac{|C|}{\log_{|\mA|}n_{C,f}}\geq 2.
\end{align}
In particular, for the global cut set $C=\{e_8,e_9\}$, we have $|I_C|=2$ and $n_{C,f}=2$ (cf. the first bullet in Case 2 above), so that $|C|/\log_{|\mA|}n_{C,f}=2$. Thus, we have proved \eqref{equ:upb=2}.

Toward proving Theorem~\ref{thm_butterfly_network_non-tight}, it remains to prove that $k/n<2$ for any $(k,n)$ network code that can compute $f$ on $\mN$, which will be done by contradiction. Assume that the upper bound $2$ is achievable. To be specific, for some positive integer~$n$, there exists a rate-$2$ $(2n, n)$ network code $\mbC=\{ g_{e_i}(\vec{x}_1, \vec{x}_2): 1\leq i \leq 9 \}$ that can compute the target function $f$ at the sink node $\rho$, where $\vec{x}_l\in \mA^{2n}$ stands for $2n$ symbols in $\mA$ generated by the source node $\sigma_l$, $l=1,2$, and $g_{e_i}(\vec{x}_1, \vec{x}_2)\in \mA^{n}$ is the global encoding function of $e_i$ that contains at most $n$ symbols in $\mA$ transmitted on the edge $e_i$, $1\leq i \leq 9$. For notational  simplicity, we write $g_{e_i}(\vec{x}_1, \vec{x}_2)$ as $g_{i}(\vec{x}_1, \vec{x}_2)$ for all $1\leq i \leq 9$. We may further simplify $g_{i}(\vec{x}_1, \vec{x}_2)$ to $g_{i}$ when its dependence on $\vec{x}_1$ and $\vec{x}_2$ is implicitly assumed.

Consider the edge set $C=\{e_1, e_4, e_5\}$, which is a global cut set. Since the $(2n,n)$ network code $\mbC$ can compute the target function $f$, there must exist a decoding function $\psi_C$ from $\mA^n \times \mA^n \times \mA^n$ to $\mA^{2n}$ such that
\begin{align}\label{equ:decoding_function_psi_C}
\psi_C\big(g_1(\vec{a}_1, \vec{a}_2), g_4(\vec{a}_1, \vec{a}_2), g_5(\vec{a}_1, \vec{a}_2)\big)=f(\vec{a}_1, \vec{a}_2), \quad \forall\  \vec{a}_l\in \mA^{2n},\ l=1,2.
\end{align}
With this, we split the the network $\mN$ into two sub-networks $\mN_1$ and $\mN_2$, depicted in Fig. \ref{fig:butterfly_subnetwork1} and Fig. \ref{fig:butterfly_subnetwork2}, respectively, where in $\mN_1$, the artificial sink node $\rho'$ that takes $e_1$, $e_4$, and $e_5$ as input is created.

\begin{figure}[!t]
\centering
\begin{minipage}[b]{0.5\textwidth}
\centering
 \begin{tikzpicture}[x=0.6cm]
    \draw (-3.5,0) node[vertex] (1) [label=above:$\sigma_1$] {};
    \draw ( 3.5,0) node[vertex] (2) [label=above:$\sigma_2$] {};
    \draw ( 0,-1.5) node[vertex] (3) [label=left:] {};
    \draw ( 0,-4) node[vertex] (6) [label=below:$\rho'$] {};

    \draw[->,>=latex] (1) -- (6) node[midway, auto, swap, left=0mm] {$e_1$};
    \draw[->,>=latex] (1) -- (3) node[midway, auto, right=0mm] {$e_2$};
    \draw[->,>=latex] (2) -- (3) node[midway, auto,swap, left=0mm] {$e_3$};
    \draw[->,>=latex] (2) -- (6) node[midway, auto, right=0mm] {$e_4$};
    \draw[->,>=latex] (3) -- (6) node[pos=0.4, right=-1mm] {$e_5$};
\end{tikzpicture}
\caption{The network computation $(\mN_1, f)$.}
\label{fig:butterfly_subnetwork1}
\end{minipage}%
\centering
\begin{minipage}[b]{0.5\textwidth}
\centering
\begin{tikzpicture}[x=0.6cm]
    \draw (0,0) node[vertex] (2) [label=above:$\sigma_2'$] {};
    \draw (-4,0)node[vertex] (1) [label=above:$\sigma_1'$] {};
    \draw (4,0) node[vertex] (3) [label=above:$\sigma_3'$] {};
    \draw (-2,-2) node[vertex] (1') [label=left:] {};
    \draw (2,-2) node[vertex] (2') [label=right:] {};
    \draw (0,-4) node[vertex] (0) [label=below:$\rho$] {};

    \draw[->,>=latex] (2) -- (1') node[midway, auto, swap, pos=0.3, left=0mm] {$e_6$};
    \draw[->,>=latex] (2) -- (2') node[midway, auto,pos=0.3, right=0mm] {$e_7$};
    \draw[->,>=latex] (1') -- (0) node[midway, auto, pos=0.3, left=0mm] {$e_8$};
    \draw[->,>=latex] (2') -- (0) node[midway, auto, swap, pos=0.3, right=0mm] {$e_9$};
    \draw[->,>=latex] (1) -- (1') node[midway, auto, pos=0.3, right=0mm] {$e_1$};
    \draw[->,>=latex] (3) -- (2') node[midway, auto, swap, pos=0.3, left=0mm] {$e_4$};
    \end{tikzpicture}
\caption{The network computation $(\mN_2, F)$.}
\label{fig:butterfly_subnetwork2}
\end{minipage}
\end{figure}

We first consider computing $f$ over $\mN_1$, i.e., the network computation problem $(\mN_1, f)$ depicted in Fig.~\ref{fig:butterfly_subnetwork1}. Here, $\mN_1$ contains two source nodes $\sigma_1$ and $\sigma_2$, and one sink node $\rho'$ that is required to compute the maximum function of the source messages, i.e., $f(x_1,x_2)=\max\{ x_1, x_2 \}$ with $\mA=\mO=\{0,1\}$. For the $(2n,n)$ network code $\mbC=\{ g_i(\vec{x}_1, \vec{x}_2): 1\leq i \leq 9 \}$ on $(\mN,f)$, let $\mbC_1=\{g_i(\vec{x}_1, \vec{x}_2): 1\leq i \leq 5\}$ and we see that $\mbC_1$ is a $(2n,n)$ network code induced on $(\mN_1, f)$.

On the other hand, we consider another network computation problem $(\mN_2, F)$, where the network $\mN_2$ is depicted in Fig.~\ref{fig:butterfly_subnetwork2} and the target function $F$, which is induced by the rate-$2$ network code $\mbC_1$ on $(\mN_1,f)$, is given as follows. Let the alphabet of the source messages and the transmitted messages be $\mA^n$. The source node $\sigma_l'$ in $\mN_2$ generates the source vector in $\mA^n$, denoted by $\vec{y}_l$, $l=1,2,3$. 
The target function $F$ is defined as
\begin{align}\label{defn:F}
F:\ \big(\mA^n\big)\times \big(\mA^n\big) \times \big(\mA^n\big) \ \longrightarrow &\ \ \ \big(\mA^{2n})\nonumber\\
(\vec{y}_1,\vec{y}_2,\vec{y}_3)  \ \longmapsto &\ \ \ \psi_C(g_1=\vec{y}_1, g_4=\vec{y}_3, g_5=\vec{y}_2),
\end{align}
where $\psi_C$ is the decoding function of the network code $\mbC_1$ (cf. \eqref{equ:decoding_function_psi_C}). Note that the target function $F$ is defined upon the network code $\mbC_1$, and we will prove later that $F$ is indeed well-defined.

With the $(2n,n)$ network code $\mbC$ on $(\mN, f)$, let $\mbC_2=\{g_i(\vec{x}_1, \vec{x}_2): i=1,4,6,7,8,9 \}$. Then $\mbC_2$ is a $(1,1)$ network code on $(\mN_2, F)$. Here $\big(\mA^n\big)$ corresponds to $\mA$ and $\big(\mA^{2n})$ corresponds to $\mO$ in the definition of a network code in Section~\ref{sec:preliminaries}. To be specific, for source inputs $(\vec{y}_1, \vec{y}_2, \vec{y}_3)$ generated by $\sigma_1'$, $\sigma_2'$ and $\sigma_3'$, respectively, let $g_1=\vec{y}_1$, $g_4=\vec{y}_3$, $g_6=\theta_6(\vec{y}_2)$, $g_7=\theta_7(\vec{y}_2)$ (which is equivalent to letting $g_5=\vec{y}_2$), $g_8=\theta_8(g_1, g_6)$, and $g_9=\theta_9(g_4, g_7)$, where $\theta_i$ denotes the local encoding function of $e_i$ for $i=6,7,8,9$ in $\mbC$. The construction of $\mbC_2$ implies that $\mC(\mN_2, F)\geq 1$.

However, we will prove in the rest of the section that for any function $F$ induced by a rate-$2$ network code on $(\mN_1, f)$, the rate $1$ is not achievable on $(\mN_2, F)$, i.e., $\mC(\mN_2, F)<1$. This immediately leads to a contradiction, which implies that $2$ is not achievable.

\subsection{The Network Computation Problem $(\mN_1,f)$}

\begin{figure}[!t]
\centering
\begin{minipage}[b]{0.5\textwidth}
\centering
 \begin{tikzpicture}[x=0.6cm]
    \draw (-3,0) node[vertex] (1) [label=above:$\sigma_1$] {};
    \draw ( 3,0) node[vertex] (2) [label=above:$\sigma_2$] {};
    \draw ( 0,-1.5) node[vertex] (3) [label=left:] {};
    \draw ( 0,-4) node[vertex] (6) [label=below:$\rho'$] {};

    \draw[->,>=latex] (1) -- (6) node[midway, auto, swap, left=0mm] {$g_1$};
    \draw[->,>=latex] (1) -- (3) node[midway, auto, right=0mm] {$g_2$};
    \draw[->,>=latex] (2) -- (3) node[midway, auto,swap, left=0mm] {$g_3$};
    \draw[->,>=latex] (2) -- (6) node[midway, auto, right=0mm] {$g_4$};
    \draw[->,>=latex] (3) -- (6) node[pos=0.4, right=-1mm] {$g_5$};
\end{tikzpicture}
\label{fig:butterfly_network_Part1_a}
\end{minipage}%
\begin{minipage}[b]{0.5\textwidth}
\centering
\begin{tabular}{p{5mm}p{2.1cm}}
\hline
\specialrule{0em}{1pt}{1pt}
$g_1$: & $x_{11}$\\
\specialrule{0em}{1pt}{1pt}
$g_2$: & $x_{12}$\\
\specialrule{0em}{1pt}{1pt}
$g_3$: & $x_{22}$\\
\specialrule{0em}{1pt}{1pt}
$g_4$: & $x_{21}$\\
\specialrule{0em}{1pt}{1pt}
$g_5$: & $\max\{x_{12}, x_{22}\}$\\
\specialrule{0em}{2pt}{2pt}
\hline
\end{tabular}
\vspace{3mm}
\end{minipage}
\caption{A rate-$2$ $(2,1)$ network code $\{g_i:1\leq i \leq 5\}$ on $(\mN_1, f=\max)$, where $\vec{x}_1=(x_{11}, x_{12})^\top$ and $\vec{x}_2=(x_{21}, x_{22})^\top$ are source vectors generated by $\sigma_i$, $i=1,2$, respectively.}
\label{fig:butterfly_network_Part1_b}
\end{figure}
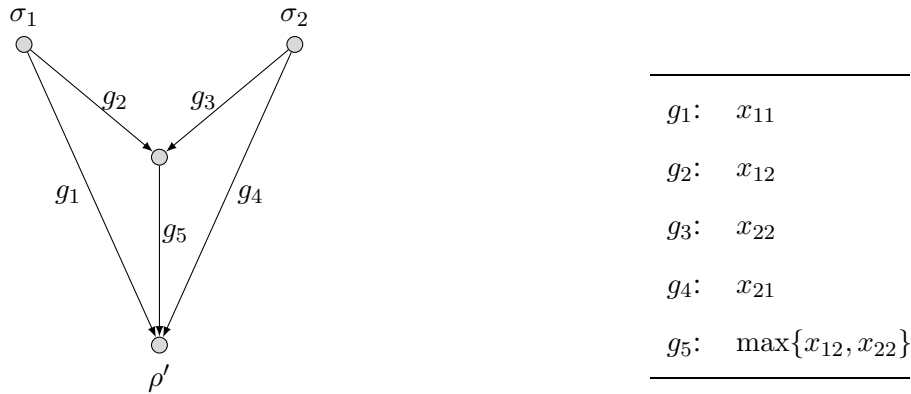

In this subsection, we will give some necessary properties that all rate-$2$ network codes on $(\mN_1, f)$ must satisfy. First, we have $\mC(\mN_1, f)=2$, because Theorem~\ref{thm:upper_bound} implies $\mC(\mN_1, f)\leq 2$ (for example consider $C=\{e_1,e_2\}$ and $n_{C,f}=2$) and Fig.~\ref{fig:butterfly_network_Part1_b} gives a coding scheme achieving the rate $2$.

In general, we let $\mbC_1=\{g_i(\vec{x}_1, \vec{x}_2): 1\leq i \leq 5\}$ be a $(2n, n)$ network code on $\mN_1$ with respect to $f$, where $n$ is a positive integer. Since $K_{\{e_1\}}=\{\sigma_1\}$ (cf. \eqref{def_K_C}) in $\mN_1$, the global encoding function $g_1(\vec{x}_1,\vec{x}_2)$ only depends on the source inputs $\vec{x}_1$ of $\sigma_1$ and hence we write $g_1(\vec{x}_1,\vec{x}_2)$ as $g_1(\vec{x}_1)$, a function from $\mA^{2n}$ to $\mA^n$. In fact, $g_1$ is the local encoding function $\theta_1$ of the edge $e_1$ (cf. \eqref{defn_local_function} for the definition), i.e., $\theta_1(\vec{x}_1)=g_1(\vec{x}_1)$. Similarly, we can write $g_2(\vec{x}_1, \vec{x}_2)$, $g_3(\vec{x}_1, \vec{x}_2)$, and $g_4(\vec{x}_1, \vec{x}_2)$ as $g_2(\vec{x}_1)$, $g_3(\vec{x}_2)$, and $g_4(\vec{x}_2)$, respectively. They are also the local encoding functions $\theta_2(\vec{x}_1)$, $\theta_3(\vec{x}_2)$, and $\theta_4(\vec{x}_2)$ corresponding to the edges $e_2$, $e_3$, and $e_4$, respectively. For the edge $e_5$, since $K_{\{e_5\}}=\{\sigma_1, \sigma_2\}$ which means that $g_5(\vec{x}_1,\vec{x}_2)$ possibly is affected by both $\vec{x}_1$ and $\vec{x}_2$, we keep $g_5(\vec{x}_1,\vec{x}_2)$. Then
\begin{align}
g_5(\vec{a}_1, \vec{a}_2)=\theta_5\big(g_2(\vec{a}_1), g_3(\vec{a}_2)\big), \quad \forall\ \vec{a}_l\in \mA^{2n},\ l=1,2,
\end{align}
where $\theta_5$ is the local encoding function of the edge $e_5$. With the above, we rewrite the network code $\mbC_1=\{g_i(\vec{x}_1, \vec{x}_2): 1\leq i \leq 5\}$ as
\begin{align*}
\mbC_1=\big\{g_1(\vec{x}_1), g_2(\vec{x}_1), g_3(\vec{x}_2), g_4(\vec{x}_2), g_5(\vec{x}_1, \vec{x}_2)\big\},
\end{align*}
or $\mbC_1=\{g_i: 1\leq i \leq 5\}$ for simplicity.

\begin{lemma}\label{lemma1:non_tight}
Let $\mbC_1=\{g_i: 1\leq i \leq 5\}$ be a $(2n,n)$ network code on $(\mN_1, f=\max)$. Then for any two distinct vectors $\vec{a}$ and $\vec{b}$ in $\mA^{2n}$,
\begin{align}
\big( g_1(\vec{a}), g_2(\vec{a}) \big)&\neq
\big( g_1(\vec{b}), g_2(\vec{b}) \big),\label{equ:1_lemma7}\\
\big( g_3(\vec{a}), g_4(\vec{a}) \big)&\neq
\big( g_3(\vec{b}), g_4(\vec{b}) \big).\label{equ:2_lemma7}
\end{align}
In other words, $\big( g_1(\vec{x}_1), g_2(\vec{x}_1) \big)$ (resp. $\big( g_3(\vec{x}_2), g_4(\vec{x}_2) \big)$), regarded as a function from $\mA^{2n}$ to $\mA^n\times \mA^n$, is a bijection.
\end{lemma}
\begin{IEEEproof}
We first prove by contradiction that \eqref{equ:1_lemma7} holds for any two distinct vectors $\vec{a}$ and $\vec{b}$ in $\mA^{2n}$. Assume the contrary that there exist two distinct vectors $\vec{a}$ and $\vec{b}$ in $\mA^{2n}$ such that
\begin{align}
\big( g_1(\vec{a}), g_2(\vec{a}) \big)=
\big( g_1(\vec{b}), g_2(\vec{b}) \big),
\end{align}
i.e., $g_1(\vec{a})=g_1(\vec{b})$ and $g_2(\vec{a})=g_2(\vec{b})$.

Let $\vec{x}_2=\vec{0}$, the all-zero $2n$-vector in $\mA^{2n}$. By $g_2(\vec{a})=g_2(\vec{b})$, we obtain
\begin{align}
g_5\big(\vec{a}, \vec{0}\big)
=\theta_5\big(g_2(\vec{a}), g_3(\vec{0})\big)
=\theta_5\big(g_2(\vec{b}), g_3(\vec{0})\big)=g_5\big(\vec{b}, \vec{0}\big).
\end{align}
Together with $g_1(\vec{a})=g_1(\vec{b})$,  we immediately have
\begin{align}\label{equ:non_tight_lemma1_1}
\big(g_1(\vec{a}), g_4(\vec{0}), g_5(\vec{a}, \vec{0}) \big)
=\big(g_1(\vec{b}), g_4(\vec{0}), g_5(\vec{b}, \vec{0}) \big),
\end{align}
i.e., for the distinct source inputs $(\vec{a}, \vec{0})$ and $(\vec{b}, \vec{0})$, the two corresponding messages transmitted on $C=\{e_1,e_4,e_5\}$ are the same.

Since the network code $\mbC_1$ can compute $f$ with zero error and the cut set $C$ is global, we obtain
\begin{align}
\vec{a}=f(\vec{a}, \vec{0})=\psi_C\big(g_1(\vec{a}), g_4(\vec{0}), g_5(\vec{a}, \vec{0})\big)=\psi_C\big(g_1(\vec{b}), g_4(\vec{0}), g_5(\vec{b}, \vec{0})\big)=f(\vec{b}, \vec{0})=\vec{b},
\end{align}
where $\psi_C$ is the decoding function of $\mbC_1$. This contradicts the assumption that $\vec{a}\neq\vec{b}$.

The same result for $(g_3, g_4)$ can be proved by using the same argument. The proof is  completed.
\end{IEEEproof}

We now introduce some notations below that will be used frequently in the sequel:
\begin{itemize}
  \item Denote by $g_i(\mA^{2n})$ the {\em image} of $\mA^{2n}$ under $g_i$ for $i=1,2,3,4$, i.e.,
\begin{align}\label{notation_image_set_g_i}
g_i(\mA^{2n})=\big\{g_i(\vec{a}):\ \vec{a}\in \mA^{2n}\big\}\subseteq \mA^n, \ i=1,2,3,4.
\end{align}
Similarly, denote by $g_5(\mA^{2n}, \mA^{2n})$ the {\em image} of $\mA^{2n}\times\mA^{2n}$ under $g_5$, i.e.,
\begin{align}\label{equ:35}
g_5(\mA^{2n}, \mA^{2n})=\big\{g_5(\vec{a}, \vec{b}):\ (\vec{a}, \vec{b})\in \mA^{2n}\times\mA^{2n} \big\}\subseteq \mA^n.
\end{align}
  \item Let $\vec{\gamma}\in \mA^n$. For $1\leq i \leq 5$, denote by $g_i^{-1}(\vec{\gamma})$ the {\em inverse image} of $\vec{\gamma}$ under $g_i$, i.e.,
\begin{align}
    g_i^{-1}(\vec{\gamma})&=\big\{\vec{a}\in \mA^{2n}:\ g_i(\vec{a})=\vec{\gamma} \big\}\subseteq \mA^{2n},\quad i=1,2,3,4; \\
    g_5^{-1}(\vec{\gamma})&=\big\{(\vec{a}, \vec{b}) \in \mA^{2n}\times \mA^{2n}:\ g_5(\vec{a}, \vec{b})=\vec{\gamma} \big\}\subseteq \mA^{2n}\times\mA^{2n}.
\end{align}
\end{itemize}

\begin{lemma}\label{thm:non-tight}
Let $\mbC_1=\{g_i: 1\leq i \leq 5\}$ be a $(2n,n)$ network code on $(\mN_1, f=\max)$.
Then
\begin{enumerate}
  \item All global encoding functions $g_i$, $1\leq i \leq 5$, are surjective, i.e.,
\begin{align*}
g_i(\mA^{2n})=g_5(\mA^{2n}, \mA^{2n})=\mA^n,\quad i=1,2,3,4.
\end{align*}
  \item For every $\vec{\gamma}\in \mA^n$ and each $i=1,2,3,4$,
\begin{align*}
\left|g_i^{-1}(\vec{\gamma})\right|=2^n.
\end{align*}
In other words, $\left\{g_i^{-1}(\vec{\gamma}):\ \vec{\gamma}\in \mA^n \right\}$ forms an equipartition of $\mA^{2n}$.
\end{enumerate}
\end{lemma}
\begin{IEEEproof}
We first prove $g_i(\mA^{2n})=\mA^n$ for $i=1,2,3,4$. Since $\big( g_1(\vec{x}_1), g_2(\vec{x}_1) \big)$ (resp. $\big( g_3(\vec{x}_2), g_4(\vec{x}_2) \big)$) is a bijection from $\mA^{2n}$ to $\mA^n\times \mA^n$ by Lemma~\ref{lemma1:non_tight}, we obtain
\begin{align}\label{equ:cal_g1_g2}
g_1(\mA^{2n})=g_2(\mA^{2n})=\mA^n\quad (\text{resp. } g_3(\mA^{2n})=g_4(\mA^{2n})=\mA^n).
\end{align}

Before proving $g_5(\mA^{2n}, \mA^{2n})=\mA^n$, we first prove 2) in Lemma~\ref{thm:non-tight} that $\big|g_i^{-1}(\vec{\gamma})\big|=2^n$, $\forall~\vec{\gamma}\in \mA^n$ for $i=1,2,3,4$. Consider $g_1$ and assume that there exists $\vec{\gamma}$ in $\mA^n$ such that $\big|g_1^{-1}(\vec{\gamma})\big|\neq 2^n$. We further assume that $\big|g_1^{-1}(\vec{\gamma})\big|>2^n$, which does not lose any generality. This is explained as follows. Since $g_1(\mA^{2n})=\mA^n$ by \eqref{equ:cal_g1_g2}, we obtain that $g_1^{-1}(\vec{\gamma})\neq \emptyset$, $\forall~\vec{\gamma}\in \mA^n$ and so $\big\{ g_1^{-1}(\vec{\gamma}):~\vec{\gamma}\in \mA^n \big\}$ constitutes a partition of $\mA^{2n}$ that contains $|\mA^n|=2^n$ blocks.\footnote{The definition of a partition requires that every subset of a partition is nonempty and these subsets are called {\em blocks}.} Hence, if $\big|g_1^{-1}(\vec{\gamma})\big|<2^n$, there must exist another $\vec{\gamma}' \in \mA^{n}$ such that $\big|g_1^{-1}(\vec{\gamma}')\big|> 2^n$ because otherwise $\sum_{\vec{\gamma} \in \mA^{n}}|g_1^{-1}(\vec{\gamma})\big|<|\mA|^{2n}$, a contradiction.

Now, since $\big|g_1^{-1}(\vec{\gamma})\big|> 2^n$ and $\big|g_2(\mA^{2n})\big|=2^n$ by \eqref{equ:cal_g1_g2}, there exist two distinct $2n$-column vectors $\vec{a}, \vec{b} \in g_1^{-1}(\vec{\gamma})$ such that
\begin{align}\label{equ:non_tight_thm1_2}
g_2(\vec{a})=g_2(\vec{b}).
\end{align}
Consider $\vec{x}_2=\vec{0}$. By \eqref{equ:non_tight_thm1_2}, we see that
\begin{align}\label{equ:non_tight_thm1_3}
g_5\big( \vec{a}, \vec{0} \big)=\theta_5\big( g_2(\vec{a}), g_3(\vec{0}) \big)=\theta_5\big( g_2(\vec{b}), g_3(\vec{0}) \big)=g_5\big( \vec{b}, \vec{0} \big).
\end{align}
Together with $g_1(\vec{a})=g_1(\vec{b})=\vec{\gamma}$, this immediately implies that for $C=\{e_1, e_4, e_5 \}$,
\begin{align}\label{equ85}
g_C\big(\vec{a}, \vec{0}\big)=\big( g_1(\vec{a}), g_4(\vec{0}), g_5(\vec{a}, \vec{0}) \big)=\big( g_1(\vec{b}), g_4(\vec{0}), g_5(\vec{b}, \vec{0}) \big)=g_C\big( \vec{b}, \vec{0} \big).
\end{align}
Since $C$ is global ($I_C=S$) and $\psi_C$ is the decoding function of the network code $\mbC_1$, we obtain by \eqref{equ85} that
\begin{align}
\vec{a}=f(\vec{a}, \vec{0})&=\psi_C\big(g_C(\vec{a}, \vec{0})\big)=\psi_C\big(g_C(\vec{b}, \vec{0})\big)=f(\vec{b}, \vec{0})=\vec{b},\label{equ:non_tight_thm1_4}
\end{align}
a contradiction to $\vec{a}\neq \vec{b}$. Thus, we have proved that $\big|g_1^{-1}(\vec{\gamma})\big|=2^n$, $\forall~\vec{\gamma}\in \mA^n$.

By a symmetrical argument, we can prove that $\big|g_2^{-1}(\vec{\gamma})\big|=2^n$, $\forall~\vec{\gamma}\in \mA^n$. Similarly, we can prove that  $\big|g_3^{-1}(\vec{\gamma})\big|=\big|g_4^{-1}(\vec{\gamma})\big|=2^n$, $\forall~\vec{\gamma}\in \mA^n$.



Now, we proceed to prove that $g_5(\mA^{2n}, \mA^{2n})=\mA^n$. For any $n$-vector $\vec{\gamma}$ in $\mA^n$, we will prove that
\begin{align}\label{equ:pf1}
 \{ g_5(\vec{a}, \vec{0}):\ \vec{a}\in g_1^{-1}(\vec{\gamma})\} = \mA^n,
\end{align}
which, together with $g_5(\mA^{2n}, \mA^{2n})\subseteq \mA^n$ implies that $g_5(\mA^{2n}, \mA^{2n})=\mA^n$.

We assume the contrary of \eqref{equ:pf1}, or equivalently,
\begin{align}\label{equ:pf2}
\big|\{ g_5(\vec{a}, \vec{0}):\ \vec{a}\in g_1^{-1}(\vec{\gamma}) \} \big|<2^n.
\end{align}
Since we have proved that $\big|g_1^{-1}(\vec{\gamma})\big|=2^n$, by \eqref{equ:pf2} there exist two distinct vectors $\vec{a}$ and $\vec{b}$ in $g_1^{-1}(\vec{\gamma})$ such that
\begin{align}\label{equ:non_tight_thm1_6}
g_5(\vec{a}, \vec{0})=g_5(\vec{b}, \vec{0}).
\end{align}
By comparing \eqref{equ:non_tight_thm1_6} with \eqref{equ:non_tight_thm1_3} and applying the argument following \eqref{equ:non_tight_thm1_3}, we obtain \eqref{equ:non_tight_thm1_4}, a contradiction to $\vec{a}\neq \vec{b}$. Hence, we have proved \eqref{equ:pf1}. This completes the proof of the lemma.
\end{IEEEproof}

\begin{removeEX4}

We can immediately obtain the following corollary from Lemmas~\ref{lemma1:non_tight}~and~\ref{thm:non-tight}.


\begin{cor}\label{cor_non-tight-2}
Let $\mbC_1=\{g_i: 1\leq i \leq 5\}$ be a $(2n,n)$ network code on $(\mN_1, f=\max)$. For each $\vec{\gamma}\in \mA^n$,
\begin{align*}
\left\{ g_1(\vec{a}):\ \vec{a}\in g_2^{-1}(\vec{\gamma})\right\}=\left\{ g_2(\vec{a}):\  \vec{a}\in g_1^{-1}(\vec{\gamma})\right\}=\mA^n,
\end{align*}
and
\begin{align*}
\left\{ g_3(\vec{a}):\ \vec{a}\in g_4^{-1}(\vec{\gamma})\right\}=\left\{ g_4(\vec{a}):\ \vec{a}\in g_3^{-1}(\vec{\gamma})\right\}=\mA^n.
\end{align*}
\end{cor}
\begin{IEEEproof}
For each $\vec{\gamma}\in \mA^n$, it follows from Lemma~\ref{lemma1:non_tight} that $g_1(\vec{a})$, $\vec{a}\in g_2^{-1}(\vec{\gamma})$ are distinct.
Together with $\big|g_2^{-1}(\vec{\gamma})\big|=2^n$ by Lemma~\ref{thm:non-tight}, we obtain that $g_1(\vec{a})$, $\vec{a}\in g_2^{-1}(\vec{\gamma})$ are $2^n$ distinct vectors in $\mA^n$, i.e., $\left\{ g_1(\vec{a}):\ \vec{a}\in g_2^{-1}(\vec{\gamma})\right\}=\mA^n$. By the same argument, we can prove that
\begin{align*}
\left\{ g_2(\vec{a}):\  \vec{a}\in g_1^{-1}(\vec{\gamma})\right\}=\left\{ g_3(\vec{a}):\ \vec{a}\in g_4^{-1}(\vec{\gamma})\right\}=\left\{ g_4(\vec{a}):\ \vec{a}\in g_3^{-1}(\vec{\gamma})\right\}=\mA^n,
\end{align*}
and so the proof is accomplished.
\end{IEEEproof}

In the following example, we use the coding scheme depicted in Fig.~\ref{fig:butterfly_network_Part1_b} to illustrate the concepts and conclusions mentioned above.

\begin{example}\label{eg:N_1}
The coding scheme depicted in Fig.~\ref{fig:butterfly_network_Part1_b} is a $(2,1)$ network code for the network computation problem $(\mN_1, f)$. First, we see that $\big( g_1(\vec{x}_1), g_2(\vec{x}_1) \big)=(x_{11}, x_{12})$ and $\big( g_3(\vec{x}_2), g_4(\vec{x}_2) \big)=(x_{22}, x_{21})$ are two bijections from $\mA^2$ to $\mA \times \mA$, where $\vec{x}_1=(x_{11}, x_{12})^\top$ and $\vec{x}_2=(x_{21}, x_{22})^\top$. Also,
\begin{align*}
g_1(\mA^2)&=\{ g_1(\left[\begin{smallmatrix}0\\0\end{smallmatrix}\right])
           =g_1(\left[\begin{smallmatrix}0\\1\end{smallmatrix}\right])=0,
            g_1(\left[\begin{smallmatrix}1\\0\end{smallmatrix}\right])
           =g_1(\left[\begin{smallmatrix}1\\1\end{smallmatrix}\right])=1\}=\mA,\\
g_2(\mA^2)&=\{ g_2(\left[\begin{smallmatrix}0\\0\end{smallmatrix}\right])
           =g_2(\left[\begin{smallmatrix}1\\0\end{smallmatrix}\right])=0,
            g_2(\left[\begin{smallmatrix}0\\1\end{smallmatrix}\right])
           =g_2(\left[\begin{smallmatrix}1\\1\end{smallmatrix}\right])=1\}=\mA,\\
g_3(\mA^2)&=\{ g_3(\left[\begin{smallmatrix}0\\0\end{smallmatrix}\right])
           =g_3(\left[\begin{smallmatrix}1\\0\end{smallmatrix}\right])=0,
            g_3(\left[\begin{smallmatrix}0\\1\end{smallmatrix}\right])
           =g_3(\left[\begin{smallmatrix}1\\1\end{smallmatrix}\right])=1\}=\mA,\\
g_4(\mA^2)&=\{ g_4(\left[\begin{smallmatrix}0\\0\end{smallmatrix}\right])
           =g_4(\left[\begin{smallmatrix}0\\1\end{smallmatrix}\right])=0,
            g_4(\left[\begin{smallmatrix}1\\0\end{smallmatrix}\right])
           =g_4(\left[\begin{smallmatrix}1\\1\end{smallmatrix}\right])=1\}=\mA,\\
g_5(\mA^2, \mA^2)&=\{\max\{x_{12}, x_{22}\}: x_{12}\in \mA, x_{22}\in \mA\}=\mA.
\end{align*}
In addition, we see that
\begin{align*}
g_1^{-1}(0)=\{\left[\begin{smallmatrix}0\\0\end{smallmatrix}\right], \left[\begin{smallmatrix}0\\1\end{smallmatrix}\right]\},\ g_1^{-1}(1)=\{\left[\begin{smallmatrix}1\\0\end{smallmatrix}\right], \left[\begin{smallmatrix}1\\1\end{smallmatrix}\right]\},
\end{align*}
immediately,
$|g_1^{-1}(0)|=|g_1^{-1}(1)|=|\mA|=2$. We also have
\begin{align*}
|g_i^{-1}(0)|=|g_i^{-1}(1)|=|\mA|=2, \quad i=2,3,4.
\end{align*}
Then, the $(2,1)$ network code in Fig.~\ref{fig:butterfly_network_Part1_b} for $(\mN_1,f)$ qualifies Lemmas~\ref{lemma1:non_tight}~and~\ref{thm:non-tight}, and Corollary~\ref{cor_non-tight-2}.
\end{example}

\end{removeEX4}
\subsection{The Network Computation Problem $(\mN_2,F)$}

\begin{figure}[t]
  \centering
{
 \begin{tikzpicture}[x=0.7cm]
    \draw (0,0) node[vertex] (2) [label=above:$\sigma_2'$] {};
    \draw (-2,-1.5) node[vertex] (1') [label=left:] {};
    \draw (2,-1.5) node[vertex] (2') [label=right:] {};
    \draw (0,-3) node[vertex] (0) [label=below:$\rho$] {};
    \draw[->,>=latex] (2) -- (1') node[midway, auto, swap, pos=0.3, left=0mm] {$e_6$};
    \draw[->,>=latex] (2) -- (1') node[midway, auto, pos=0.7, right=0mm] {$g_6$};
    \draw[->,>=latex] (2) -- (2') node[midway, auto,pos=0.3, right=0mm] {$e_7$};
    \draw[->,>=latex] (2) -- (2') node[midway, auto, swap, pos=0.7, left=0mm] {$g_7$};
    \draw[->,>=latex] (1') -- (0) node[midway, auto, swap, pos=0.7, left=0mm] {$g_8$};
    \draw[->,>=latex] (1') -- (0) node[midway, auto, pos=0.3, right=0mm] {$e_8$};
    \draw[->,>=latex] (2') -- (0) node[midway, auto, pos=0.7, right=0mm] {$g_9$};
    \draw[->,>=latex] (2') -- (0) node[midway, auto, swap, pos=0.3, left=0mm] {$e_9$};

    \draw node[vertex,label=above:$\sigma_1'$] at (-4,0) (1) {};
    \draw[->,>=latex] (1) -- (1') node[midway, auto,swap,pos=0.7, left=0mm] {$g_1$};
    \draw[->,>=latex] (1) -- (1') node[midway, auto, pos=0.3, right=0mm] {$e_1$};
    \draw node[vertex,label=above:$\sigma_3'$] at (4,0) (3) {};
    \draw[->,>=latex] (3) -- (2') node[midway, auto, pos=0.7, right=0mm] {$g_4$};
    \draw[->,>=latex] (3) -- (2') node[midway, auto, swap, pos=0.3, left=0mm] {$e_4$};
    \end{tikzpicture}
}
\caption{The network $\mathcal{N}_2=(G_2, S_2=\{\sigma_1', \sigma_2', \sigma_3' \}, \rho)$.}
  \label{fig:2}
\end{figure}
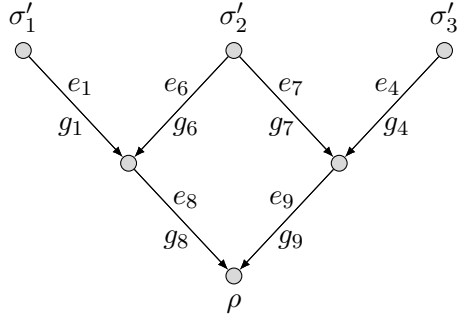

From the first paragraph of the last subsection, we have $\mC(\mN_1,f)=2$. Then, let $\mbC_1=\{g_i: 1\leq i \leq 5\}$ be an arbitrary rate-$2$ $(2n,n)$ network code on $(\mN_1,f)$, where $n$ is a positive integer. Consider the target function $F$, induced by the network code $\mbC_1$ as given in \eqref{defn:F}, which is required to be computed on the network $\mN_2$ (see Fig.~\ref{fig:2}).

To show that the target function $F$ is well-defined, we need to show that $F(\vec{y}_1,\vec{y}_2,\vec{y}_3)$ is defined for every input $(\vec{y}_1,\vec{y}_2,\vec{y}_3)\in(\mA^n)\times (\mA^n) \times (\mA^n)$. To see this, we only need to observe that $g_1(\mA^{2n})=g_5(\mA^{2n}, \mA^{2n})=g_4(\mA^{2n})=\mA^n$ by Lemma~\ref{thm:non-tight} and thus the domain of $F$ is $(\mA^n) \times (\mA^n) \times (\mA^n)$.

The following theorem asserts that for any target function $F$ induced by a rate-$2$ network code $\mbC_1$ for $(\mN_1, f)$, it is impossible for $(\mN_2, F)$ to achieve the rate $1$, i.e., $\mC(\mN_2, F)<1$.

\begin{thm}\label{thm:Cap_N_2_F}
Let $F$ be a target function induced by a rate-$2$ network code on $(\mN_1, f=\max)$ as given in \eqref{defn:F}. Then $\mC(\mN_2, F)<1$.
\end{thm}


To prove Theorem~\ref{thm:Cap_N_2_F}, we first prove Lemma~\ref{thm:no_equ_classes} after explicitly characterizing two equivalence relations. Consider a global cut set $\widehat{C}=\{ e_8, e_9 \}$ in the network $\mN_2$. Denote $I_{\widehat{C}}$ and $J_{\widehat{C}}$ by $I$ and $J$, respectively for notational simplicity. Then, $I=S_2=\{\sigma_1', \sigma_2', \sigma_3'\}$, the set of source nodes in $\mN_2$, and $J=\emptyset$ so that $\vec{a}_J$ is an empty vector. Hence, the $(I, \vec{a}_J)$-equivalence relation is given as follows (see Definition~\ref{def:ec}):
$\vec{\alpha}_{S_2}$ and $\vec{\beta}_{S_2}$ in $(\mA^n)^3$ are $(I,\vec{a}_J)$-equivalent, if
\begin{align}\label{equ_F_A}
F(\vec{\alpha}_{S_2})=F(\vec{\beta}_{S_2}),\quad \text{ or equivalently, }\quad  \psi_C(\vec{\alpha}_{S_2})=\psi_C(\vec{\beta}_{S_2}).
\end{align}

For an $(I,\vec{a}_J)$-equivalence class, let $\vec{m}$ be the common value of $F(\vec{\alpha}_{S_2})$ for all $\vec{\alpha}_{S_2}$ in the equivalence class. Then we see that the equivalence class is uniquely identified by $\vec{m}$. We claim that $\forall~\vec{m}\in \mA^{2n}$,
\begin{align*}
F^{-1}(\vec{m})\triangleq \big\{ \vec{\alpha}_{S_2}\in (\mA^n)\times (\mA^n) \times (\mA^n): F(\vec{\alpha}_{S_2})=\vec{m} \big\}\neq \emptyset.
\end{align*}
It then follows that the total number of $(I,\vec{a}_J)$-equivalence classes is $|\mA^{2n}|=2^{2n}$.

Consider a fixed $\vec{m}\in\mA^{2n}$. Note that $f(\vec{0},\vec{m})=\max\{\vec{0}, \vec{m}\}=\vec{m}$, and it follows from \eqref{defn:F} that
\begin{align*}
F\big(\vec{y}_1=g_1(\vec{0}), \vec{y}_2=g_5(\vec{0},\vec{m}), \vec{y}_3=g_4(\vec{m})\big)=\vec{m}.
\end{align*}
This shows that $F^{-1}(\vec{m})\neq \emptyset$, proving the claim.


Next, we consider the partition equivalence relation (see Definition~\ref{defn_P_E_Relation}) with respect to $\widehat{C}$. The unique nontrivial (strong) partition of $\widehat{C}$ is $\big\{\widehat{C}_1=\{ e_8 \}, \widehat{C}_2=\{ e_9 \}\big\}$, denoted by $\mP_{\widehat{C}}$, and $I_{\widehat{C}_1}=\{\sigma_1'\}$, $I_{\widehat{C}_2}=\{\sigma_3'\}$, and $I\setminus(I_{\widehat{C}_1}\cup I_{\widehat{C}_2})=\{\sigma_2'\}$. Let $I_1=I_{\widehat{C}_1}$, $I_{2}=I_{\widehat{C}_2}$, and $L=I\setminus(I_{1}\cup I_{2})=\{\sigma_2'\}$. For $\vec{y}_L=(\vec{y}_i: \sigma_i'\in L)=\vec{y}_2=\vec{\gamma}$, an arbitrary vector in $(\mA^n)^{L}=\mA^n$, by Definition~\ref{defn_P_E_Relation}, we say that $\vec{\alpha}$ and $\vec{\beta}$ in $(\mA^n)^{I_1}=\mA^n$ are $(I_1, \vec{\gamma}, \vec{a}_J)$-equivalent if for each $\vec{\eta}\in (\mA^n)^{I_2}=\mA^n$,
\begin{align}\label{equ:F_B}
F(\vec{y}_1=\vec{\alpha}, \vec{y}_2=\vec{\gamma}, \vec{y}_3=\vec{\eta})=
F(\vec{y}_1=\vec{\beta}, \vec{y}_2=\vec{\gamma}, \vec{y}_3=\vec{\eta}),
\end{align}
or equivalently,
\begin{align}
\psi_C(g_1=\vec{\alpha}, g_4=\vec{\eta}, g_5=\vec{\gamma})=
\psi_C(g_1=\vec{\beta}, g_4=\vec{\eta}, g_5=\vec{\gamma}),
\end{align}
where as given in \eqref{defn:F}, $g_1=\vec{y}_1$, $g_4=\vec{y}_3$, and $g_5=\vec{y}_2$. Similarly, we can define the {\em $(I_2, \vec{\gamma}, \vec{a}_J)$-equivalence relation}.

\begin{lemma}\label{thm:no_equ_classes}
For every $\vec{\gamma} \in (\mA^n)^L=\mA^n$, the total number of $(I_l, \vec{\gamma}, \vec{a}_J)$-equivalence classes is $2^n$, and every vector $\vec{\alpha}$ in $(\mA^n)^{I_l}=\mA^n$ by itself forms an $(I_l, \vec{\gamma}, \vec{a}_J)$-equivalence class, $l=1,2$.
\end{lemma}
\begin{IEEEproof}
By symmetry, we only need to prove the lemma for $l=1$.

Fix $\vec{\gamma} \in \mA^n$. We will prove that $(\mA^n)^{I_1}=\mA^n$ is partitioned into $2^n$ $(I_1, \vec{\gamma}, \vec{a}_J)$-equivalence classes. Equivalently, we will prove that any two distinct vectors $\vec{\alpha}$ and $\vec{\beta}$ in $(\mA^n)^{I_1}=\mA^n$ are not $(I_1, \vec{\gamma}, \vec{a}_J)$-equivalent, i.e.,
\begin{align*}
\exists\ \vec{\eta}\in (\mA^n)^{I_2}=\mA^n,\ \text{  s.t.  }\
\psi_C(g_1=\vec{\alpha}, g_4=\vec{\eta}, g_5=\vec{\gamma})\neq
\psi_C(g_1=\vec{\beta}, g_4=\vec{\eta}, g_5=\vec{\gamma}).
\end{align*}
Let $\mL$ be the set of all possible image values under the local encoding function $\theta_5$, i.e.,
\begin{align}\label{eqU:95}
\mL = \{ \theta_5(\vec{\xi},\vec{\eta}):~(\vec{\xi}, \vec{\eta})\in {\mA^n}\times \mA^n \}.
\end{align}
Since $g_2(\mA^{2n})=g_3(\mA^{2n})=\mA^{n}$ by Lemma~\ref{thm:non-tight}, it follows from \eqref{eqU:95} that
\begin{align}
\mL=\big\{ \theta_5\big(g_2(\vec{a}_1),g_3(\vec{a}_2)\big):~(\vec{a}_1, \vec{a}_2)\in {\mA^{2n}}\times \mA^{2n} \big\}=g_5(\mA^{2n}, \mA^{2n})=\mA^n,
\end{align}
which the last equality follows from Lemma~\ref{thm:non-tight}.

Now, we let
\begin{align}
\mL_1=\big\{ \theta_5\big(g_2=\vec{\xi}, g_3(\vec{0})\big):\ \vec{\xi} \in \mA^n \big\},
\end{align}
which is a subset of $\mL$ such that $\vec{\eta}=g_3(\vec{0})\in \mA^n$ with $\vec{0}$ being the all-zero vector in $\mA^{2n}$. In the following, we prove by contradiction that $\mL_1=\mL=\mA^n$. Assume otherwise. Then there exist two distinct $\vec{\xi}_1, \vec{\xi}_2 \in \mA^n$ such that
$$\theta_5\big(\vec{\xi}_1, g_3(\vec{0})\big)=\theta_5\big(\vec{\xi}_2, g_3(\vec{0})\big).$$
Now, for any $\vec{\alpha}\in \mA^n$, we have
\begin{align}
\Big( g_1=\vec{\alpha}, g_4(\vec{0}), \theta_5\big(\vec{\xi}_1, g_3(\vec{0})\big) \Big)
=\Big( g_1=\vec{\alpha}, g_4(\vec{0}), \theta_5\big(\vec{\xi}_2, g_3(\vec{0})\big) \Big),
\end{align}
and hence
\begin{align}\label{equ1:thm:no_equ_classes}
\psi_C\Big( g_1=\vec{\alpha}, g_4(\vec{0}), \theta_5\big(\vec{\xi}_1, g_3(\vec{0})\big) \Big)
=\psi_C\Big( g_1=\vec{\alpha}, g_4(\vec{0}), \theta_5\big(\vec{\xi}_2, g_3(\vec{0})\big) \Big).
\end{align}
By Lemma~\ref{lemma1:non_tight}, we let $g_1^{-1}(\vec{\alpha})\cap g_2^{-1}(\vec{\xi}_1)=\{\vec{a}_1\}$ and $g_1^{-1}(\vec{\alpha})\cap g_2^{-1}(\vec{\xi}_2)=\{\vec{a}_2\}$, where $\vec{a}_1\neq \vec{a}_2$. Together with \eqref{equ1:thm:no_equ_classes}, we obtain
\begin{align}
\vec{a}_1=f(\vec{a}_1, \vec{0})=&~\psi_C\Big( g_1=\vec{\alpha}, g_4(\vec{0}), \theta_5\big(\vec{\xi}_1, g_3(\vec{0})\big) \Big)\\
=&~\psi_C\Big( g_1=\vec{\alpha}, g_4(\vec{0}), \theta_5\big(\vec{\xi}_2, g_3(\vec{0})\big) \Big)=f(\vec{a}_2, \vec{0})=\vec{a}_2,
\end{align}
a contradiction. Thus, we have proved that
\begin{align}\label{equ2:thm:no_equ_classes}
\mL_1=\Big\{ \theta_5\big(\vec{\xi}, g_3(\vec{0})\big):\ \vec{\xi}\in \mA^n \Big\}=\mA^n=\mL.
\end{align}

Now, by \eqref{equ2:thm:no_equ_classes}, we see that $\theta_5\big(\cdot, g_3(\vec{0})\big)$ is a bijection from $\mA^n$ to $\mA^n$. Hence, for the fixed $\vec{\gamma}$ in $\mA^n=\mL$, there exists exactly one $\vec{\xi}$ in $\mA^n$ such that $\theta_5\big(\vec{\xi}, g_3(\vec{0})\big)=\vec{\gamma}.$

Next, we prove that any two distinct $\vec{\alpha}$ and $\vec{\beta}$ in $\mA^n$ are not $(I_1, \vec{\gamma}, \vec{a}_J)$-equivalent. By Lemma~\ref{lemma1:non_tight}, let
\begin{align}
g_1^{-1}(\vec{\alpha})\cap g_2^{-1}(\vec{\xi})=\{\vec{b}_1\},\quad g_1^{-1}(\vec{\beta})\cap g_2^{-1}(\vec{\xi})=\{\vec{b}_2\},
\end{align}
where $\vec{b}_1\neq \vec{b}_2$. With this, we obtain that
\begin{align}
&\psi_C\Big(g_1(\vec{b}_1)=\vec{\alpha}, g_4(\vec{0}), g_5(\vec{b}_1, \vec{0})=\theta_5\big(g_2(\vec{b}_1)=\vec{\xi}, g_3(\vec{0})\big)=\vec{\gamma} \Big)=f(\vec{b}_1, \vec{0})=\vec{b}_1\\
& \neq \vec{b}_2=f(\vec{b}_2, \vec{0})=\psi_C\Big(g_1(\vec{b}_2)=\vec{\beta}, g_4(\vec{0}), g_5(\vec{b}_2, \vec{0})=\theta_5\big(g_2(\vec{b}_2)=\vec{\xi}, g_3(\vec{0})\big)=\vec{\gamma}\Big),
\end{align}
which implies that $\vec{\alpha}$ and $\vec{\beta}$ are not $(I_1, \vec{\gamma}, \vec{a}_J)$-equivalent. Immediately, we see that every vector $\vec{\alpha}$ in $(\mA^n)^{I_1}=\mA^n$ by itself forms an $(I_1, \vec{\gamma}, \vec{a}_J)$-equivalence class. The proof is accomplished.
\end{IEEEproof}

\begin{IEEEproof}[Proof of Theorem~\ref{thm:Cap_N_2_F}]
We proceed to prove that $\mC(\mN_2, F)<1$ by applying our improved upper bound in Theorem~\ref{thm:upper_bound}.

Recall the discussion following Theorem~\ref{thm:Cap_N_2_F} and the definition of $N\big({a}_{L}, \Cl[{a}_J]\big)$ in \eqref{no_finer_eq_cl}. Here, $a_L$ corresponds to $\vec{y}_2$ and we let $\vec{y}_2=\vec{\gamma}$ in $\mA^n$, and each $(I, \vec{a}_J)$-equivalence class $\Cl[{a}_J]$ can be indexed by one and only one image value $\vec{m}\in\mA^{2n}$ under the function $F$ (see the second paragraph before Lemma~\ref{thm:no_equ_classes}). So we write $N\big({a}_{L}, \Cl[{a}_J]\big)$ as $N(\vec{\gamma}, \vec{m})$ in the sequel to simplify notation. By Lemma~\ref{thm:no_equ_classes}, \eqref{no_finer_eq_cl} and \eqref{equ:F_B}, for every $\vec{\gamma}\in \mA^n$ and every image value $\vec{m} \in \mA^{2n}$ under the target function $F=\psi_C$, we have
\begin{align}\label{equ:N_gamma_f}
 N(\vec{\gamma}, \vec{m})=&\#\left\{ (\vec{\alpha}, \vec{\beta})\in \mA^{n}\times \mA^{n}:\ \psi_C\big(g_1=\vec{\alpha},g_4=\vec{\beta}, g_5=\vec{\gamma} \big)=\vec{m} \right\}.
\end{align}
Similar to $N\big(\Cl[{a}_J]\big)$ (see \eqref{equ:N_Cl}), we let
\begin{align}\label{equ:max_N_gamma_f}
N(\vec{m})=&\max_{\vec{\gamma}\in\mA^{n}} N(\vec{\gamma}, \vec{m}).
\end{align}

Next, we will evaluate the value of $N(\vec{m})$. For each $\vec{m}\in \mA^{2n}$, there exists at least one inverse image $(\vec{\alpha}, \vec{\beta}, \vec{\gamma})\in (\mA^n)\times (\mA^n)\times (\mA^n)$ of $\vec{m}$ under $F$, i.e., $F(\vec{\alpha}, \vec{\beta}, \vec{\gamma})=\vec{m}$ (cf. the second paragraph before Lemma~\ref{thm:no_equ_classes}). This implies
\begin{align}\label{equ:N_m_geq_1}
N(\vec{m})\geq 1,\ \forall\ \vec{m}\in \mA^{2n}.
\end{align}

Consider the image value $\vec{m}=\vec{0}$, the all-zero $2n$-vector in $\mA^{2n}$. Clearly, the unique inverse image of $\vec{0}$ under the function $f=\max$ is $(\vec{x}_1=\vec{0}, \vec{x}_2=\vec{0})$. This implies that the inverse image of $\vec{0}$ under the function $F=\psi_C$ is also unique and the unique inverse image is $$\big( g_1(\vec{x}_1=\vec{0}), g_4(\vec{x}_2=\vec{0}), g_5(\vec{x}_1=\vec{0}, \vec{x}_2=\vec{0}) \big),$$
or equivalently,
$$\big( g_1(\vec{x}_1=\vec{0}), g_4(\vec{x}_2=\vec{0}), \theta_5\big(g_2(\vec{x}_1=\vec{0}), g_3(\vec{x}_2=\vec{0})\big) \big).$$


Now, we let $\vec{\gamma}^*=\theta_5\big(g_2(\vec{0}), g_3(\vec{0})\big)$. Then, for each $\vec{\gamma}$ in $\mA^n$ such that $\vec{\gamma}\neq \vec{\gamma}^*$,
\begin{align}\label{equ:110}
\psi_C\big(g_1=\vec{\alpha}, g_4=\vec{\beta}, g_5=\vec{\gamma} \big)\neq \vec{0},
\quad \forall\ (\vec{\alpha}, \vec{\beta})\in \mA^n \times \mA^n,
\end{align}
and so the set
\begin{align*}
\Big\{ \psi_C\big(g_1=\vec{\alpha}, g_4=\vec{\beta}, g_5=\vec{\gamma} \big):\ \forall\ (\vec{\alpha}, \vec{\beta})\in \mA^n \times \mA^n \Big\}\subsetneq \mA^{2n},
\end{align*}
because it does not contain $\vec{0}$. Hence,
\begin{align}
\#\Big\{\psi_C\big(g_1=\vec{\alpha}, g_4=\vec{\beta}, g_5=\vec{\gamma} \big):\ \forall\ (\vec{\alpha}, \vec{\beta})\in \mA^n \times \mA^n \Big\}<2^{2n}.
\end{align}
Together with $|\mA^n\times \mA^n|=2^{2n}$, there must exist two distinct pairs $(\vec{\alpha}_1, \vec{\beta}_1)$ and $(\vec{\alpha}_2, \vec{\beta}_2)$ in $\mA^n\times \mA^n$ such that
\begin{align}\label{equ:psi_C_value}
\psi_C\big(g_1=\vec{\alpha}_1, g_4=\vec{\beta}_1, g_5=\vec{\gamma} \big)=\psi_C\big(g_1=\vec{\alpha}_2, g_4=\vec{\beta}_2, g_5=\vec{\gamma} \big)
\neq \vec{0},
\end{align}
(cf.~\eqref{equ:110}). Denote the common value of $\psi_C$ in \eqref{equ:psi_C_value} by $\vec{m}'$. Immediately, we obtain that $N(\vec{\gamma}, \vec{m}')\geq 2$ (cf.~\eqref{equ:N_gamma_f}), which, together with \eqref{equ:max_N_gamma_f}, further implies that
\begin{align}\label{equ:N_m'}
N(\vec{m}')\geq 2.
\end{align}

Consequently, by \eqref{eq:sum_N_Cl}-\eqref{n_C_Parti_1st} and $\mP_{\widehat{C}}=\big\{\widehat{C}_1=\{ e_8 \}, \widehat{C}_2=\{ e_9 \}\big\}$, we have
\begin{align}\label{defn:n_psi_C}
n_{\widehat{C}}(\mP_{\widehat{C}})=\sum_{\vec{m}\in \mA^{2n}} N(\vec{m}).
\end{align}
Hence, by combining \eqref{equ:N_m'} with \eqref{equ:N_m_geq_1}, it follows from \eqref{defn:n_psi_C} that
\begin{align}
n_{\widehat{C}}(\mP_{\widehat{C}})> 2^{2n}.
\end{align}
From the definition of $n_{C,f}$ in \eqref{n_C_f_1st}, here we obtain $n_{\widehat{C},F}\geq n_{\widehat{C}}(\mP_{\widehat{C}})$. It then follows from our improved upper bound in Theorem~\ref{thm:upper_bound} that
\begin{align}
\mC(\mN_2, F)\leq \dfrac{|\widehat{C}|}{\log_{|\mA^n|}n_{\widehat{C},F}}\leq \dfrac{|\widehat{C}|}{\log_{|\mA^n|}n_{\widehat{C}}(\mP_{\widehat{C}})}
<\dfrac{2}{\log_{2^{n}}2^{2n}}=1.
\end{align}
Therefore, the theorem is proved.
\end{IEEEproof}

In the following example, we use the rate-$2$ network code for $(\mN_1,f)$ depicted in Fig.~\ref{fig:butterfly_network_Part1_b}
 to induce a network computation problem $(\mN_2, F)$, and then illustrate that the computing capacity $\mC(\mN_2,F)$ is strictly smaller than $1$.

\begin{example}
Consider the $(2,1)$ network code depicted in Fig.~\ref{fig:butterfly_network_Part1_b}. According to \eqref{defn:F}, the target function $F$ ($=\psi_C$) induced by the network code is given as follows:
\begin{align*}
F:\ \{0,1\}^3 \ \longrightarrow &\ \ \ \{0,1\}^2\\
({y}_1,{y}_2,{y}_3)  \ \longmapsto &\ \ \
\begin{bmatrix}
\max\{y_1,y_3\}\\ y_2
\end{bmatrix}.
\end{align*}
Clearly, we see that
$g_1(\mA^2)=g_5(\mA^2, \mA^2)=g_4(\mA^2)=\mA$, namely, the domain of $F$ is $\mA \times \mA \times \mA$. Hence, $F$ is well-defined.

For the network computation problem $(\mN_2, F)$ by \eqref{equ_F_A}, the $(I, {a}_J)$-equivalence classes are:
\begin{align*}
\Cl_1=&\left\{ ({y}_1,{y}_2,{y}_3)\in \mA\times\mA\times\mA: F({y}_1,{y}_2,{y}_3)=\left[\begin{smallmatrix}0\\0\end{smallmatrix}\right]   \right\}=\{(0,0,0)\},\\
\Cl_2=&\left\{ ({y}_1,{y}_2,{y}_3)\in \mA\times\mA\times\mA: F({y}_1,{y}_2,{y}_3)=\left[\begin{smallmatrix}0\\1\end{smallmatrix}\right]   \right\}=\{(0,1,0)\},\\
\Cl_3=&\left\{ ({y}_1,{y}_2,{y}_3)\in \mA\times\mA\times\mA: F({y}_1,{y}_2,{y}_3)=\left[\begin{smallmatrix}1\\0\end{smallmatrix}\right]   \right\}=\{(0,0,1),(1,0,0),(1,0,1)\},\\
\Cl_4=&\left\{ ({y}_1,{y}_2,{y}_3)\in \mA\times\mA\times\mA: F({y}_1,{y}_2,{y}_3)=\left[\begin{smallmatrix}1\\1\end{smallmatrix}\right]   \right\}=\{(0,1,1),(1,1,0),(1,1,1)\}.
\end{align*}
Further, by \eqref{equ:N_gamma_f}, the values of $N(\vec{\gamma}, \vec{m})$ for all $\vec{\gamma}\in \mA$ and $\vec{m}\in \mA^2$ are:
\begin{align*}
N\left(0, \left[\begin{smallmatrix}0\\0\end{smallmatrix}\right]\right)
&=\#\left\{({y}_1,{y}_3)\in \mA\times\mA: F({y}_1,{y}_2=0,{y}_3)=\left[\begin{smallmatrix}0\\0\end{smallmatrix}\right]   \right\}=|\{(0,0)\}|=1,\\
N\left(0, \left[\begin{smallmatrix}0\\1\end{smallmatrix}\right]\right)
&=\#\left\{({y}_1,{y}_3)\in \mA\times\mA: F({y}_1,{y}_2=0,{y}_3)=\left[\begin{smallmatrix}0\\1\end{smallmatrix}\right]   \right\}=0,\\
N\left(0, \left[\begin{smallmatrix}1\\0\end{smallmatrix}\right]\right)
&=\#\left\{({y}_1,{y}_3)\in \mA\times\mA: F({y}_1,{y}_2=0,{y}_3)=\left[\begin{smallmatrix}1\\0\end{smallmatrix}\right]   \right\}=|\{(0,1),(1,0),(1,1)\}|=3,\\
N\left(0, \left[\begin{smallmatrix}1\\1\end{smallmatrix}\right]\right)
&=\#\left\{({y}_1,{y}_3)\in \mA\times\mA: F({y}_1,{y}_2=0,{y}_3)=\left[\begin{smallmatrix}1\\1\end{smallmatrix}\right]   \right\}=0,\\
N\left(1, \left[\begin{smallmatrix}0\\0\end{smallmatrix}\right]\right)
&=N\left(1, \left[\begin{smallmatrix}1\\0\end{smallmatrix}\right]\right)=0,\
N\left(1, \left[\begin{smallmatrix}0\\1\end{smallmatrix}\right]\right)=1,\
N\left(1, \left[\begin{smallmatrix}1\\1\end{smallmatrix}\right]\right)=3.
\end{align*}
By \eqref{equ:max_N_gamma_f} and \eqref{defn:n_psi_C}, we obtain
\begin{align*}
N\left(\left[\begin{smallmatrix}0\\0\end{smallmatrix}\right]\right)
=N\left(0, \left[\begin{smallmatrix}0\\0\end{smallmatrix}\right]\right)=1,\ &\
N\left(\left[\begin{smallmatrix}0\\1\end{smallmatrix}\right]\right)
=N\left(1, \left[\begin{smallmatrix}0\\1\end{smallmatrix}\right]\right)=1,\\
N\left(\left[\begin{smallmatrix}1\\0\end{smallmatrix}\right]\right)
=N\left(0, \left[\begin{smallmatrix}1\\0\end{smallmatrix}\right]\right)=3,\ &\
N\left(\left[\begin{smallmatrix}1\\1\end{smallmatrix}\right]\right)
=N\left(1, \left[\begin{smallmatrix}1\\1\end{smallmatrix}\right]\right)=3,
\end{align*}
and consequently,
\begin{align*}
n_{\widehat{C}}(\mP_{\widehat{C}})=1+1+3+3=8,
\end{align*}
which implies that
\begin{align*}
\mC(\mN_2, F)\leq \dfrac{|\widehat{C}|}{\log_{|\mA^n|}n_{\widehat{C},F}} = \frac{2}{\log_2 8}=\frac{2}{3}<1.
\end{align*}
\end{example}

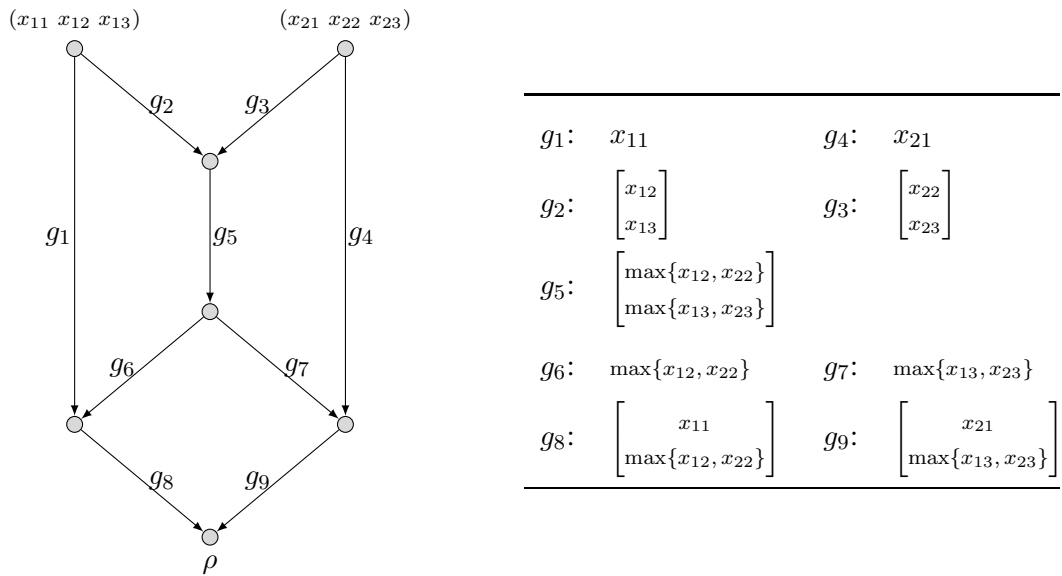
\begin{figure}[!t]
\centering
\begin{minipage}[h]{0.4\textwidth}
 \begin{tikzpicture}[x=0.6cm]
    \draw (-3,0) node[vertex] (1) [label=above:{\scriptsize $(x_{11}\ x_{12}\ x_{13})$}] {};
    \draw ( 3,0) node[vertex] (2) [label=above:{\scriptsize $(x_{21}\ x_{22}\ x_{23})$}] {};
    \draw ( 0,-1.5) node[vertex] (3) [label=left:] {};
    \draw (-3,-5) node[vertex] (4) [label=left:] {};
    \draw ( 3,-5) node[vertex] (5) [label=right:] {};
    \draw ( 0,-3.5) node[vertex] (6) [label=right:] {};
    \draw ( 0,-6.5) node[vertex] (7) [label=below: $\rho$] {};

    \draw[->,>=latex] (1) -- (4) node[midway, auto,swap, left=-1mm] {$g_1$};
    \draw[->,>=latex] (1) -- (3) node[midway, auto, right=-0.5mm] {$g_2$};
    \draw[->,>=latex] (2) -- (3) node[midway, auto,swap, left=-0.5mm] {$g_3$};
    \draw[->,>=latex] (2) -- (5) node[midway, auto, right=-1mm] {$g_4$};
    \draw[->,>=latex] (3) -- (6) node[midway, auto, right=-1mm] {$g_5$};
    \draw[->,>=latex] (6) -- (4) node[midway, auto, left=-0.5mm] {$g_6$};
    \draw[->,>=latex] (6) -- (5) node[midway, auto,swap, right=-0.5mm] {$g_7$};
    \draw[->,>=latex] (4) -- (7) node[midway, auto,swap, right=-0.5mm] {$g_8$};
    \draw[->,>=latex] (5) -- (7) node[midway, auto, left=-0.5mm] {$g_9$};
    \end{tikzpicture}
\end{minipage}%
\quad
\begin{minipage}{0.4\textwidth}
\centering
\begin{tabular}{p{5mm}p{2cm}p{0mm}p{5mm}p{2cm}}
\hline
\specialrule{0em}{2pt}{2pt}
$g_1$: & {$\small x_{11}$} & & $g_4$: & {$\small x_{21}$}\\
\specialrule{0em}{2pt}{2pt}
$g_2$: & ${\scriptsize \begin{bmatrix} x_{12} \\ x_{13} \end{bmatrix}}$ & & $g_3$: & ${\scriptsize \begin{bmatrix} x_{22} \\ x_{23} \end{bmatrix}}$\\
\specialrule{0em}{2pt}{2pt}
$g_5$: & {\scriptsize $\begin{bmatrix} \max\{x_{12}, x_{22}\} \\ \max\{x_{13}, x_{23}\} \end{bmatrix}$} & & \\
\specialrule{0em}{2pt}{2pt}
$g_6$: & {\scriptsize $\max\{x_{12}, x_{22}\}$} & & $g_7$: & {\scriptsize $\max\{x_{13}, x_{23}\}$}\\
\specialrule{0em}{2pt}{2pt}
$g_8$: & {\scriptsize $\begin{bmatrix} x_{11} \\ \max\{x_{12}, x_{22}\} \end{bmatrix}$} & & $g_9$: & {\scriptsize $\begin{bmatrix} x_{21} \\ \max\{x_{13}, x_{23}\} \end{bmatrix}$}\\
\specialrule{0em}{2pt}{2pt}
\hline
\end{tabular}
\end{minipage}
\caption{A coding scheme of the computing rate $3/2$ to compute the binary maximum function of the source messages, i.e., $f(x_1,x_2)=\max\{ x_1, x_2 \}$, where $\mA=\mO=\{0,1\}$.}
  \label{butfly_net_rate_3_over_2}
\end{figure}

\begin{remark}
For the original network computation problem $(\mN, f=\max)$, we give a coding scheme (see Fig.~\ref{butfly_net_rate_3_over_2}) achieving the computing rate $3/2$. Together with Theorem~\ref{thm_butterfly_network_non-tight}, we obtain that for any $(k,n)$ network code that can compute $\max$ over $\mN$, the achievable rate $k/n$ satisfies
\begin{align*}
\frac{3}{2}\leq \frac{k}{n} < 2.
\end{align*}
\end{remark}

\begin{remark}
By symmetry, we have $\mC(\mN, \max)=\mC(\mN, \min)$, where $\min$ is the binary minimum function. So, for any $(k,n)$ network code computing $\min$ over $\mN$, $3/2 \leq k/n < 2$. Note that the function $\min$ is in fact equivalent to the multiplication over $\mathbb{F}_2$, i.e., $f(x_1,x_2)=x_1\cdot x_2$. On the other hand, we note that if $f$ is the summation over $\mathbb{F}_2$, i.e., $f(x_1,x_2)=x_1+x_2$, $\mC(\mN, f)$ can be determined \cite{Koetter-CISS2004}. Therefore, for the function computation over a network, there is an intrinsic difference between addition and multiplication.
\end{remark}

\begin{journalonly}
\section{Future Works}
By a further observation, we believe that if the cuts $C_1$ and $C_2$ can be partitioned further by the method we used herein, better upper bounds are possibly obtained. Thus, an interesting problem is that by using the same analysis approach, what the best we can achieve. Besides that, notice that all cuts $C\in \Lambda(\mN)$ are considered for the later two non-trivial upper bounds, which results in high complexity for searching. Hence, another research problem is how to reduce the search scope largely meanwhile the minimum is still obtained.
\end{journalonly}

\section{Conclusion}\label{sec:concl}
In this paper, we have proved a new upper bound on the computing capacity in network function computation which can be applied to arbitrary target functions and arbitrary network topologies. Our bound not only is a strict improvement over the previous ones, but also is the first tight upper bound on the computing capacity for computing an arithmetic sum over a certain ``non-tree'' network. Previously, only upper bounds for general target functions and network topologies that are tight only for tree networks have been reported.

On the other hand, we have shown that our improved upper bound is in general not achievable. Specifically, the bound is not achievable for computing the binary maximum function over the reverse butterfly network. However, whether the bound is in general asymptotically achievable remains open.

\section*{Acknowledgement}
This work was partially supported by NSFC Grant (Nos. 61771259 and 61471215), the University Grants Committee of the Hong Kong SAR, China (Project No. AoE/E-02/08), and the Vice-Chancellor's One-off Discretionary Fund of CUHK (Project Nos. VCF2014030 and VCF2015007).

\end{document}